\documentclass{aa}  

\usepackage{graphicx}
\usepackage{amsmath,amssymb} 
\usepackage{natbib}
\usepackage{soul}
\usepackage{txfonts}
\usepackage{hyperref}
\usepackage[caption=false]{subfig}
\usepackage[symbol]{footmisc}
\def\oiii{[\ion{O}{iii}]}
\def\sii{[\ion{S}{ii}]}
\def\oii{[\ion{O}{ii}]}
\def\nii{[\ion{N}{ii}]}
\def\hii{\ion{H}{ii}}

\begin{document}

   \title{Spatially resolved electron density in the Narrow Line Region of z<0.02 radio AGNs}

   \author{D. Kakkad
          \inst{1,2,3}
          \thanks{ESO Fellow}
          \and
          B. Groves\inst{4}
          \and
          M. Dopita\inst{4}
          \and
          Adam D. Thomas\inst{4,5}
          \and
          Rebecca L. Davies\inst{6}
          \and
          V. Mainieri\inst{1}
          \and
          Preeti Kharb\inst{7}
          \and
          J. Scharw\"{a}chter\inst{8}
          \and
          E. J. Hampton\inst{4}
          \and
          I-Ting Ho\inst{9}
          }

   \institute{European Southern Observatory, Karl-Schwarzschild-Str. 2, 85748, Garching bei M\"unchen, Germany\\             
   			 \email{dkakkad@eso.org}
         \and
             Ludwig Maximilian Universit\"at, Professor-Huber-Platz 2, 80539, M\"unchen, Germany
         \and
         	European Southern Observatory, Alonso de Cordova 3107, Vitacura, Casilla 19001, Santiago de Chile, Chile
          \and
        	Research School for Astronomy \& Astrophysics, Australian National University, Cotter Road, Weston Creek, ACT 2611, Australia 
        	\and
        	ARC center of Excellence for All Sky Astrophysics in 3 Dimensions (ASTRO 3D), Australia
        	\and
        	Max-Planck-Institut f\"{u}r Extraterrestrische Physik, Giessenbachstrasse, D-85748 Garching, Germany.
        	\and
        	National Centre for Radio Astrophysics, Tata Institute of Fundamental Research, S. P. Pune University Campus, Post Bag 3, Ganeshkhind, Pune 411 007, India.
        	\and
        	Gemini Observatory, Northern Operations Center, 670 N. A’ohoku Pl., Hilo, Hawaii, 96720, USA
        	\and
        	Max Planck Institute for Astronomy, K\"{o}nigtuhl, 17, 69117 Heidelberg, Germany
             }

   \date{Received ?; accepted ?}

 
  \abstract
   {Although studying outflows in the host galaxies of Active Galactic Nuclei (AGN) have become the forefront of extra-galactic astronomy in recent years, estimating the energy associated with these outflows have been a major challenge. Determination of the energy associated with an outflow often involves an assumption of uniform density in the Narrow Line Region (NLR), which span a wide range in literature leading to large systematic uncertainties in energy estimation. }
   { In this paper, we present electron density maps for a sample of outflowing and non-outflowing Seyfert galaxies at z<0.02 drawn from the Siding Spring Southern Seyfert Spectroscopic Snapshot Survey (S7) and understand the origin and values of the observed density structures to reduce the systematic uncertainties in outflow energy estimation.}
   {We use the ratio of the \sii $\lambda$6716,6731 emission lines to derive spatially resolved electron densities ($\lesssim$50-2000 cm$^{-3}$). Using optical Integral Field Unit observations from the WiFeS instrument, we are able to measure densities across the central 2-5 kpc of the selected AGN host galaxies. We compare the density maps with the positions of the \hii ~regions derived from the narrow H$\alpha$ component, ionization maps from \oiii, and spatially resolved BPT diagrams, to infer the origin of the observed density structures. We also use the electron density maps to construct density profiles as a function of distance from the central AGN.}
   {We find a spatial correlation between the sites of high star formation and high electron density for targets without an active ionized outflow. The non-outflowing targets also show an exponential drop in the electron density as a function of distance from the center, with a mean exponential index of $\sim$0.15. The correlation between the star forming sites and electron density ceases for targets with an outflow. The density within the outflowing medium is not uniform and shows both low and high density sites, most likely due to the presence of shocks and highly turbulent medium. We compare these results in the context of previous results obtained from fiber and slit spectra.}
   {}
   \maketitle
%

\section{Introduction}        \label{sect1}

Ionizing radiation from accreting Super-massive black holes (SMBH) at the center of galaxies is believed to illuminate the narrow line regions in the host galaxies of Active Galactic Nuclei (AGN). Unlike the centrally concentrated Broad Line Region (BLR), the Narrow Line Region (NLR) can extend up to kilo-parsec scales \citep[e.g.][]{greene12, sun17}. 

This offers a platform to study the ISM properties of the AGN host galaxies such as electron density, electron temperature and the shape of the ionizing radiation field using rest frame optical emission lines arising from the ionized gas \citep[e.g.][]{kewley02, dopita06, bennert06a, kewley13, Rdavies17}. For low redshift galaxies the spatial extent means the NLR structure can be resolved and hence, the NLR properties can be studied as a function of distance from the galactic nucleus \citep[e.g.][]{heckman90,bennert06a,bennert06b,bennert06c}. 

Outflows in the ionized gas phase in AGN host galaxies are known to exist at both low as well as high redshift \citep[e.g.][]{perna15, villar-martin16, zakamska16, toba17}. Calculation of the energy associated with such outflows often requires an assumption on the value of the electron density in the NLR, the value of which spans between <10-10,000 cm$^{-3}$  \citep[e.g.][]{nesvadba06, liu13, harrison14}. This represents the largest fraction of the systematic uncertainty budget in the outflow energy estimate in the ionized gas phase \citep[e.g.][]{muller-sanchez11,kakkad16,rose17, perna17} and consequently the coupling between the kinetic power of the outflowing gas and the physical properties of the central AGN or star formation remains unclear. This further impedes us from knowing the source of such outflows i.e. whether they are driven by the AGN itself or star bursts. Therefore, it is crucial to understand the electron density values in the ISM of an AGN host galaxy and how the presence of an outflow affects it. 

Most existing electron density studies in galaxies are based on the optical lines \sii $\lambda$6716, 6731 (hereafter \sii ~doublet) or \oii $\lambda$3726, 3729 (hereafter \oii ~doublet) emission lines, as these lines are easily accessible across a wide range of redshift. Both set of lines arise from two closely spaced ``meta-stable" energy levels due to which the relative flux values only depend on the electron density occupying these levels and any degeneracy due to different ions is also removed \citep{osterbrock06}. Although the temperature dependence of electron density is non-negligible, the measurement errors are usually larger than the errors introduced due to the temperature assumption \citep[see also][]{sanders16}. Due to the small wavelength separation, the usage of \oii ~doublet to determine the electron density is limited by the spectral resolution of most of the instruments (including the Wide Field Spectrograph used in this paper, see Sect. \ref{sect3}). Hence, the optical \sii ~doublet is more commonly used to measure electron densities. The \sii ~doublet is sensitive to densities between $\sim$50-2000 cm$^{-3}$, which falls within the typically estimated range of densities in the extended NLR (see Sect. \ref{sect3}). Below and above these densities, the flux ratio of the doublet becomes saturated and are hence unreliable.

Investigation of electron densities have been conducted in both star forming as well as AGN host galaxies in the past decade \citep[e.g.][]{bennert06a, xu07, hainline09, shirazi14, darvish15, sanders16, rose17, perna17, comerford17} most of which claim elevated electron densities linked to the higher star formation rates \citep[e.g.][]{shimakawa15, kaasinen17},  although there have been counter-arguments to this result \citep[e.g.][]{darvish15}. Electron density obtained from one dimensional fiber and two-dimensional slit spectra for high redshift star forming galaxies show an increase in the density by an order of magnitude compared to the low redshift studies \citep[e.g.][]{sanders16, kaasinen17}. Also, recent studies targeting optically selected AGNs from SDSS \citep{perna17} suggested an enhancement of electron density in the outflowing medium.

However, most of these studies at high redshift (z$\sim$1-2) have the observational limitation in resolving the central and the outer regions of the galaxy. The electron density values for these high redshift galaxies are based on "average" density values which have been obtained from integrated spectrum using one-dimensional fiber or two-dimensional slit spectroscopy. The central regions of a galaxy are expected to have a higher density compared to the outer NLR \citep[e.g][]{bennert06a}, and therefore estimation of density from an integrated spectrum has the limitation that they could potentially be contaminated by the higher density nuclear regions of the galaxies.

Galaxies at low-redshift, on the other hand, overcome the limitation of low spatial resolution and therefore, are ideal to perform spatially resolved electron density studies. By combining electron density maps with information from other emission lines such as H$\alpha$ (tracing current star formation) and \oiii ~(tracing AGN ionization), we can understand if and how the density morphology is impacted by star formation or AGN processes or outflows. In this paper, therefore, we focus on the low redshift AGN sample. 

\begin{figure}
\centering
\includegraphics[width=6cm]{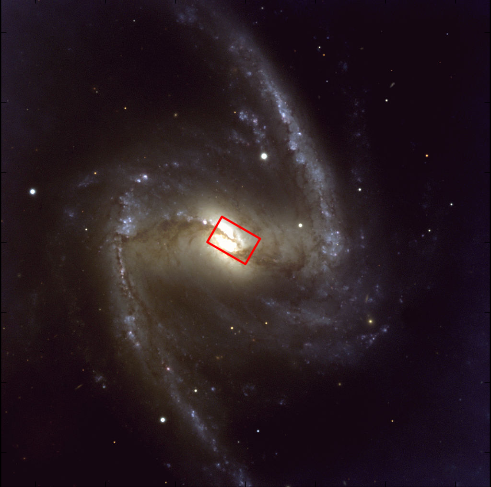}
\caption{A representative figure of the field-of-view (FOV) covered by WiFeS instrument used in the S7 survey. The figure shows a 7$\times$7 arcmin$^{2}$ FORS1/VLT image of NGC 1365, one of the galaxies targeted with the S7 survey, as an example. The WiFeS FOV (38$\times$25 arcsec$^{2}$) is overlaid in red. The S7 FOV for the rest of the targets are presented in \citet{dopita15} and \citet{thomas17}. North is up and East is towards left. Image Credit: ESO. \label{fig:hst_s7_view}}
\end{figure}

There are numerous studies which have presented spatial information on electron density in nearby galaxies \citep[e.g.][]{bennert06a, storchi-bergmann09, sharp10, westmoquette11, westmoquette13, freitas18}. These studies include a combination of starburst galaxies, AGN hosts and Ultra-luminous Infrared Galaxies (ULIRGs) with a fraction of them hosting fast velocity outflows of $\sim$1000 km/s. \citet{bennert06a} presented the electron density as a function of radius from the central AGN in a sample of low redshift Seyfert-1 and Seyfert-2 galaxies using single slit spectroscopy from FORS1/VLT\footnote{FOcal Reducer/low dispersion Spectrograph} and EMMI/NTT\footnote{ESO Multi-Mode Instrument}. It was observed that the density measured using the optical \sii ~doublet saturate at the low limit (<50 cm$^{-3}$) for distance greater than 500 pc from the central AGN. Although \citet{bennert06a} work was based on targets which did not show the presence of an outflow, the drop in density is supported in targets which show an outflow as evident from most of the density maps presented in \citet{sharp10}, \citet{cresci15b} and \citet{freitas18} for instance. Also, the electron densities probed in these maps are consistent with the typical density in the NLR ranging from tens to a few thousand cm$^{-3}$ . 

An important addition to these previous works is to determine the density in the outflowing medium, which otherwise has been measured from the total flux of the \sii ~doublet ratio potentially including contributions from the non-outflowing material. Moreover, it is essential to know whether the density values and morphology change due to the presence of an ionized outflow. To understand the impact of an outflow on the electron density, it is necessary to perform similar spatially resolved analysis on multiple targets and compare the results for targets with and without an active ionized outflow, which forms the basis of this paper. Note that we refer to the density of the warm ionized gas component which is traced by the optical emission lines as \oiii$\lambda$5007 and \sii. Numerous other tracers for electron density exists such as trans-auroral emission lines \citep{rose17,spence18} and FeII doublet \citep{storchi-bergmann09} which trace a different gas phase and are sensitive to regions with higher density (> 10$^{4}$ cm$^{-3}$). 

Today, large samples of optical and near-Infrared IFU spectra of extra-galactic sources building up through IFU surveys such as SAMI \citep{croom12, green17}, MaNGA \citep{bundy15, wake17}, S7 \citep{dopita15,thomas17}, KASHz \citep{harrison15}, KMOS$^{3D}$ \citep{wisnioski15} and KDS \citep{turner17} allow us to do such a study on a large sample of galaxies. In the following paper, we present the electron density distribution for a sub-sample of low redshift AGN host galaxies derived from the S7 survey with the aim to derive their electron density profiles as a function of radius and identify the mechanisms- Star formation, AGN ionization or outflows- which drive the observed density patterns.

The paper is structured as follows: In Sect. \ref{sect2}, we briefly summarize the properties of S7 sub-sample used in this paper, its observations and data reduction procedures. Sect. \ref{sect3} gives a comprehensive overview of the analysis procedures namely line fitting, map construction and error estimation. The resulting spectra, star formation and extinction maps, \oiii ~flux maps and electron density profiles have been reported in Sect. \ref{sect4}. We discuss the processes leading to observed electron density patterns and the implications of these results in Sect. \ref{sect5}. The main conclusions of the paper are presented in Sect. \ref{sect6}.

Throughout this paper, we adopt a $\Lambda$-CDM cosmology with H$_{0}$=70 km/s/Mpc, $\Omega_{m}$ = 0.3, $\Omega_{\Lambda}$ = 0.7 and $\Omega_{r}$=0.0.

\section{Sample: the Siding Spring Southern Seyfert Spectroscopic Snapshot Survey} \label{sect2} 

\begin{figure*}
\centering
\includegraphics[width=11.5cm, height=6cm]{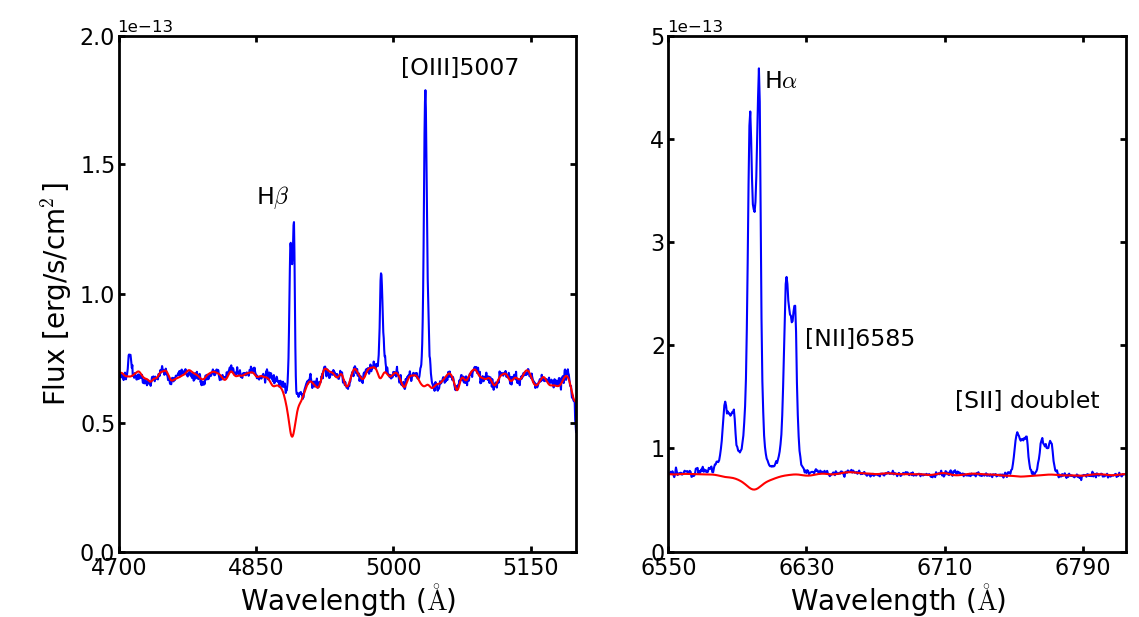}
\caption{An example of stellar continuum fitting in the ``blue'' and ``red'' spectrum of NGC 1365, one of the galaxies targeted with S7 survey. The flux calibrated raw spectrum is shown in blue colour and the stellar continuum fit to the spectrum using pPXF is shown in red colour. The left panel shows the integrated ``blue" spectrum extracted from the entire S7 FOV and zoomed in the region around H$\beta$ and \oiii$\lambda$4959, 5007 emission lines and the right panel shows the spectra in the ``red" cube zoomed in the region around \nii$\lambda$6549, 6585, H$\alpha$ and the \sii ~doublet. The X-axis in both the panels shows the observed wavelength. \label{fig:stellar_cont}}
\end{figure*} 

Our sample is drawn from the {\it Siding Spring Southern Seyfert Spectroscopic Snapshot Survey} \citep[S7,][]{dopita15, thomas17}, which is an Integral Field Spectroscopic survey of nearby Seyfert and LINER galaxies at z$<$0.02. Below we briefly describe the basic characteristics of this survey. We refer the reader to the data release papers, \citet{dopita15} and \citet{thomas17} for further details on the sample selection, observing strategy and object-by-object description of the survey. 

The observations for the S7 survey were carried out using the Wide Field Spectrograph (WiFeS) on ANU 2.3 m telescope at the Siding Spring Observatory. The field of view is 38 $\times$ 25 arcsec$^{2}$ with the nucleus centered on the field of view and orientation close to the major axis of the galaxy (Fig. \ref{fig:hst_s7_view} shows NGC 1365 as an example). The spectra have a resolution of R=3000 for 340$<\lambda <$560 nm and a moderately high spectral resolution of R=7000 ($\sim$50 km/s) in the wavelength range 530$<\lambda <$710 nm (simply referred to as "blue" and "red" spectrum respectively, hereafter). The average spatial resolution for the observations is $\sim$1-2 arcsec corresponding to the seeing, with exposure times in the range 800-1000 s. The data reduction was performed using Python pipeline PYWIFES \citep{childress14} which provides calibrated sky subtracted data cubes re-sampled on a 1$\times$1 arsec$^{2}$ pixel scale. The error budget on flux calibration is up to 4\%. The survey is aimed at probing the ionizing mechanism of the extended narrow line region (ENLR) of nearby radio-selected AGNs with 20 cm radio flux density greater than 20 mJy. Given the redshift distribution of the sample, the spatial resolution of the data is $\lesssim$400 pc/arcsec and diagnostic lines such as H$\beta$, \oiii$\lambda$5007, H$\alpha$, \nii$\lambda$6585 and \sii$\lambda$6716,6731 used in this paper are well within the spectral range of WiFeS.  Such a wide range in the coverage of emission lines makes this survey extremely useful to disentangle the contribution to these line fluxes due to star formation or AGN or both using spatially resolved diagnostic diagrams.

The S7 sub-sample used in this paper is chosen to cover a wide range in morphology of star forming regions as traced by the H$\alpha$ emission line and NLRs, as traced by the \oiii ~emission. The sample includes a mixture of Seyfert 1s, Seyfert 2s and AGN-Starburst (SB) composite systems. We also filter out targets in which the \sii ~doublet is contaminated by skylines, stellar continuum and atmospheric absorption features and where the S/N of the \sii ~emission line is less than 5. This is because a contamination of the \sii ~lines and/or poor data quality can significantly affect the density measurement as explained in Sect. \ref{sect3}. The 13 targets presented in this paper lie in the low redshift range of 0.0038-0.0164 which probes a maximum physical scale of $\sim$5 kpc corresponding to the WiFeS FOV. The average seeing of all the targets used in the paper lie between 1.1-2.2 arcsec which corresponds to 1-2 pixels. These properties have been summarized in Table \ref{table:sample}. Such spatial resolution and the FOV therefore allow us to probe the density structures within the AGN host galaxy from scales of $\sim$100 pc-5 kpc. 

Three of the targets, namely NGC 1068, NGC 1266 and NGC 2110, are classified as having an ionized outflow based on the presence of broad components in the \oiii 5007 profile of the S7 data. The broad components of the \oiii ~profiles in the three targets show fast outflows with FWHM > 500 km/s compared to the other targets which require only narrow components with FWHM < 200 km/s (See Sect. \ref{sect3.1} for more details on emission line fitting). A number of works in literature report the presence of outflows in other targets as well. The difference in the classification of the presence of outflows presented in this paper from the previous works may lie in the different aperture probed by the S7 data (the central 35$\times$25 arcsec$^{2}$) and the above mentioned definition of outflow. Hereafter, we refer to the presence of ionized outflows based on the definition mentioned above as interpreted from the S7 data. This paper focuses on the analysis of the warm ionized gas in the NLR traced by the \oiii ~line. A comparison of electron density values between the outflowing and non-outflowing targets would also help us infer if the presence of outflows makes any difference to the density morphology and their values.

\begin{table*}
\centering                        
\begin{tabular}{c c c c c c}     
\hline\hline        
Target & RA & DEC & $z$ & Type$^{a}$ & Seeing FWHM\\
& (J2000) & (J2000) & & & (arcsec) \\
\hline          
   NGC 613     & 01:34:18.10 & -29:25:03.00 & 0.0049 & Seyfert 2 & 2.2\\
   NGC 1365   & 03:33:36.41 & -36:08:24.00 & 0.0056 & Seyfert 1 & 1.1\\     
   NGC 1672   & 04:45:42.19 & -59:14:51.00 & 0.0046 & SB+Seyfert 2 & 1.9\\
   NGC 4303  & 12:21:55.40 & +04:28:31.00 & 0.0053 & SB/LINER & 1.3\\
   NGC 4691   & 12:48:13.01 & -03:19:59.16 & 0.0038 & SB        & 1.4\\
   NGC 5990  & 15:46:16.49 & +02:24:56.16 & 0.0125 & SB+PSB+Seyfert2 & 1.9 \\ 
   NGC 6000  & 15:49:49:60 & +29:23:13.00 & 0.0072 & SB & 1.8\\
   NGC 6221   & 16:52:46.00 & -59:13:01.00 & 0.0047 & SB+Seyfert 2 & 1.8\\
   NGC 7469  & 23:03:16.00 & 08:52:24.50 & 0.0164 & Seyfert 1 & 1.3\\ 
   NGC 7496  & 23:09:47.3 & -43:25:40.50 & 0.0056 & SB+Seyfert 2 & 1.2\\
   NGC1068* & 02:42:40.70 & +00:00:47.16 & 0.0038 & Seyfert 2 & 1.2\\
   NGC1266* & 03:16:00.70 & -02:25:37.92 & 0.0072 & LINER & 1.0\\
   NGC2110* & 05:52:11.40 & -07:27:23.04 & 0.0078 & Seyfert 2 & 1.4\\
\hline                                  
\end{tabular}
\caption{Properties of the S7 sub-sample used in this paper. $^{a}$SB = Starburst, PSB = Post-starburst, LINER=Low-Ionization Nuclear Emitting Region, *Sample with ionized outflows inferred from the \oiii$\lambda$5007 emission line in the optical spectra of the S7 survey
\label{table:sample}}
\end{table*}

\section{Data analysis}   \label{sect3}

As described in the introduction, the \sii ~doublet and \oii ~doublet are ideal tracers of the electron density in a warm ionized medium. The spectral resolution of WiFeS does not allow us to de-blend the \oii ~doublet, hence in this paper we measure electron densities using the flux ratio of the \sii ~doublet. For the typical range of values of the electron density in the NLR i.e. $<$50-2000 cm$^{-3}$\footnote[1]{The lower and the upper values of 50 cm$^{-3}$ and  2000 cm$^{-3}$ are not strict limits and depend on the quality of the data used and can be anywhere from <50-100 cm$^{-3}$ to 10,000 cm$^{-3}$.}, the theoretical \sii ~doublet line ratio is close to unity between 0.7 and 1.45 \citep{osterbrock06}. Since the inferred electron density is highly susceptible to minute changes in the flux ratio, we applied a S/N cut of 5 in the overall spectrum as well as the individual pixels of each target during the analysis. The following section provides a detailed description of the analysis tools used in this paper. We show the spectra of NGC 1365 as an example and we follow the same procedure for the rest of the targets. 

\subsection{Emission line fitting} \label{sect3.1}

{\it Stellar continuum subtraction:} The raw data cubes include contribution from skylines and underlying stellar continuum which have to be carefully subtracted to compute accurate emission line fluxes. We use the IDL emission line fitting toolkit, LZIFU \citep{ho16,hampton17} for this purpose which also makes use of the pPXF code \citep{cappellari04} to subtract the stellar continuum. We refer the reader to \citet{ho16} and \citet{hampton17} and for further details. The stellar continuum fitting for NGC 1365 is shown as an example in Fig. \ref{fig:stellar_cont} where the raw integrated spectrum extracted from the S7 FOV for the ``red'' and the ``blue'' cubes is shown in blue colour and the stellar continuum fit is over-plotted in red colour. We follow the same procedure for the stellar continuum subtraction for the rest of the targets presented in this paper. 

The emission line fitting then follows a two-step procedure. First, we perform the line fitting in an integrated spectrum, the results of which is used as a prior for fitting across every pixel in the data cube. The robustness of the line fitting across every spaxel was checked by mapping the residuals of the fit across the wavelengths with \sii ~emission. The extraction of the integrated spectrum was done from the entire S7 field of view except the bottom two rows of pixels due to instrumental defects. This spectrum was then used to derive an ``average'' value of electron density integrated over all the pixels. We perform Gaussian line fitting primarily for H$\beta$, \oiii ~$\lambda$4959, 5007, \nii ~$\lambda$6549, 6585, H$\alpha$ and the \sii ~doublet. We used the IDL fitting routine MPFIT \citep{markwardt12} which fits emission lines based on minimizing-$\chi^2$ method, for simultaneous fitting of the nuclear continuum and the emission lines. The emission lines were reproduced using multiple Gaussian components. Initially a single Gaussian component was fitted and the addition of further Gaussian components depended on whether parameters relevant for this work (e.g. the total flux and width of \sii ~doublet) became stable within $\lesssim$10\%. We classify the individual Gaussian components as narrow or broad based on their width (FWHM). For targets without an active outflow, only narrow Gaussian components with width $\leq$200 km/s was required to fit the \sii ~emission lines while for the targets with an outflow, additional broad Gaussian components with width $\geq$500 km/s was required. The width of the narrow component is consistent with the fact that the narrow emission might originate from the disk and also similar to the values observed in literature previously \citep[e.g.][]{harrison14, brusa15a}.

\begin{figure*}
\centering
\subfloat{\includegraphics[width=6cm]{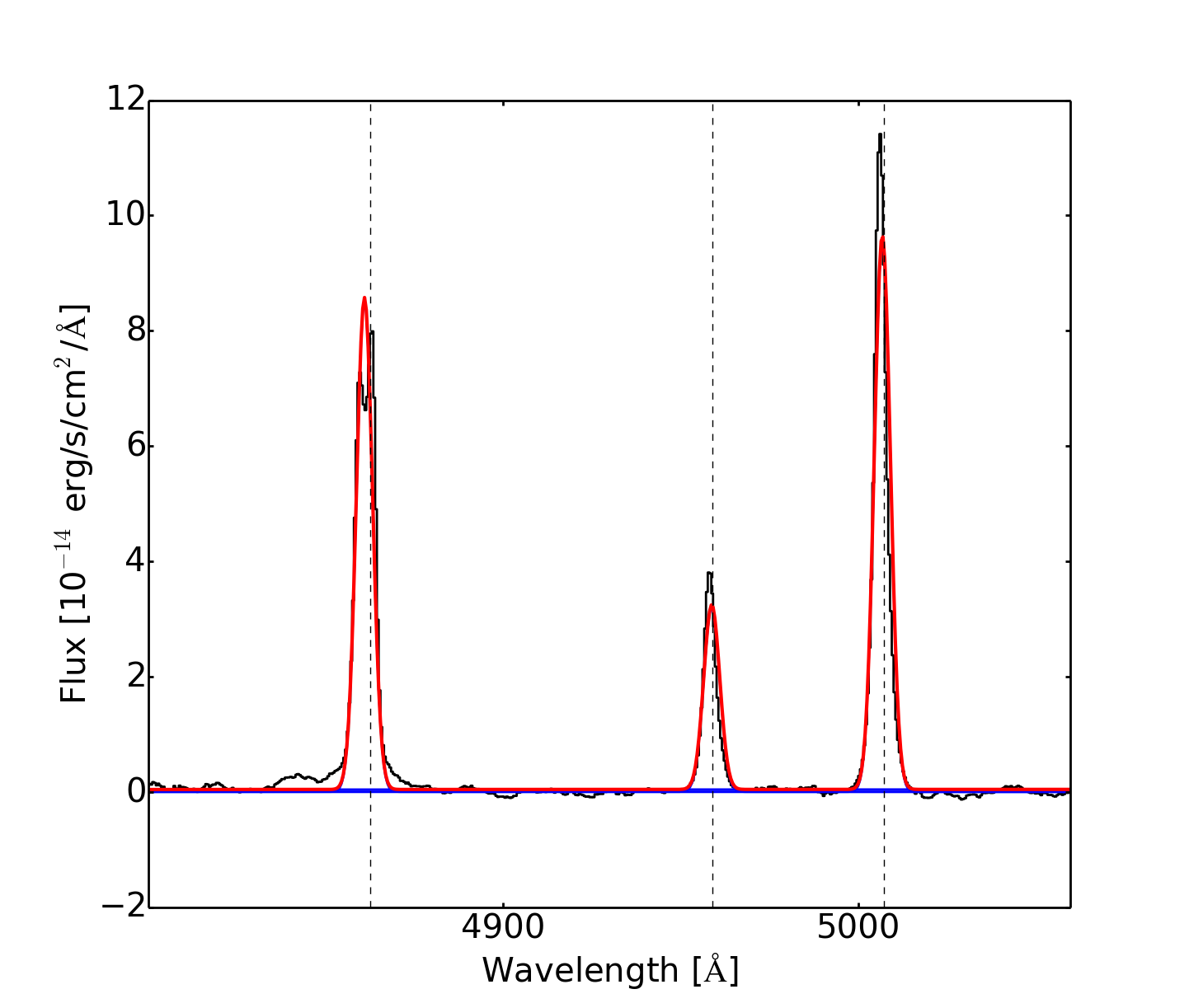}}
\subfloat{\includegraphics[width=6cm]{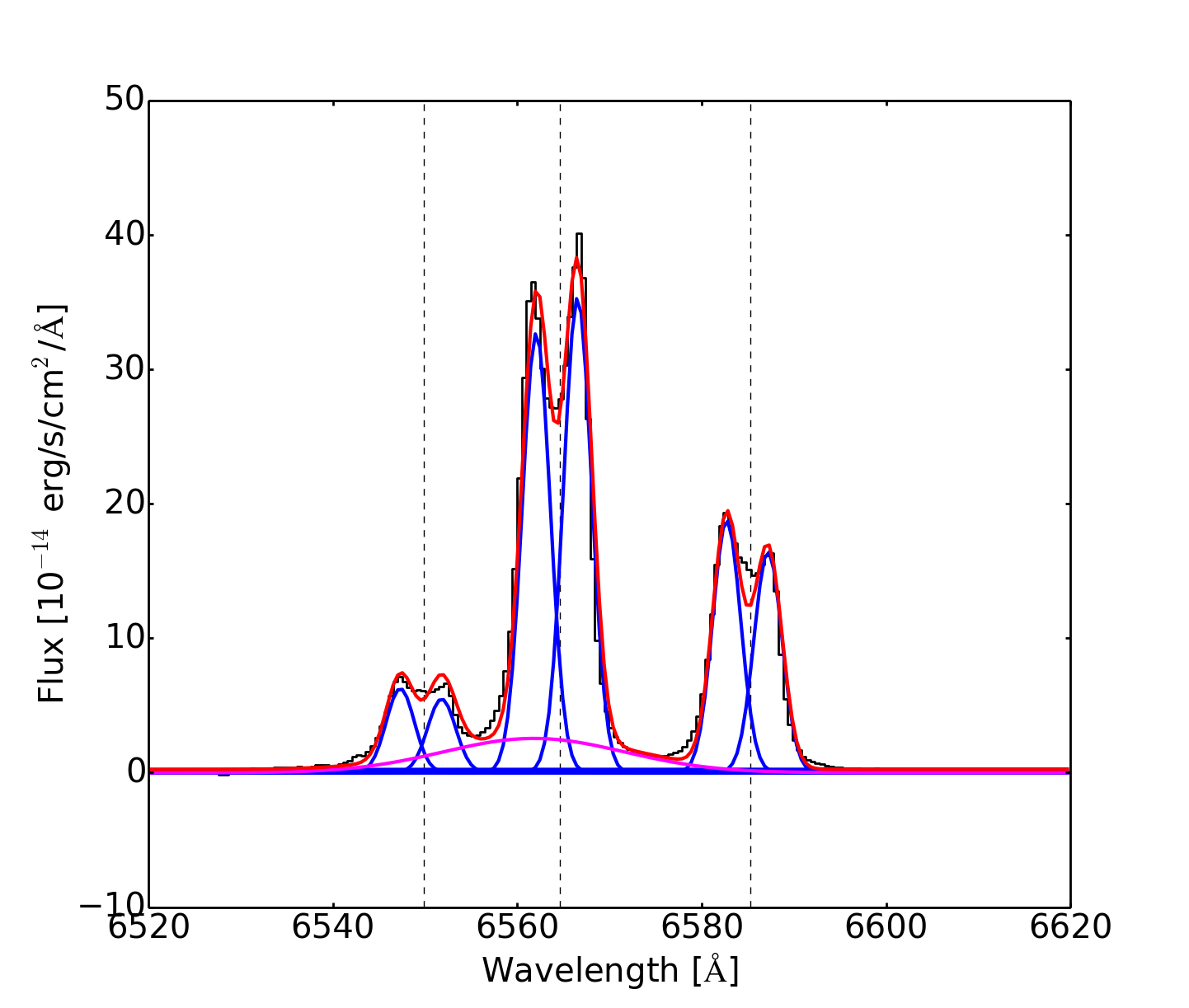}}
\subfloat{\includegraphics[width=6cm]{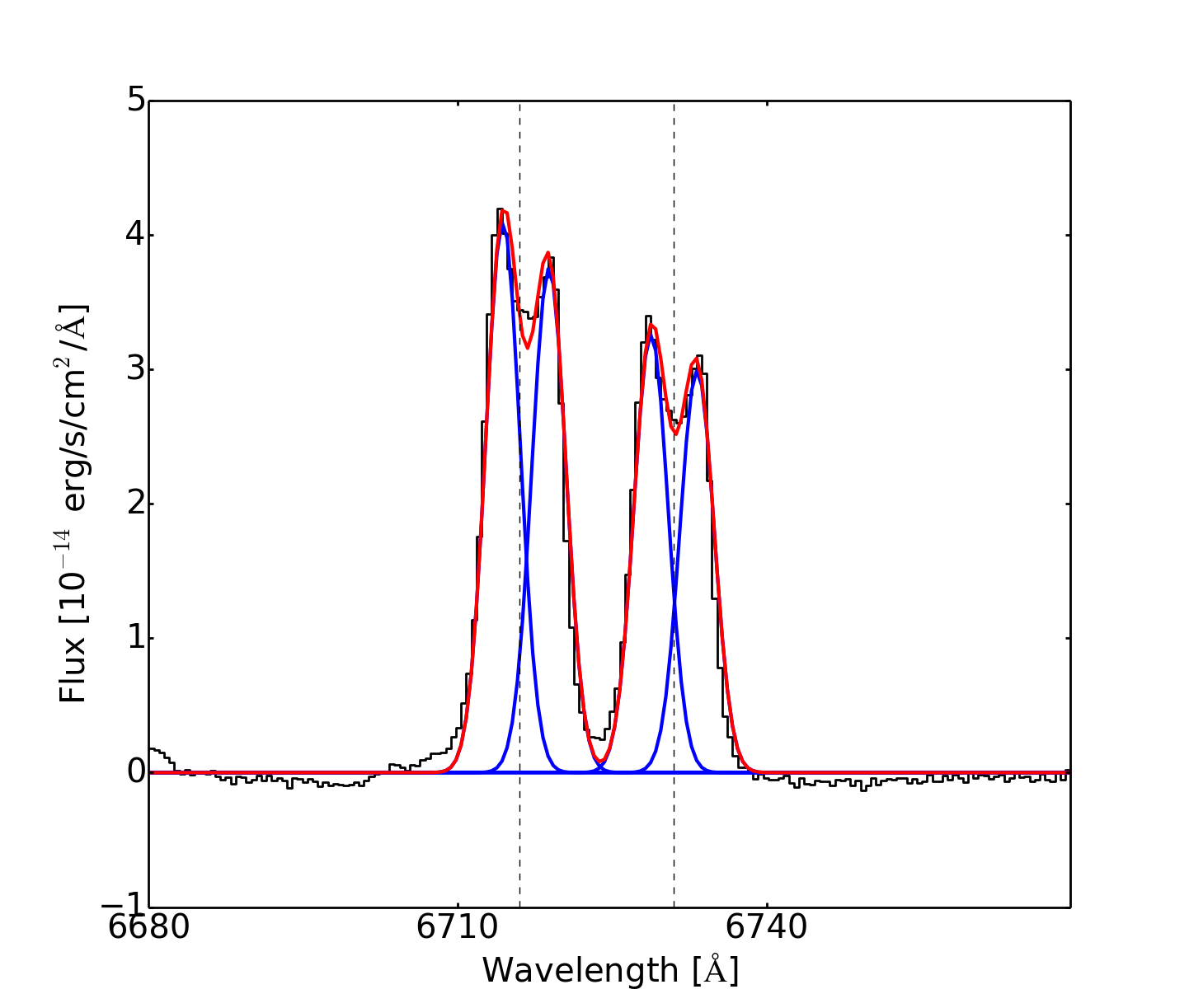}}
\caption{The stellar continuum subtracted integrated spectrum of NGC 1365 showing the H$\beta$ and \oiii$\lambda$4959,5007 emission lines (left panel),  \nii$\lambda$6548,6585 and H$\alpha$ (middle panel) and \sii ~doublet (right panel). All the wavelengths on the X-axis are the rest frame wavelengths. Underlying black curve shows the observed spectrum and the overlaid red curve shows the overall model. The blue curves show the individual narrow Gaussian components and the magenta curve shows the broad H$\alpha$ Gaussian component. The vertical dotted lines show the expected position of H$\beta$, \oiii, \nii, H$\alpha$ and \sii ~doublet emission lines inferred from the redshift of the targets.\label{fig:spectrum_NGC1365}}
\end{figure*}

\begin{figure*}
\centering
\includegraphics[width=14cm]{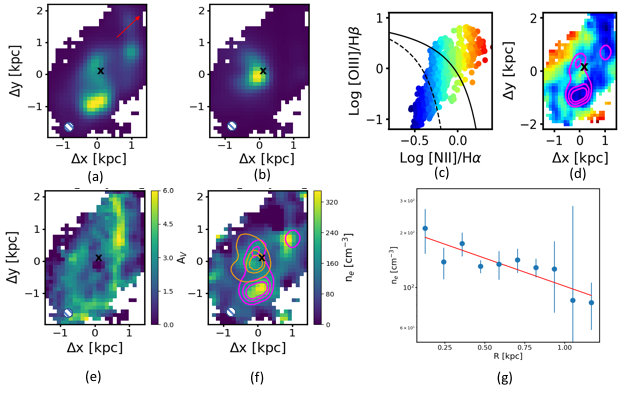}
\caption{Analysis maps of NGC 1365 shown here as an example. (a) \& (b): Narrow H$\alpha$ and \oiii$\lambda$5007 flux maps showing the star forming and NLR ionization regions in the host galaxy of NGC 1365 respectively. The red arrow indicates the North direction and the black cross in the center shows the position of the AGN.  (c) shows the BPT diagram with each point representing the data from every pixel in the S7 FOV, colour coded on the scale of the \nii/H$\alpha$ flux ratio while (d) shows the corresponding position of these data points on the image. The spatially resolved BPT map shows the regions ionized by AGN and/or shocks and star formation. The solid and the dotted black line in panel (c) correspond to the extreme starburst \citet{kewley01} line and the \citet{kauffmann03} classification respectively. The magenta contours represent the H$\alpha$ emission from panel (a) at levels 30\%, 50\% and 80\% of the peak respectively, showing that the clumps in the H$\alpha$ map are indeed sites of star formation.  (e): Extinction map derived from H$\beta$/H$\alpha$ line ratio showing the obscuring regions in NGC 1365. (f): Electron density map derived from \sii $\lambda$6716/\sii$\lambda$6731 line ratio respectively (see Sect. \ref{sect3.2} for further details). The overlaid magenta contours represent the narrow H$\alpha$ emission from panel (a) and the orange contours represent the \oiii ~emission from panel (b) at levels 30\%, 50\% and 80\% of the peak, respectively. The contour locations show that the electron density is high in sites of star formation. (g): Electron density profile as a function of radius from the central AGN (the black cross in the maps). The blue data points represent the profiles derived from the map itself and red line shows the best fit model assuming an exponential profile (see Sect. \ref{sect4} for more details). The stripped ellipse in the lower left corner of the maps in (a), (b), (e) and (f) illustrates the PSF during the observations.\label{fig:maps_NGC1365}}
\end{figure*}

 \begin{figure*}
 \centering
 \subfloat{\includegraphics[width=4.5cm, height=4cm]{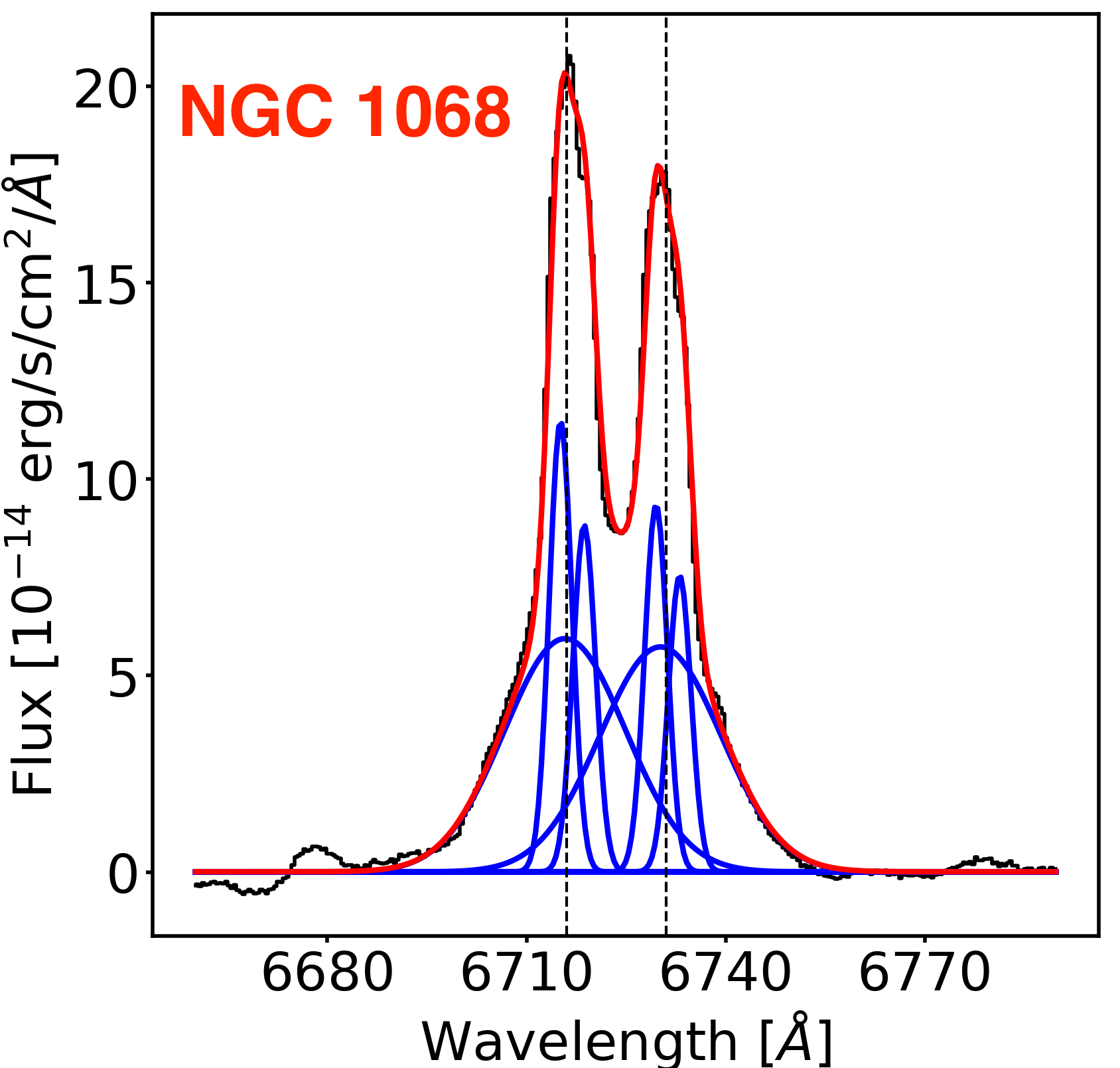}}
\subfloat{\includegraphics[width=4.5cm, height=4cm]{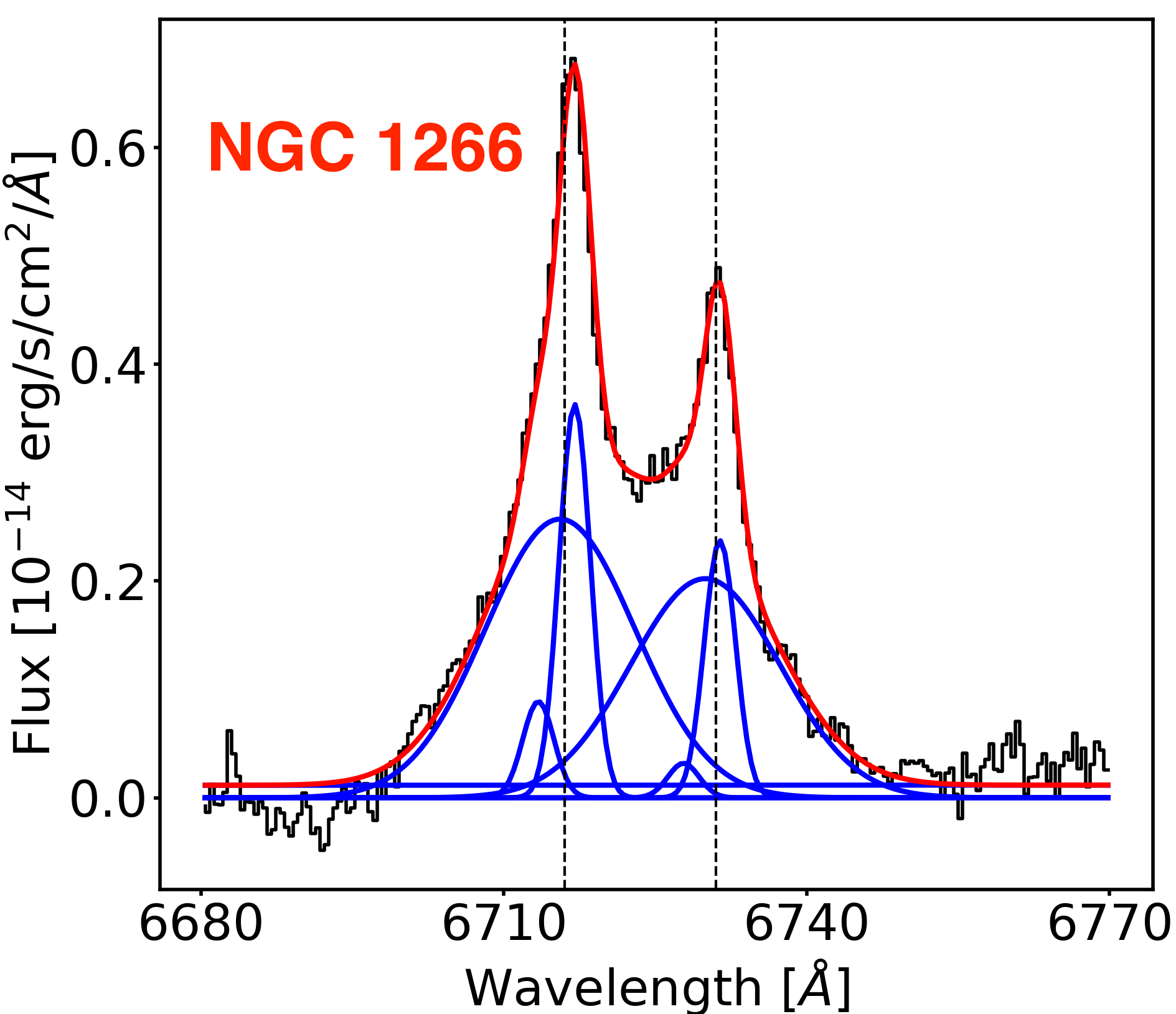}}
\subfloat{\includegraphics[width=4.5cm, height=4cm]{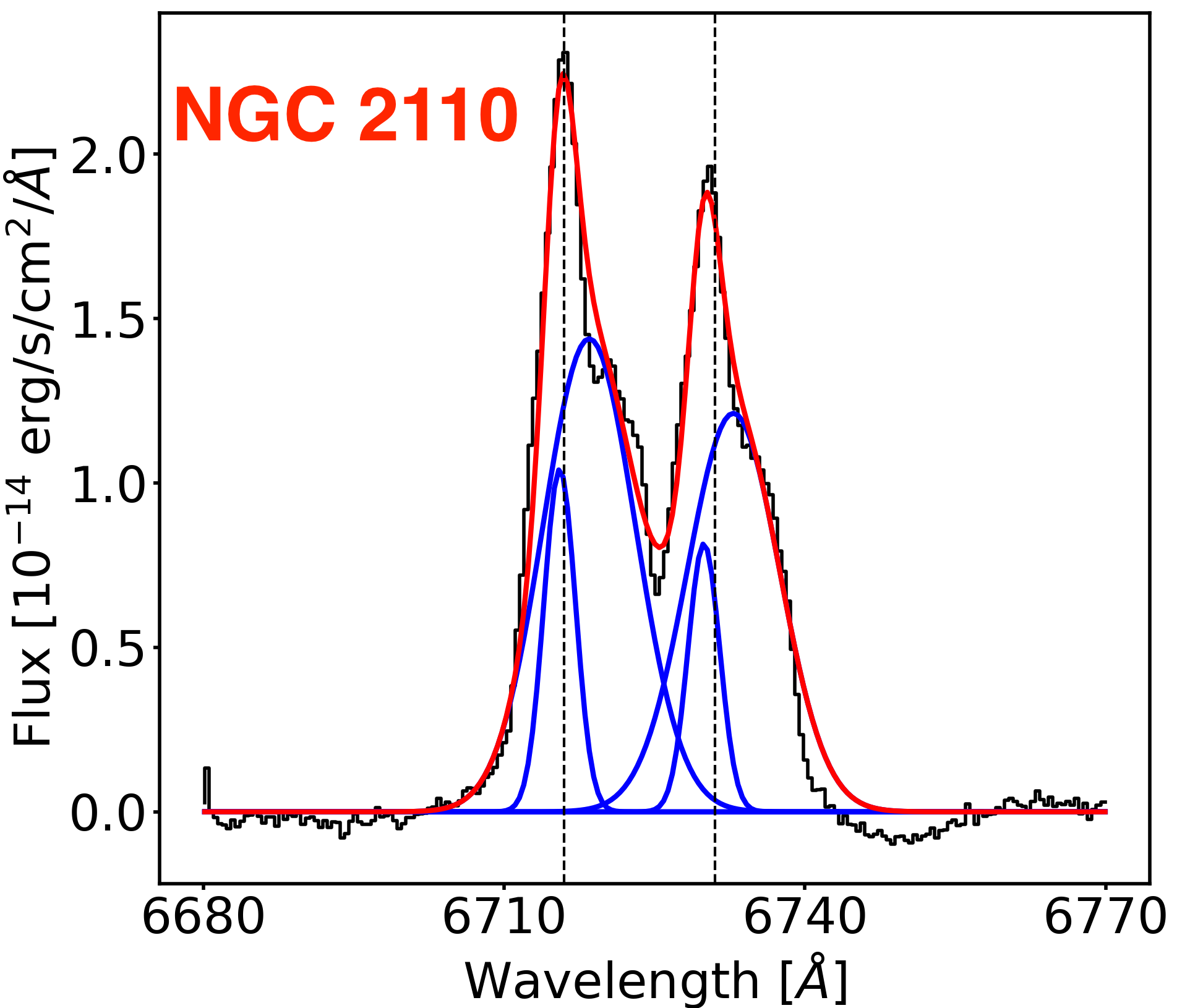}}\\
 \caption{Integrated spectrum around the \sii ~doublet of NGC 1068 (left), NGC 1266 (center) and NGC 2110 (right), the targets showing the presence of an ionized outflow as interpreted from the S7 data. The color convention is the same as in Fig. \ref{fig:spectrum_NGC1365}. \label{fig:sii_spectrum_outflow}}
\end{figure*}

\begin{table*}
\centering                        
\begin{tabular}{c c c c}     
\hline\hline        
Target & $\mathrm{n_{e}}^{a}$ & $\beta^{b}$ & Log $p/k^{c}$ \\
& (cm$^{-3}$) & & (K/cm$^{3}$)\\
\hline          
   NGC 613  & 111~$\pm$30 & 0.25~$\pm$0.09  & 6.61 \\
   NGC 1365 & 148~$\pm$15 & 0.05~$\pm$0.04  & 6.66 \\
   NGC 1672 & 163~$\pm$16 & 0.08~$\pm$0.03  & 6.74 \\
   NGC 4303 & 110~$\pm$60 & 0.15~$\pm$0.08  & 6.66 \\
   NGC 4691 & ~92~$\pm$15 & 0.12~$\pm$0.04  & 6.89 \\
   NGC 5990 & 133~$\pm$23 & 0.08~$\pm$0.07  & 6.86 \\
   NGC 6000 & 210~$\pm$40 & 0.11~$\pm$0.05   & 7.22 \\
   NGC 6221 & 168~$\pm$26 & 0.19~$\pm$0.10  & 7.10 \\
   NGC 7469 & 270~$\pm$60 & 0.22~$\pm$0.05  & 7.15 \\
   NGC 7496 & 200~$\pm$40 & 0.14~$\pm$0.11  & 6.96 \\
   NGC 1068* & 334$\pm$50 & 0.12~$\pm$0.05 & 7.56\\
   NGC 1266* & 133$\pm$95 & -0.16~$\pm$0.13 & 6.86\\ 
   NGC 2110* & 199$\pm$48 & 0.10~$\pm$0.05 & 6.89\\
\hline                                  
\end{tabular}
\caption{$^{a}$Electron density calculated using the total flux ratio of each component of the \sii ~doublet from the spectrum binned across the S7 field-of-view, $^{b}$Exponential index for the fit between electron density and distance from the center as presented in Eq. \ref{eq:power_law}. See Sect. \ref{sect4} for more details. $^{c}$The central pressure calculated using Eq. \ref{eq:pressure}. *Targets with known ionized outflows inferred from the \oiii$\lambda$5007 emission line in the optical spectra of the S7 survey.
\label{table:results}}
\end{table*}

A few constraints were employed while performing the line fitting to avoid any degeneracy in the multiple component fits. The line widths were constrained to be always greater than 100 km/s to avoid any spurious detection due to the presence of skylines. The line widths of each narrow component were coupled to each other assuming all these components originate from the same region of the host galaxy. In most of our sub-sample, the H$\alpha$ profile shows the presence of a relatively faint but a very broad component possibly originating from the BLR. These broad lines from H$\alpha$ complex are not extended enough to affect the fluxes of the \sii ~doublet \citep[see][]{thomas17}. We also coupled the relative emission line flux ratios for the doublets \oiii$\lambda$4959,5007 and \nii$\lambda$6549,6585 based on theoretical values \citep[e.g.][]{storey00, dimitrijevic07}.

\begin{figure*}[h!]
\centering
\subfloat{\includegraphics[width=7.6cm, height=4.2cm]{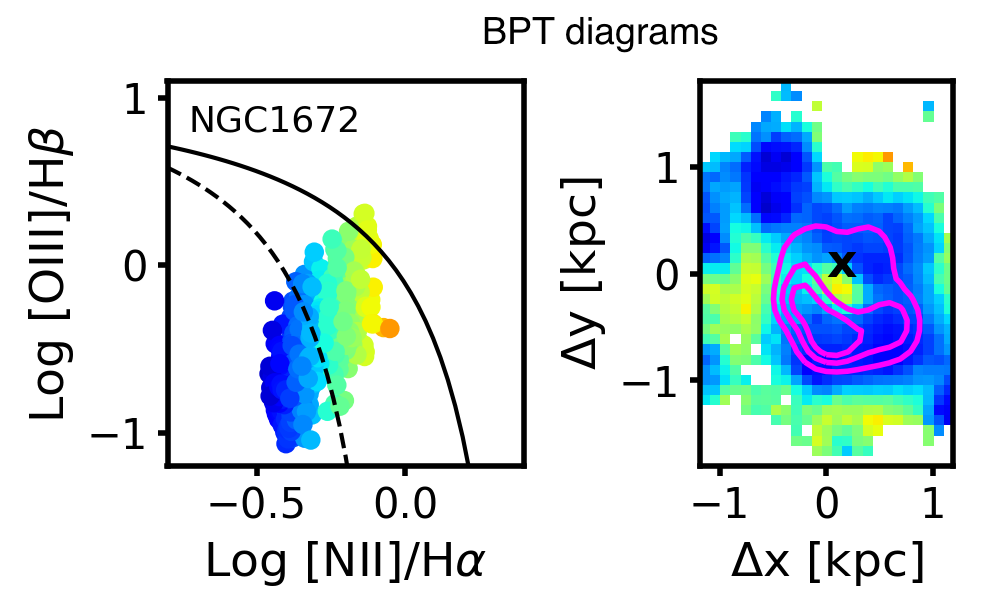}}
\subfloat{\includegraphics[width=7.8cm, height=4.3cm]{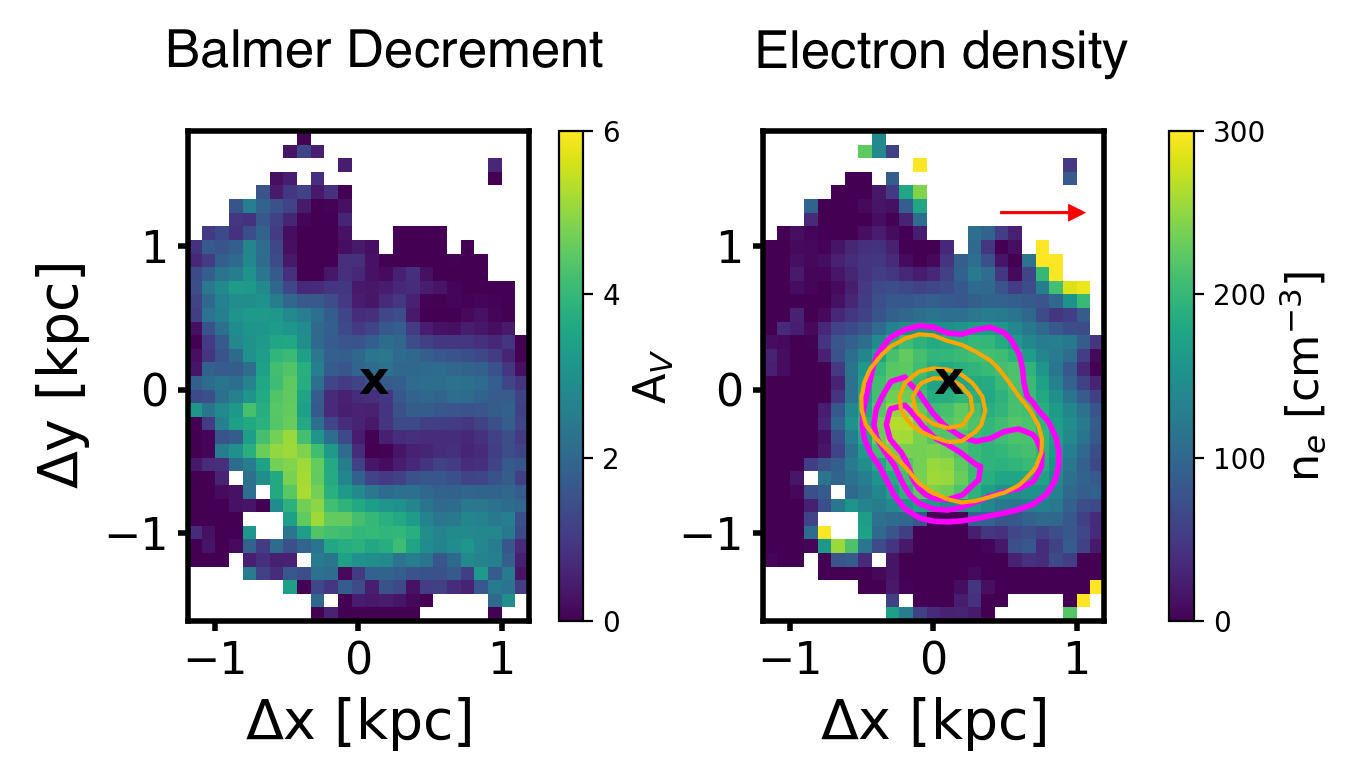}}\\
\subfloat{\includegraphics[width=7.6cm, height=3.8cm]{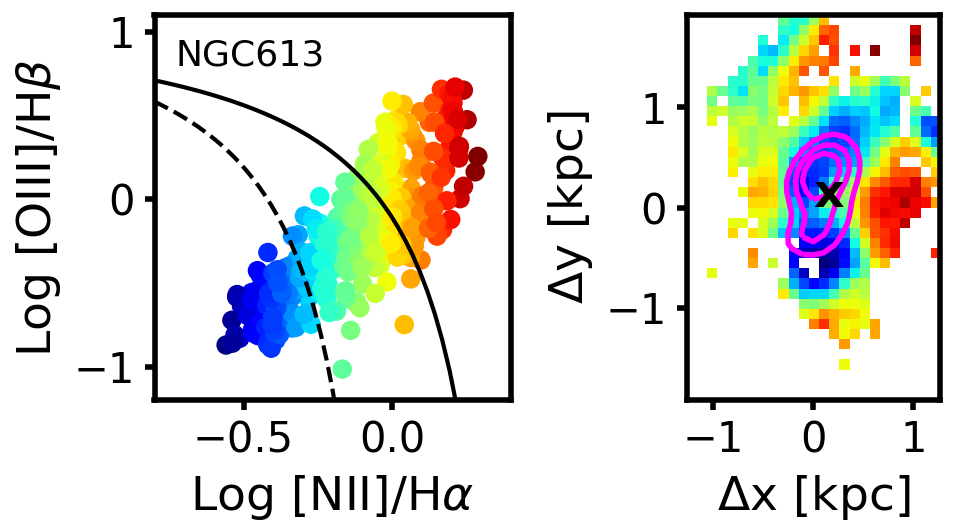}}
\subfloat{\includegraphics[width=7.8cm, height=3.8cm]{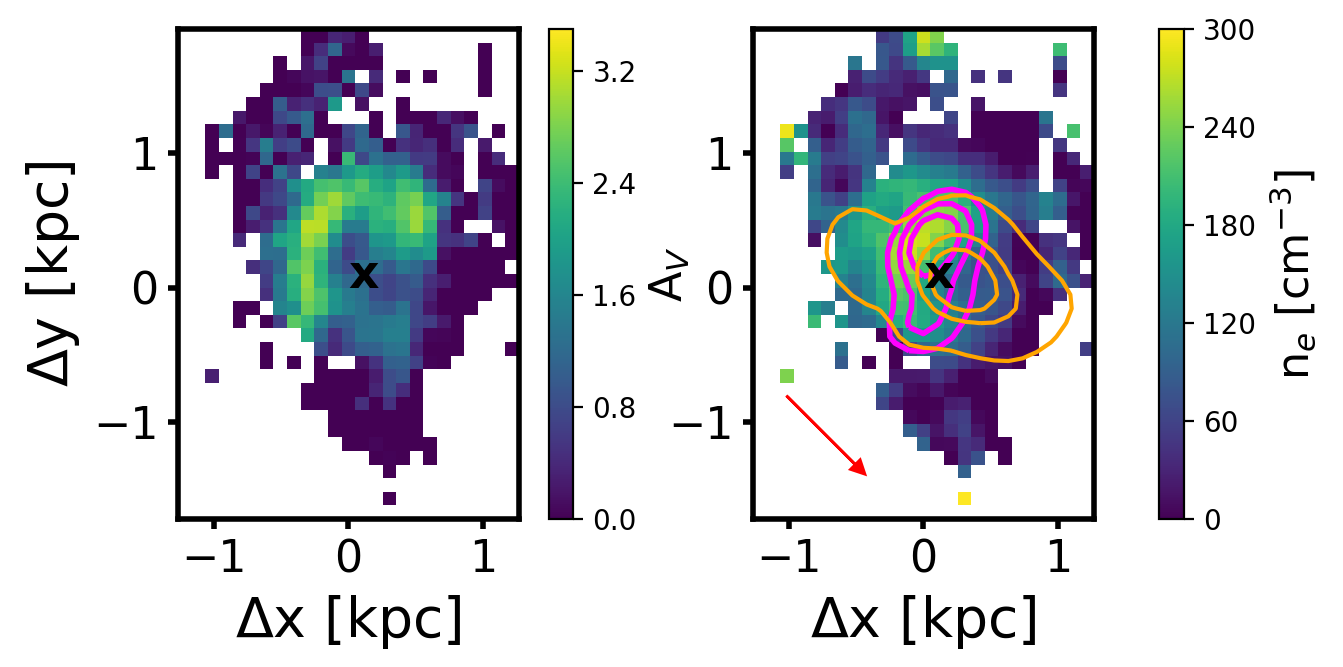}}\\
\subfloat{\includegraphics[width=7.6cm, height=3.8cm]{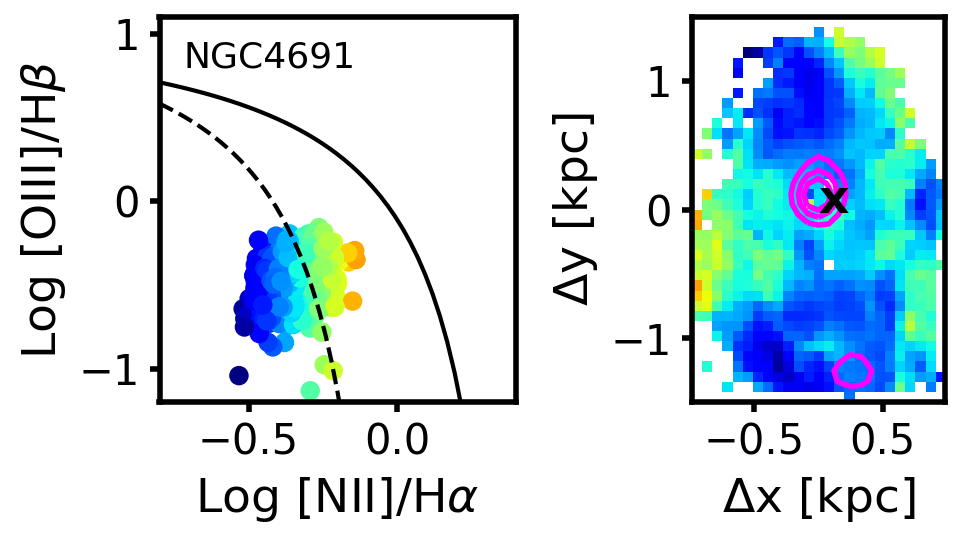}}
\subfloat{\includegraphics[width=7.8cm, height=3.8cm]{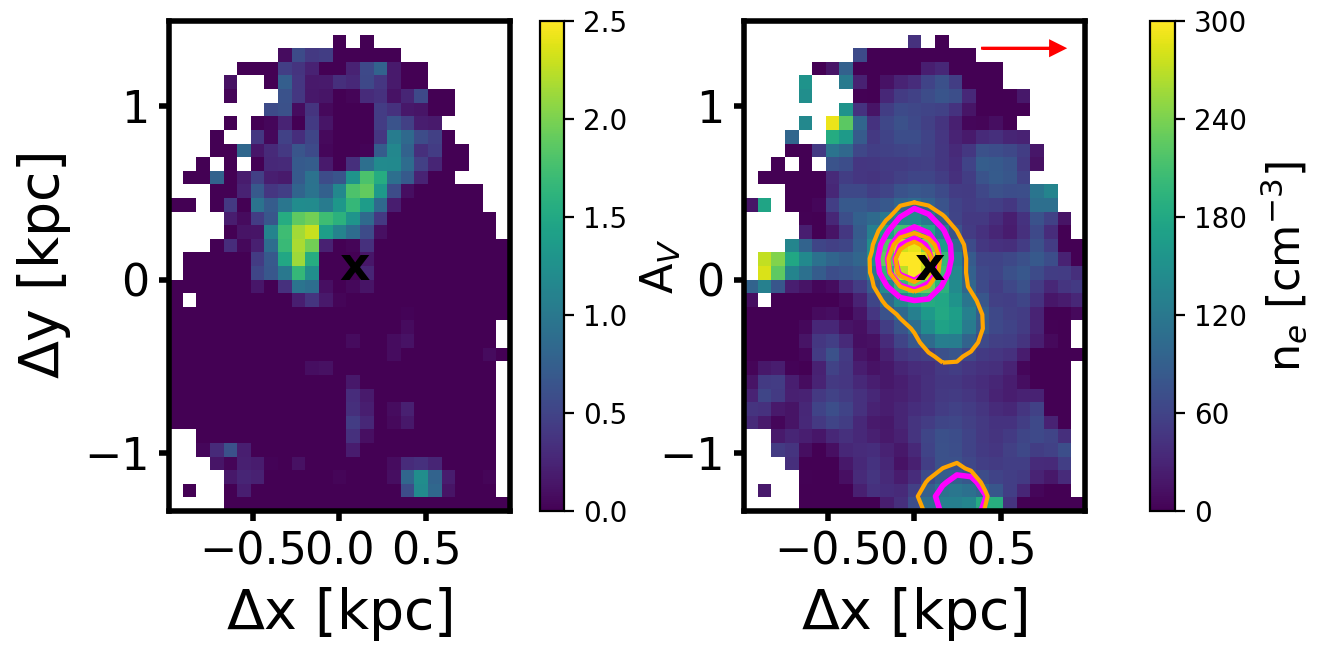}}\\
\subfloat{\includegraphics[width=7.6cm, height=3.8cm]{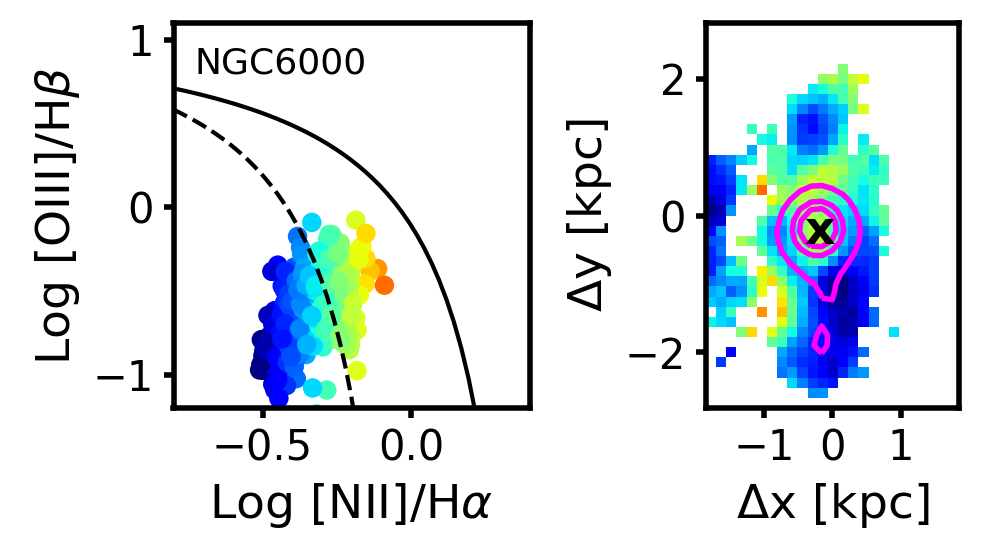}}
\subfloat{\includegraphics[width=7.8cm, height=3.8cm]{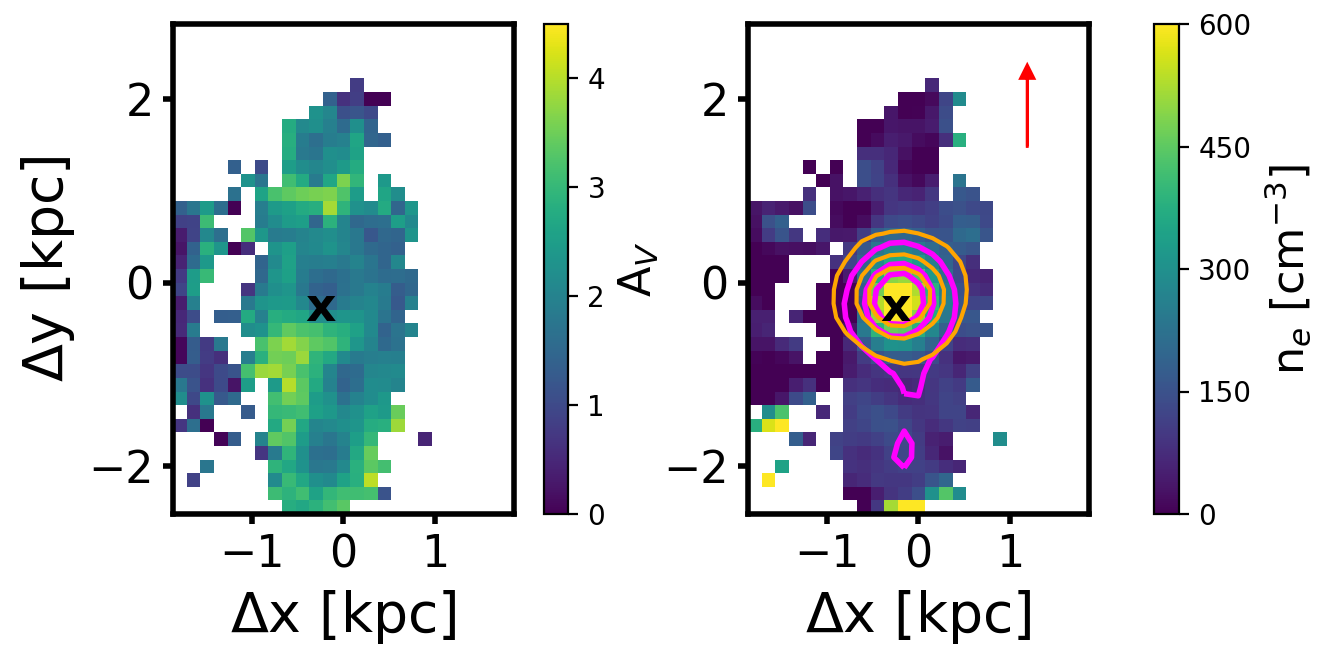}}\\
\subfloat{\includegraphics[width=7.6cm, height=3.8cm]{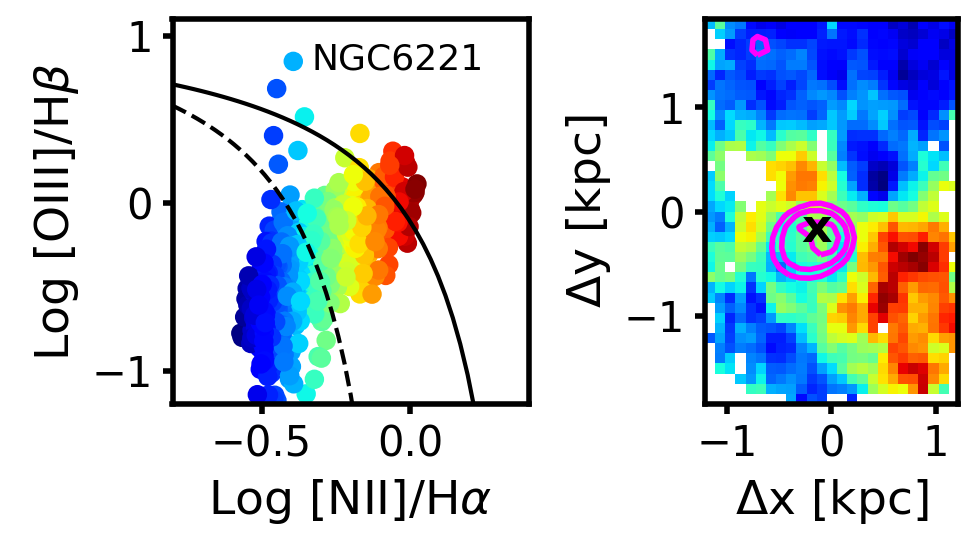}}
\subfloat{\includegraphics[width=7.8cm, height=3.8cm]{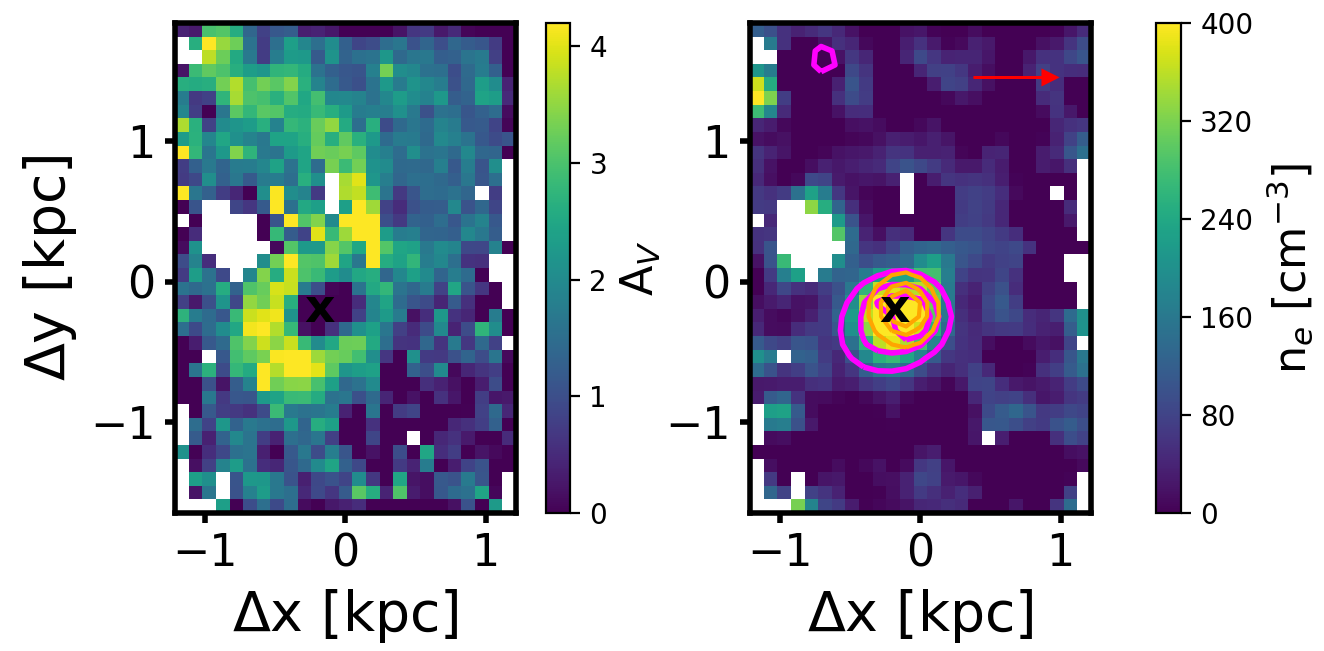}}\\
\caption{From left to right: Data points from each pixel in the S7 FOV on a BPT diagram colour coded based on the \nii/H$\alpha$ flux ratio, corresponding spatially resolved BPT map, extinction map and density map for targets (top to bottom): NGC 1672, NGC 613, NGC 4691, NGC 6000 and NGC 6221. Like in Fig. \ref{fig:maps_NGC1365}, the magenta and orange contours represent the H$\alpha$ and \oiii ~emission at levels 30\%, 50\% and 80\% of the peak, respectively. The red arrow in the density maps indicate North. In all the cases, the narrow H$\alpha$ contours in magenta almost fall under the star forming region of the BPT diagram showing that there is no strong contribution from AGN in these areas. All the targets in this figure also show a spatial correlation between sites of high star formation (from the narrow H$\alpha$ contours) and high density.  \label{fig:density_maps_correlation}}
\end{figure*}

With the emission line models for the integrated spectrum in hand, the multiple component line fitting was performed for each pixel across the FOV to create flux and density maps (explained in Sect. \ref{sect3.2}). While performing the pixel-by-pixel line fitting, we allow minor variations, with respect to the values obtained in the integrated emission line fitting, in the line width and its centroid to account for the shifts due to rotation of the host galaxy. The fluxes were always kept positive as we do not expect or assume the presence of absorption features once the stellar continuum has been subtracted. In the presence of broad lines from the BLR, we subtract this broad component (mainly in the case of H$\alpha$) with the same centroid and width as that obtained in the integrated spectrum and only allowing variations in the flux. The consistency of the results obtained using this method was checked by comparing with the maps obtained with LZIFU fitting procedure \citep[see][]{ho16, hampton17} and the final density maps (Sect. \ref{sect3.2}) do not significantly differ with both methods. 

\begin{figure*}
\centering
\subfloat{\includegraphics[width=7.6cm, height=4.2cm]{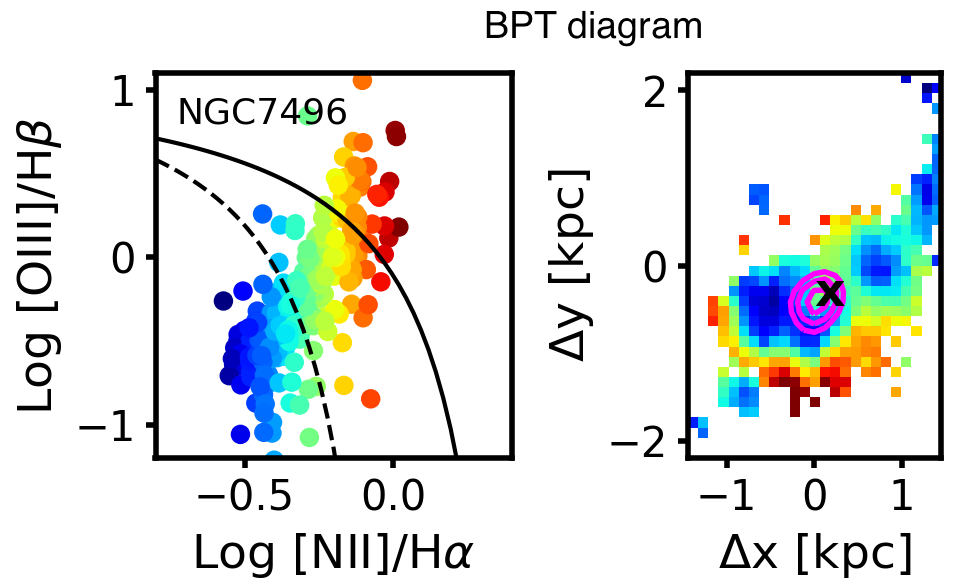}}
\subfloat{\includegraphics[width=7.8cm, height=4.2cm]{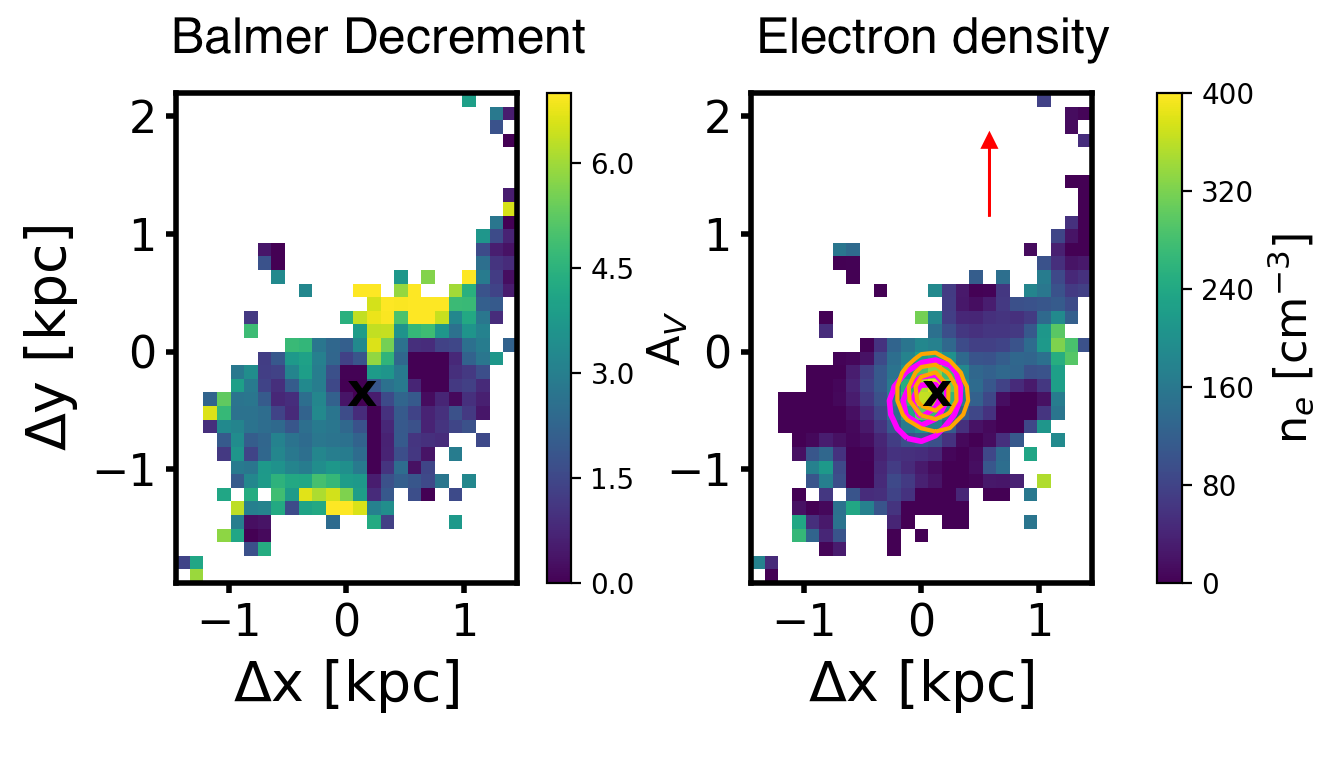}}\\
\subfloat{\includegraphics[width=7.6cm, height=3.8cm]{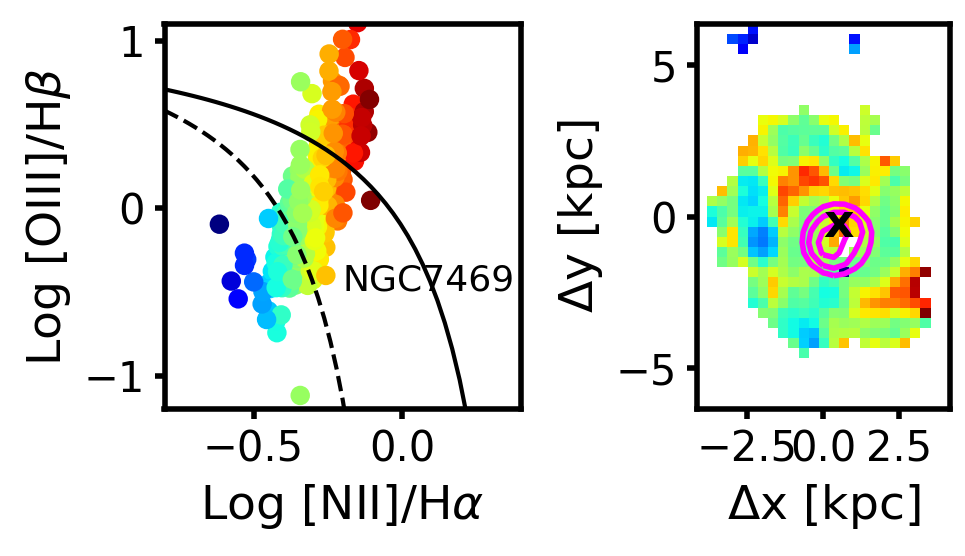}}
\subfloat{\includegraphics[width=7.8cm, height=3.8cm]{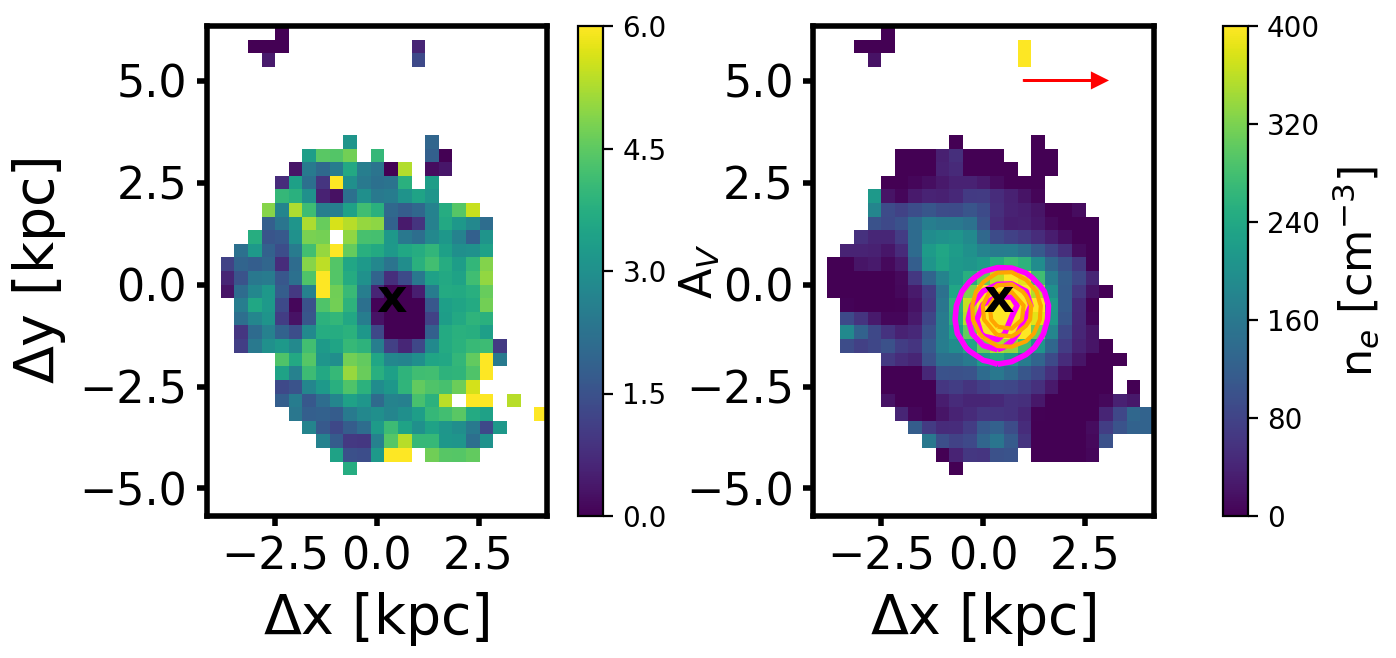}}\\
\subfloat{\includegraphics[width=7.6cm, height=3.8cm]{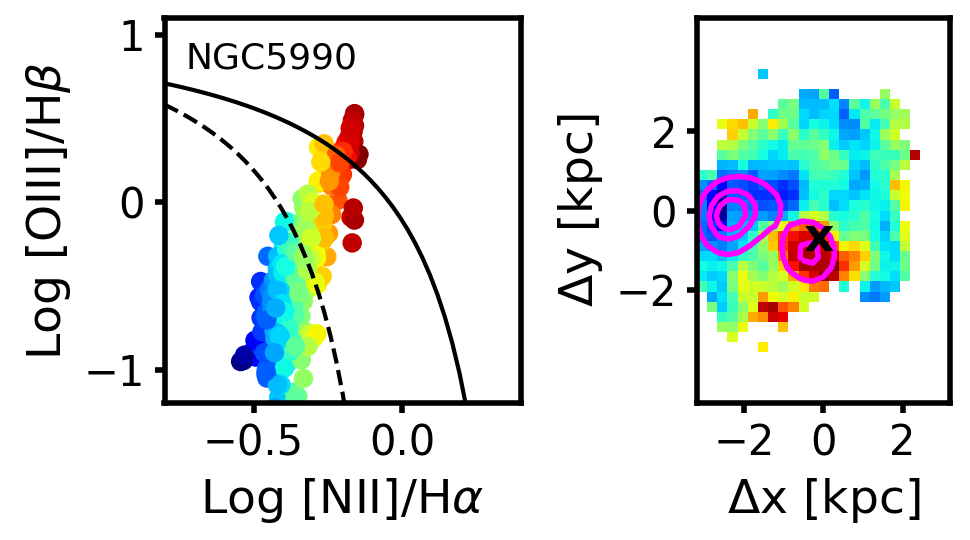}}
\subfloat{\includegraphics[width=7.8cm, height=3.8cm]{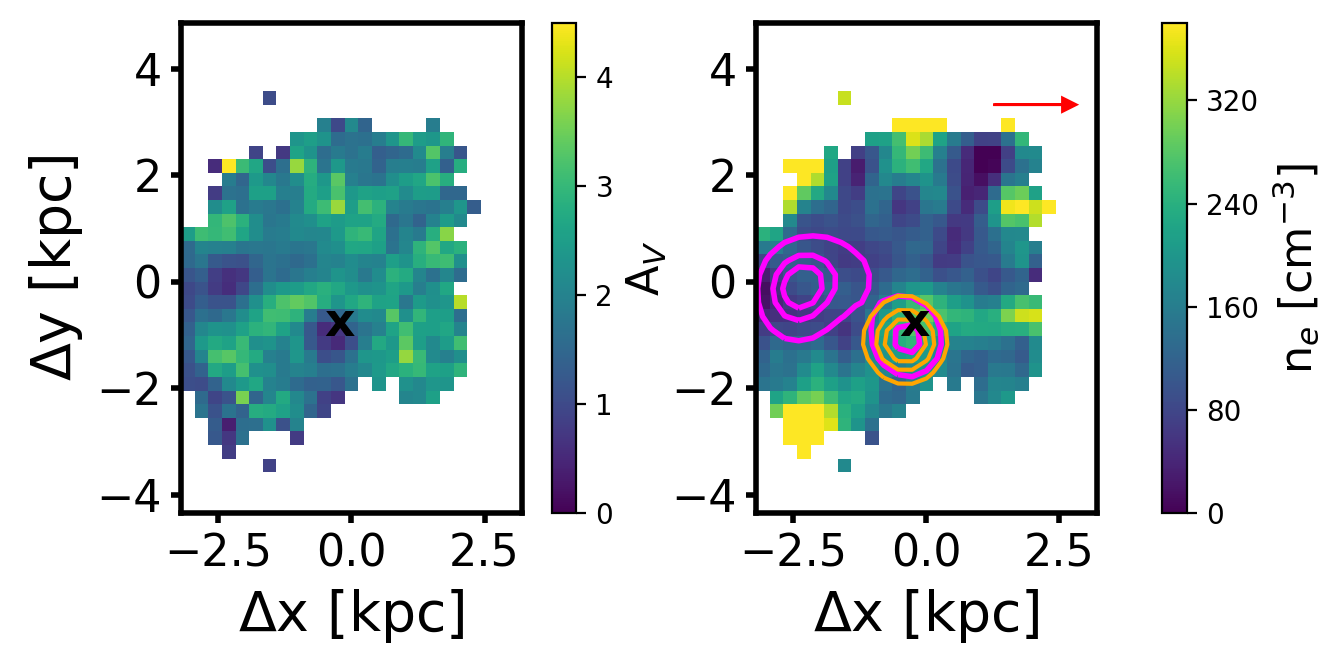}}\\
\subfloat{\includegraphics[width=7.6cm, height=3.8cm]{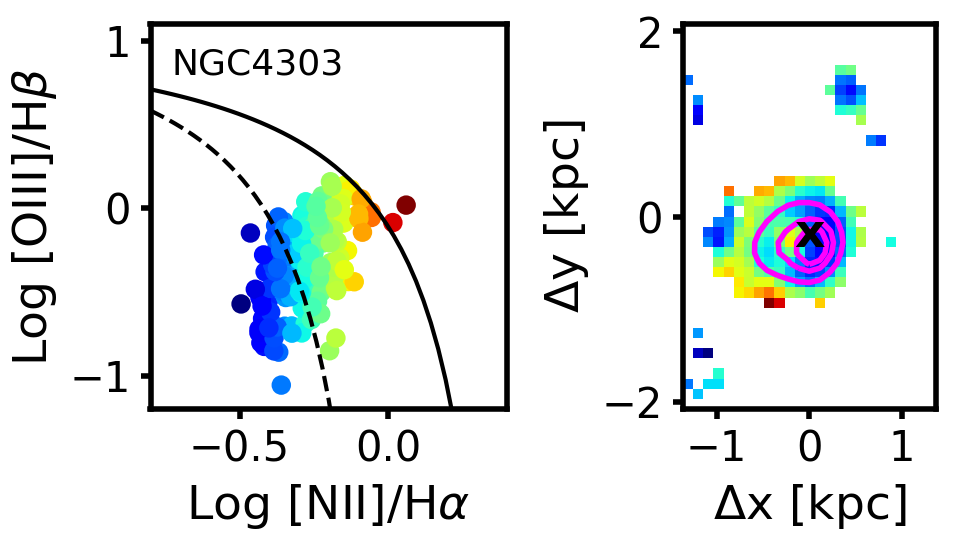}}
\subfloat{\includegraphics[width=7.8cm, height=3.8cm]{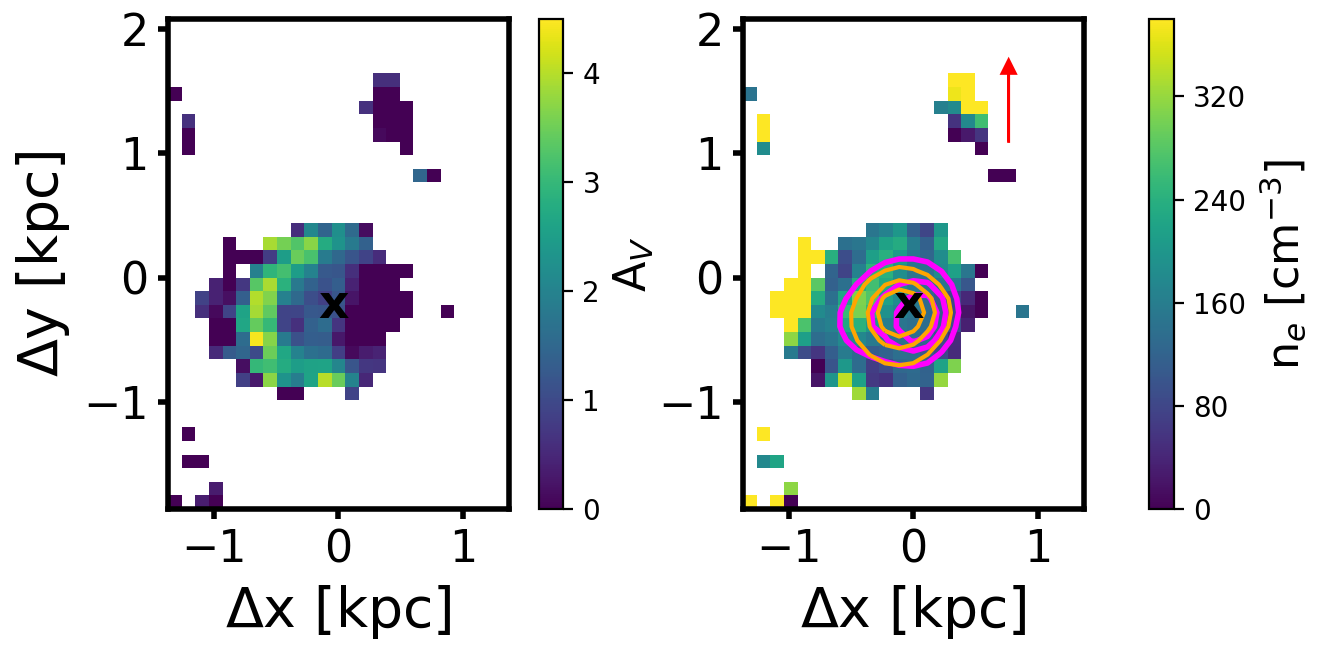}}\\
\caption{Same as Fig. \ref{fig:density_maps_correlation} for targets (top to bottom) NGC 7496, NGC 7469, NGC 5990 and NGC 4303. NGC 7496 and NGC 7469 show the same trend as the targets in Fig. \ref{fig:density_maps_correlation} i.e. the sites of high density correspond to sites of high star formation. In case of NGC 7469, the density pattern extend beyond the narrow H$\alpha$ emission in the S-E direction and a comparison with the reddening map suggests the possibility of obscured star formation. NGC 5990 and NGC 4303, on the other hand, do not show this spatial correlation. This has been further discussed in Sects. \ref{sect4} and \ref{sect5}. \label{fig:density_maps_2}}
\end{figure*}

\subsection{Determination of electron densities} \label{sect3.2}
To calculate electron densities, we use the prescription employed in \citet{sanders16}:

\begin{equation}
\mathrm{ 
n_{e}(R) = \frac{cR - ab}{a - R}
}
\label{eq:density_sii}
\end{equation}

\noindent
where the flux ratio R=f(\sii$\lambda$6716)/f(\sii$\lambda$6731]), and a=0.4315, b=2107 and c=627.1. In case of targets with an outflow, the density maps were created using the ratio of the total flux of the individual emission lines of the \sii ~doublet to determine overall density within the ionized medium as well as the flux ratio of the broad components separately to determine the density in the outflowing medium. A number of assumptions have been made to arrive at Eq. \ref{eq:density_sii} by \citet{sanders16} which we briefly summarize here \citep[see also ][]{kaasinen17}. The electron temperature is assumed to be 10,000 K and the electron densities are directly proportional to \hii ~region pressure. This may lead to an overestimation of densities in metal-rich galaxies. However, the errors due to these assumptions are significantly less than the measurement errors. For the flux ratios close to unity (0.7 < R < 1.45), the density ratio could vary between $\sim$50-2000 cm$^{-3}$. Such a large variation further reinforces the need for high S/N spectra to do the analysis.

The almost linear relation between the \sii ~doublet ratio and the electron density makes it effective in measuring the electron density in the aforementioned ranges. However the \sii ~doublet flux ratio saturates for R$>$1.45 and R$<$0.7 corresponding to the density values of $\sim$50 and $\sim$2000 cm$^{-3}$, the exact values of the lower and upper limit depends on the data quality\footnote{Note that we have made a more conservative estimate to the upper limit, unlike \citet{sanders16} where the limit goes beyond a few tens of thousands}. Therefore, we clip the minimum and maximum values to 50 and 2000 cm$^{-3}$ respectively. After arriving at the electron density value from the integrated spectrum, we calculate the density across the FOV as explained in Sect. \ref{sect3.1} to create electron density maps which are shown in Sect. \ref{sect4}. As mentioned earlier, for targets with an outflow, we create electron density maps from the total flux of the \sii ~doublet as well as the broad component to determine the density within the outflowing medium. To determine whether the density patterns observed are due to star formation or AGN, we have also created spatially resolved diagnostic diagrams which is explained in Sect. \ref{sect3.3}. 

The error cubes obtained from the PYWIFES reduction were used to create 1000 mock data cubes by the addition of Gaussian random noise from the corresponding error values. We repeated the line fitting procedure for each of these mock data cubes to obtain standard deviation of various parameters used in the analysis. We obtain similar error values in the electron density while taking into account error propagation from the flux ratio in Eq. \ref{eq:density_sii}.

\subsection{Spatially resolved BPT diagrams} \label{sect3.3}

The emission line ratios \oiii /H$\beta$ and \nii /H$\alpha$ are often used to distinguish the ionization mechanism of a nebular gas using diagnostic diagrams called as the BPT diagrams \citep[named after "Baldwin, Phillips \& Terlevich"][]{baldwin81}. The most commonly used demarcations to distinguish between the ionization by star formation or an AGN or a combination of the two is given by theoretical models of \citet{kewley01} and \citet{kauffmann03}. 

We created spatially resolved diagnostic diagrams for each target to understand whether the density morphology observed across the S7 FOV is affected by AGN or star formation processes \citep[see also][]{belfiore16,Rdavies17}. Fig. \ref{fig:maps_NGC1365}(c) shows an example for NGC 1365 where each data point corresponds to every pixel in the S7 FOV. The solid and the dashed lines are the demarcations by \citet{kewley01} and \citet{kauffmann03} where the area between the two curves correspond to the AGN-starburst composite region, and below and above these curves we would expect contribution due to star formation and AGN alone respectively. The data points have been coloured based on the \nii/H$\alpha$ line ratio which can distinguish the regions ionized by AGN and star formation in the spatially resolved \nii /H$\alpha$ map in \ref{fig:maps_NGC1365}(d) for example. We repeat a similar analysis for the rest of the targets which are collectively shown in Figs. \ref{fig:density_maps_correlation}, \ref{fig:density_maps_2} and \ref{fig:density_ngc1068}

Sect. \ref{sect4} presents the results of the analysis described in this section. 

\section{Results}      \label{sect4}

In this section, we first describe the results for a single target, NGC 1365, as a motivating example and then we collectively report the results of the rest of the targets. 

\subsection{Integrated Spectrum} \label{sect4.1}

The integrated spectrum of the stellar continuum subtracted ``blue'' and ``red'' cubes zoomed around the emission lines H$\beta$, \oiii ~$\lambda$4959,5007, H$\alpha$, \nii ~and \sii ~doublet is shown for NGC 1365 as an example in Fig. \ref{fig:spectrum_NGC1365}. NGC 1365 is a Seyfert 1 galaxy which has a strong H$\alpha$ emission line with width (FWHM) of $\sim$1100 km/s as apparent from the middle panel in Fig. \ref{fig:spectrum_NGC1365}. The narrow H$\alpha$, the \nii ~and \sii ~profiles are doubly peaked suggesting rotation drives this large line width. The fit to each individual peak has a narrower FWHM of $\sim$165 km/s. The \oiii ~profile, shown in the left panel of Fig. \ref{fig:spectrum_NGC1365}, appears symmetric, with no clear signatures for the presence of a strong ionized outflow. The \sii ~spectra for the rest of the targets are presented in Sect. \ref{appendix1}. 

\begin{figure*}
\centering
\subfloat{\includegraphics[width=7.6cm, height=4.2cm]{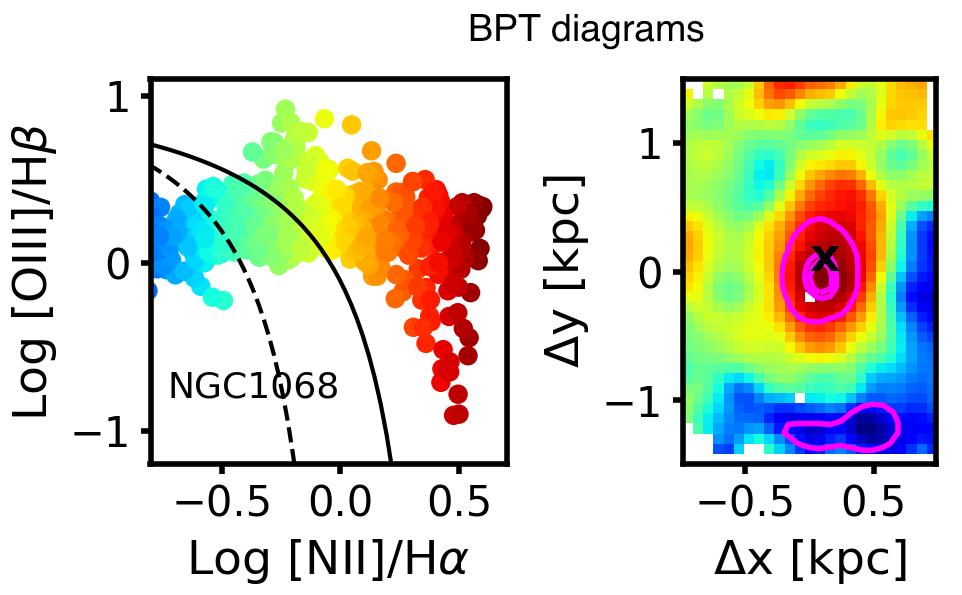}}
\subfloat{\includegraphics[width=7.8cm, height=4.2cm]{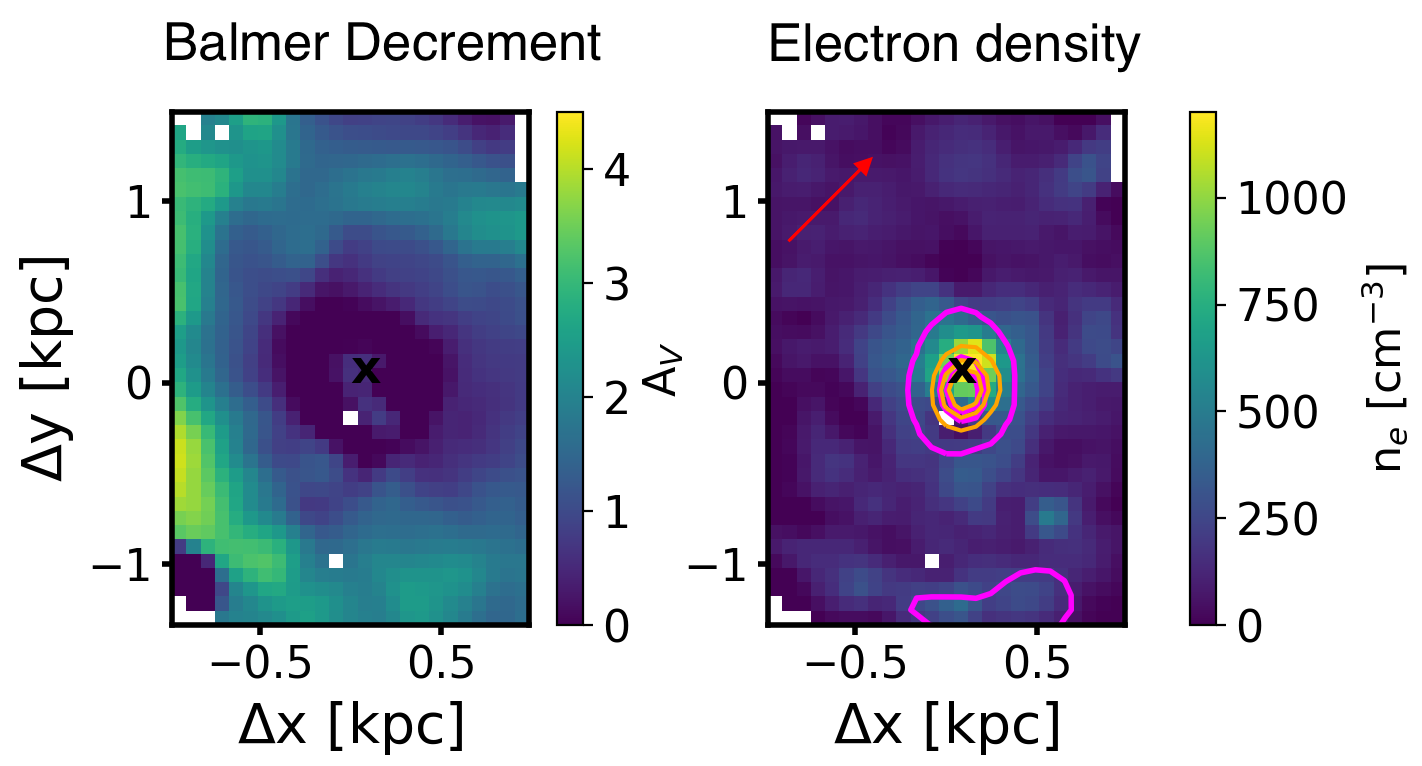}}\\
\subfloat{\includegraphics[width=7.6cm, height=3.8cm]{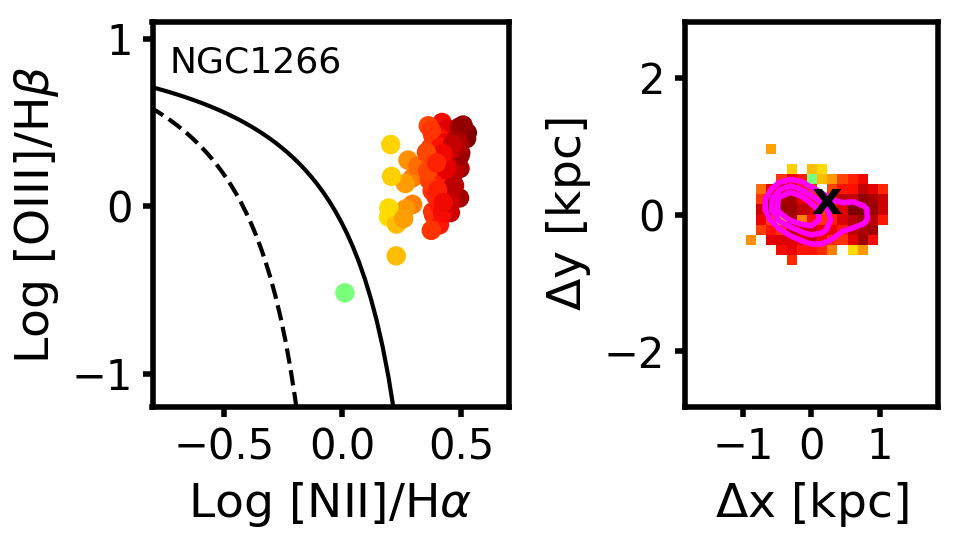}}
\subfloat{\includegraphics[width=7.8cm, height=3.8cm]{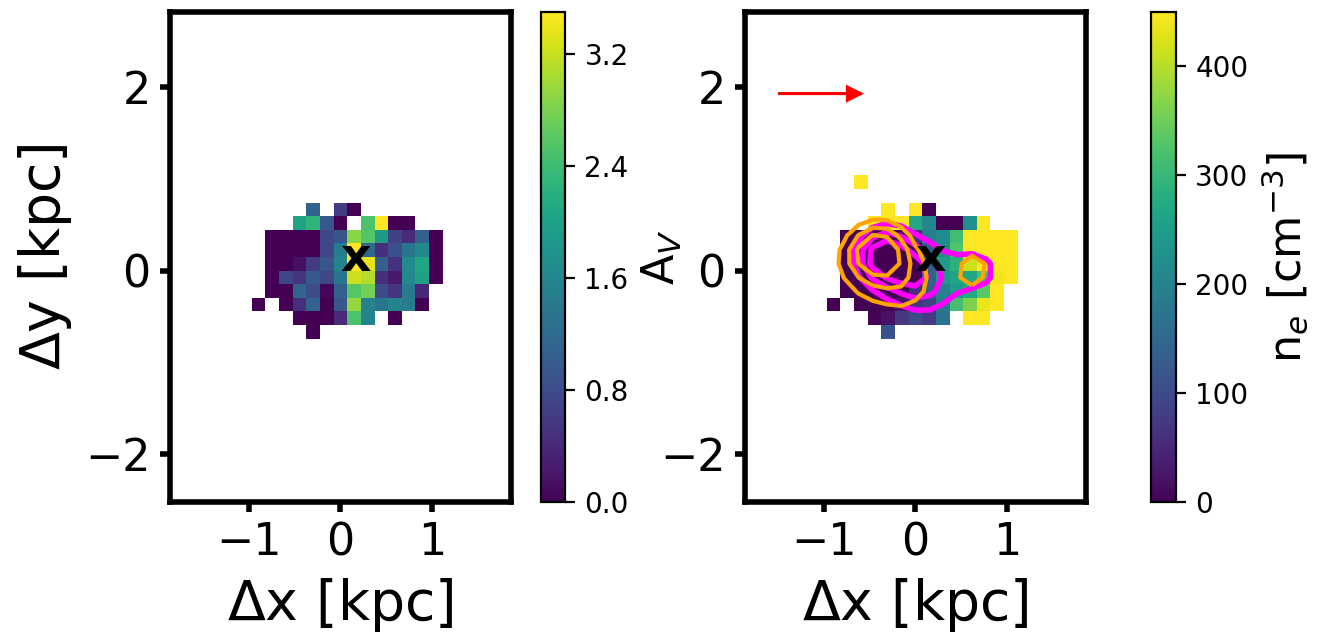}}\\
\subfloat{\includegraphics[width=7.6cm, height=3.8cm]{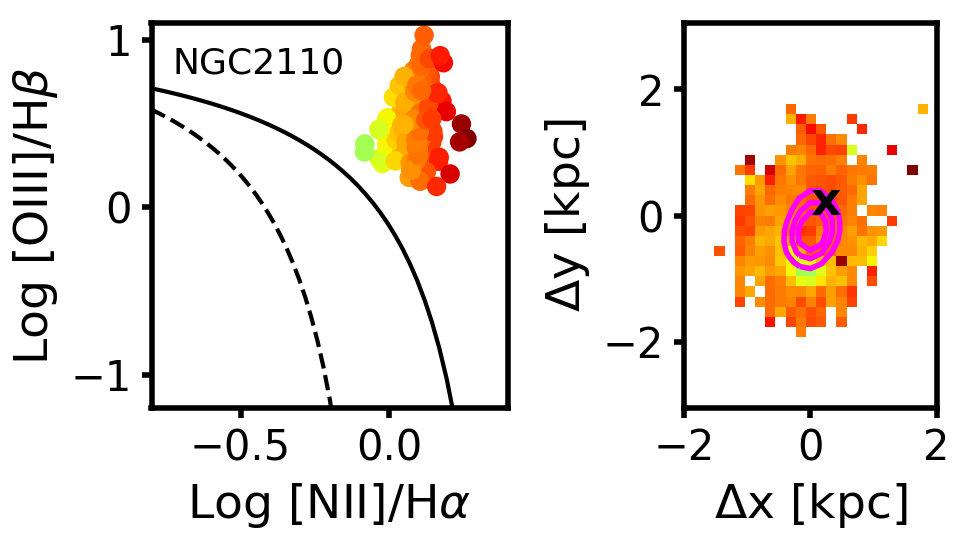}}
\subfloat{\includegraphics[width=7.8cm, height=3.8cm]{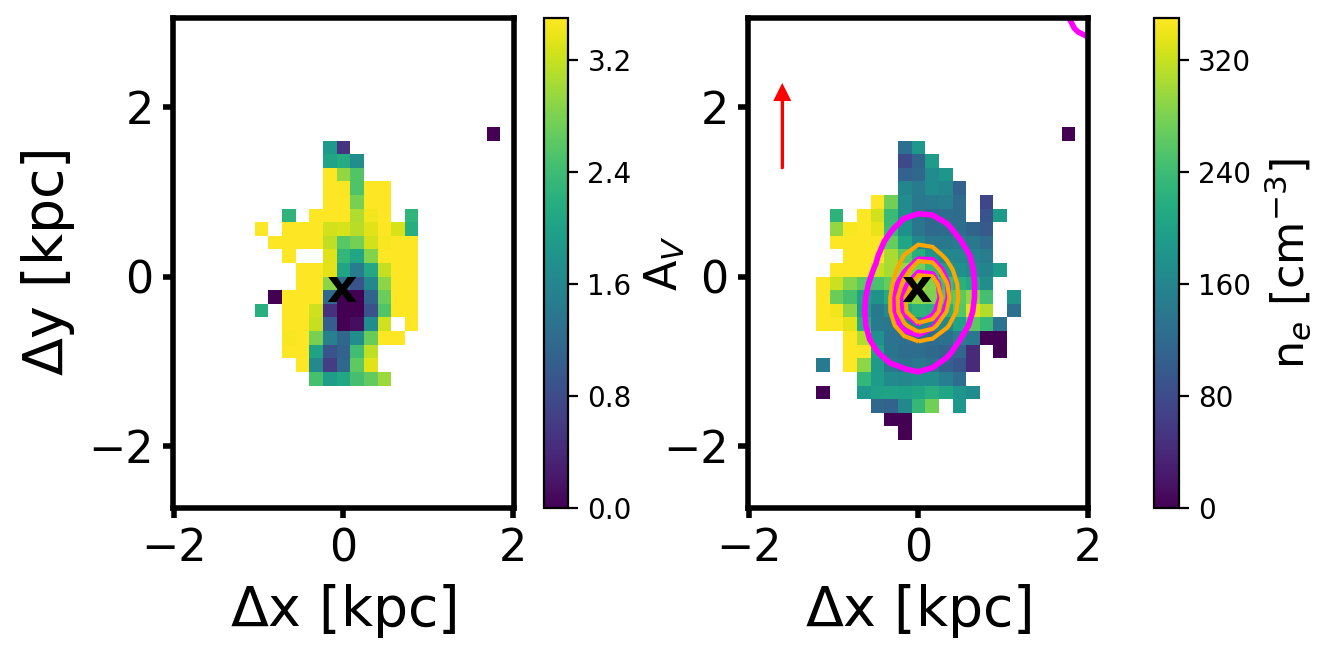}}\\
\caption{Same as Fig. \ref{fig:density_maps_correlation} but for the targets showing AGN-driven outflows: (Top to bottom) NGC 1068, NGC 1266 and NGC 2110. The density maps in this figure has been obtained from the ratio of the total flux of the individual emission lines of the \sii ~doublet, which is often used in literature. \label{fig:density_ngc1068}}
\end{figure*}

\begin{figure*}
\centering
\subfloat{\includegraphics[width=5cm, height=4.4cm]{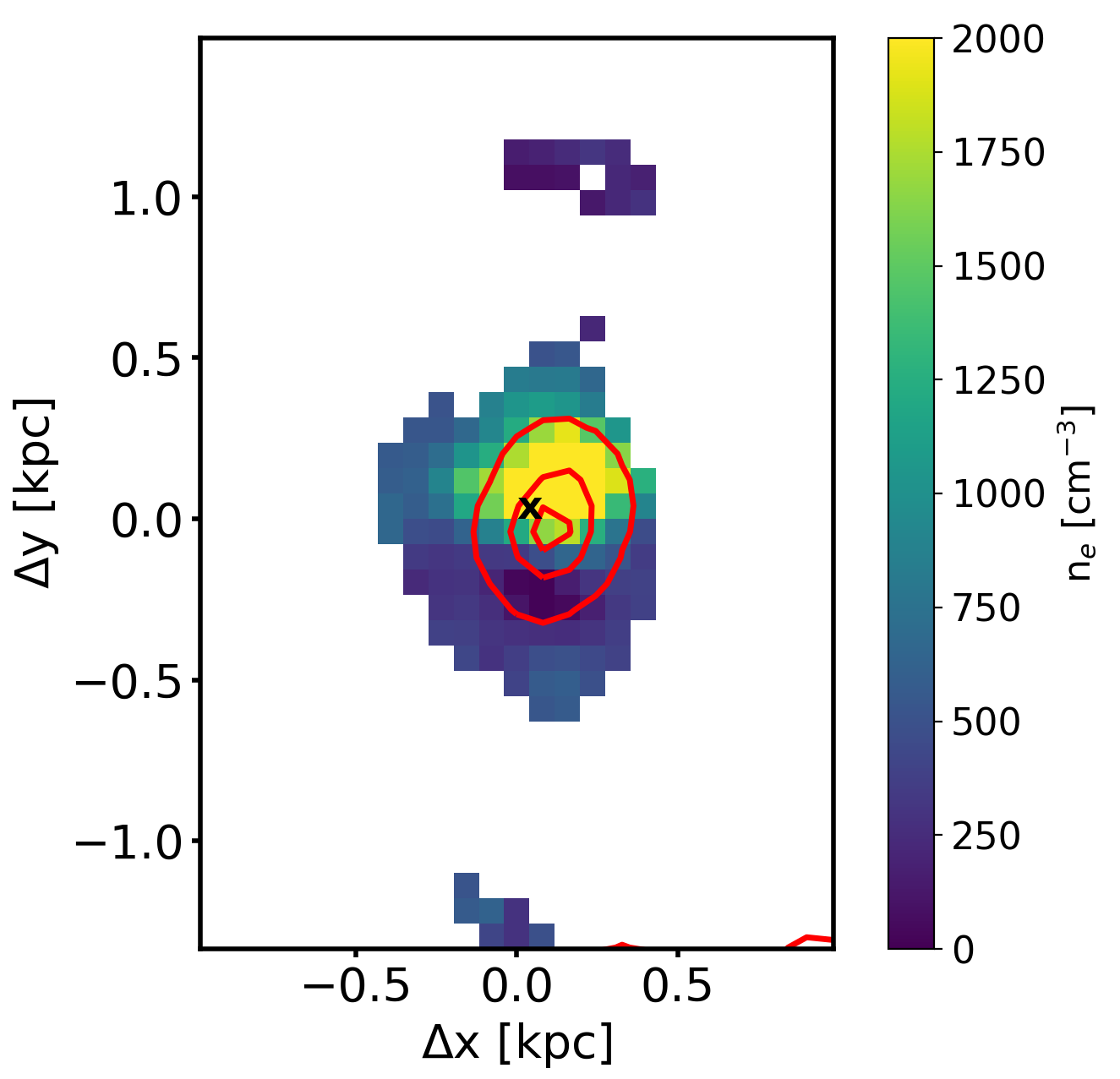}}
\subfloat{\includegraphics[width=5cm, height=4.4cm]{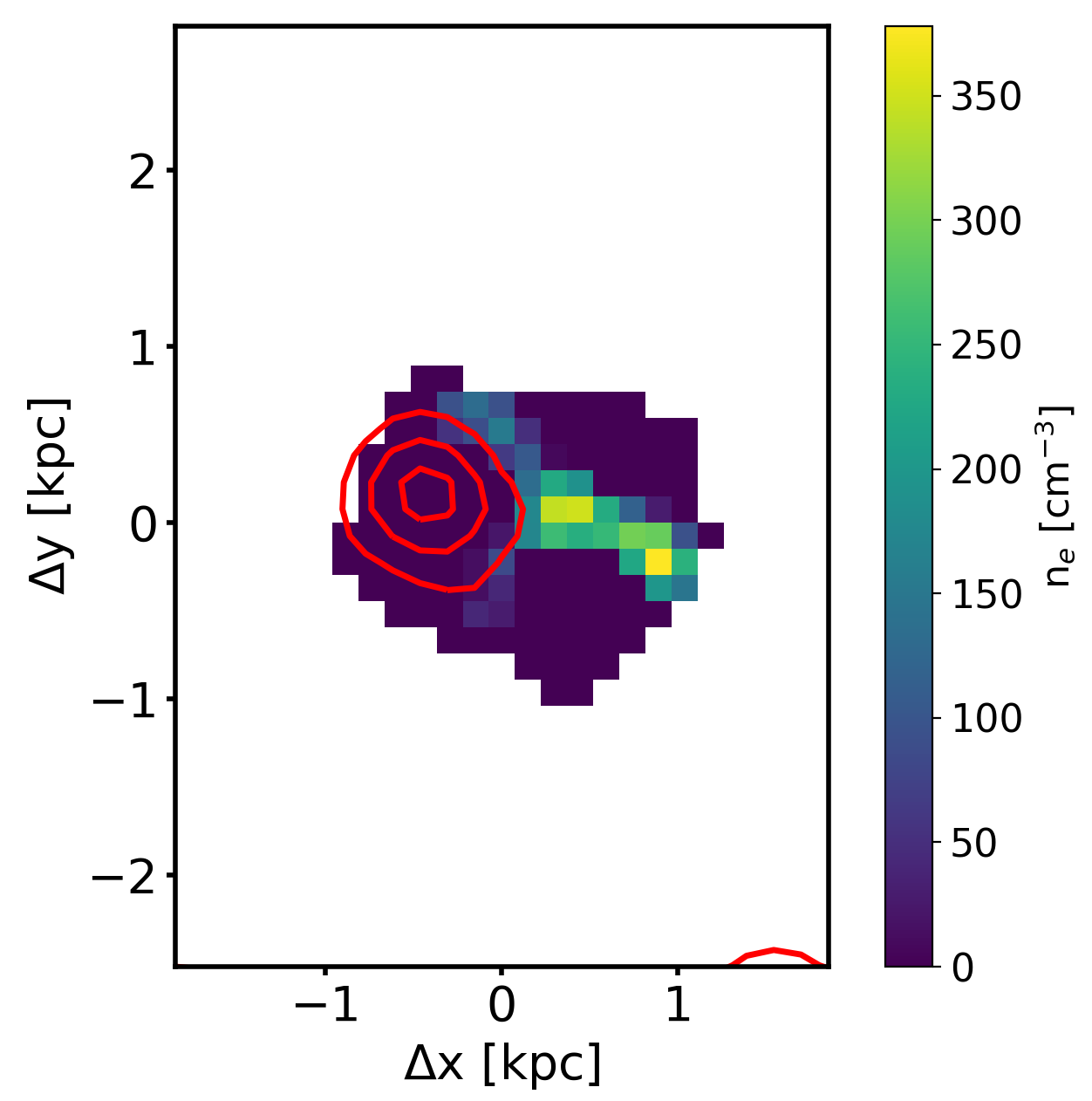}}
\subfloat{\includegraphics[width=5cm, height=4.4cm]{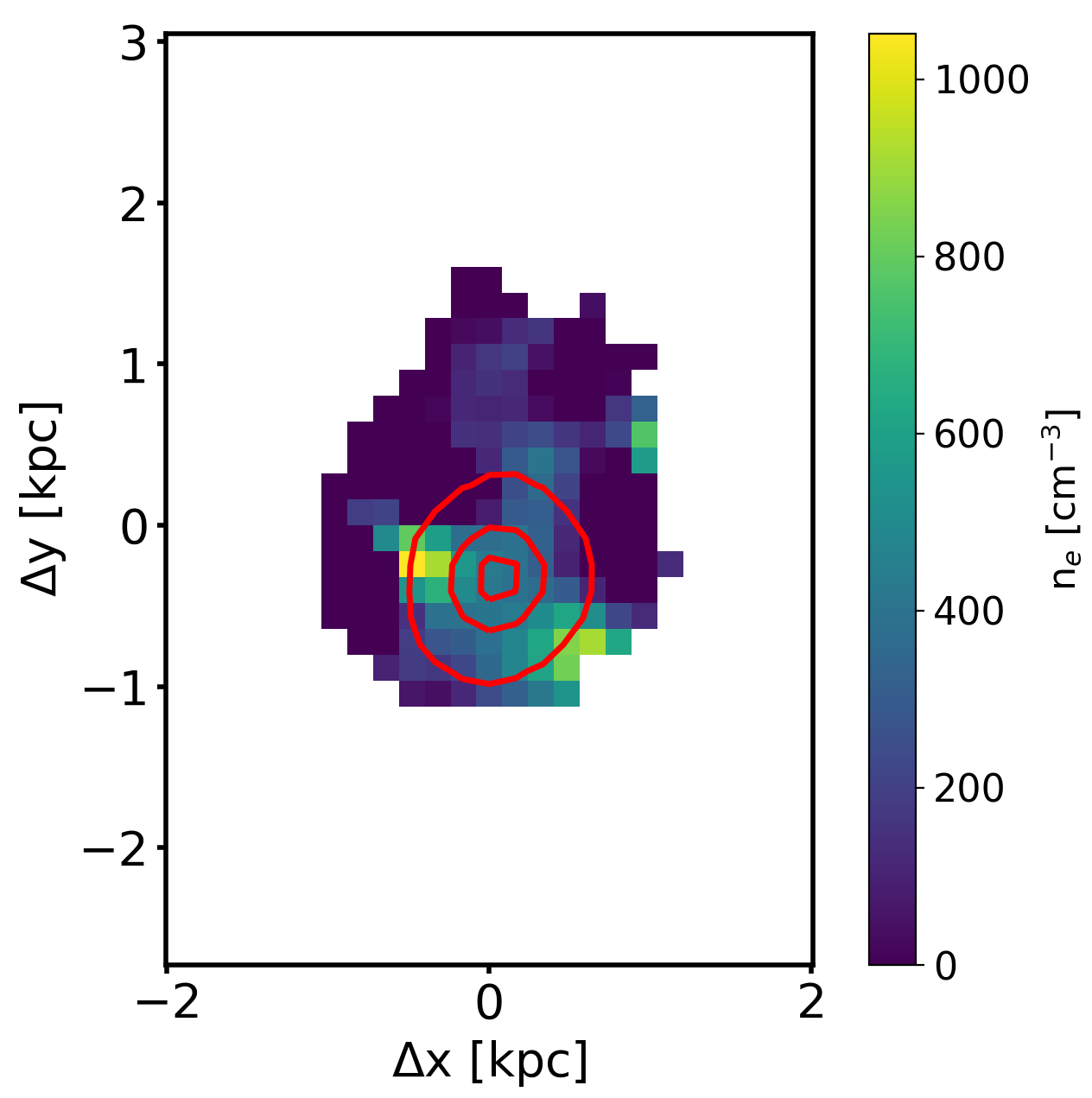}}
\caption{Electron density map in the outflowing broad component in NGC 1068 (left), NGC 1266 (middle) and NGC 2110 (right). The red contours show the location of the outflow as inferred from the broad component of \oiii. The density maps in this figure has been obtained from the ratio of the flux of the broad components of the \sii ~doublet which is believed to trace the gas in the outflowing medium.
\label{fig:density_broad_maps}}
\end{figure*}

The electron densities derived using Eq. \ref{eq:density_sii} from the integrated spectrum of NGC 1365 and the rest of the targets are reported in Table \ref{table:results}. The densities mentioned in the table are calculated from the ratio of the total flux of each component of the \sii ~doublet.  Most of the integrated density values for the targets presented in this paper lie between $\sim$100-330 cm$^{-3}$ which is comparable to the density values previously reported for the local AGN host galaxies \citep[e.g.][]{bennert06a, rupke17}. For the broad outflowing component in NGC 1068, NGC 1266 and NGC 2110, we get a density value of 726$\pm$90, 1120$\pm$295 and 237$\pm$78 cm$^{-3}$ respectively. The \sii ~spectrum showing these broad components are shown in Fig. \ref{fig:sii_spectrum_outflow}. Therefore, from the integrated spectrum analysis, the outflowing component has a significantly higher density compared to the non-outflowing components. This is in-line with the density measured with the outflowing components for SDSS targets in \citet{perna17} at high redshift (z$\sim$0.7) where the average density was $\sim$1000 cm$^{-3}$. 
 However, as explained in the following section, spatially resolved electron density maps show that the high density scenario is not always true within the outflowing medium.
 
\subsection{Electron density maps} \label{sect4.2}
Fig. \ref{fig:maps_NGC1365} shows the flux maps, resolved diagnostic maps, extinction and density maps and density profile for NGC 1365 as an example. In all maps, the black cross in the center represents the position of the AGN, the red arrow shows North, magenta contours show the narrow H$\alpha$ emission and the orange contours show the \oiii ~emission, at 30\%, 50\% and 80\% of the peak each. The beam at the bottom left of the maps in panels (a), (b), (e) and (f) shows the PSF (FWHM) during the observations.

The narrow H$\alpha$ emission in NGC 1365 (Fig. \ref{fig:maps_NGC1365}a) shows a clumpy profile in a ring like structure around the central AGN. These clumps are consistent with being star forming regions which is verified by the BPT diagrams in panels (c) and (d) in Fig. \ref{fig:maps_NGC1365}. The \oiii ~emission in NGC 1365 (Fig. \ref{fig:maps_NGC1365}b)  seems to originate from the central AGN and its morphology and spatial distribution is significantly different compared to the narrow H$\alpha$ map.  

\begin{figure*}
\centering
\subfloat{\includegraphics[width=4.5cm, height=4cm]{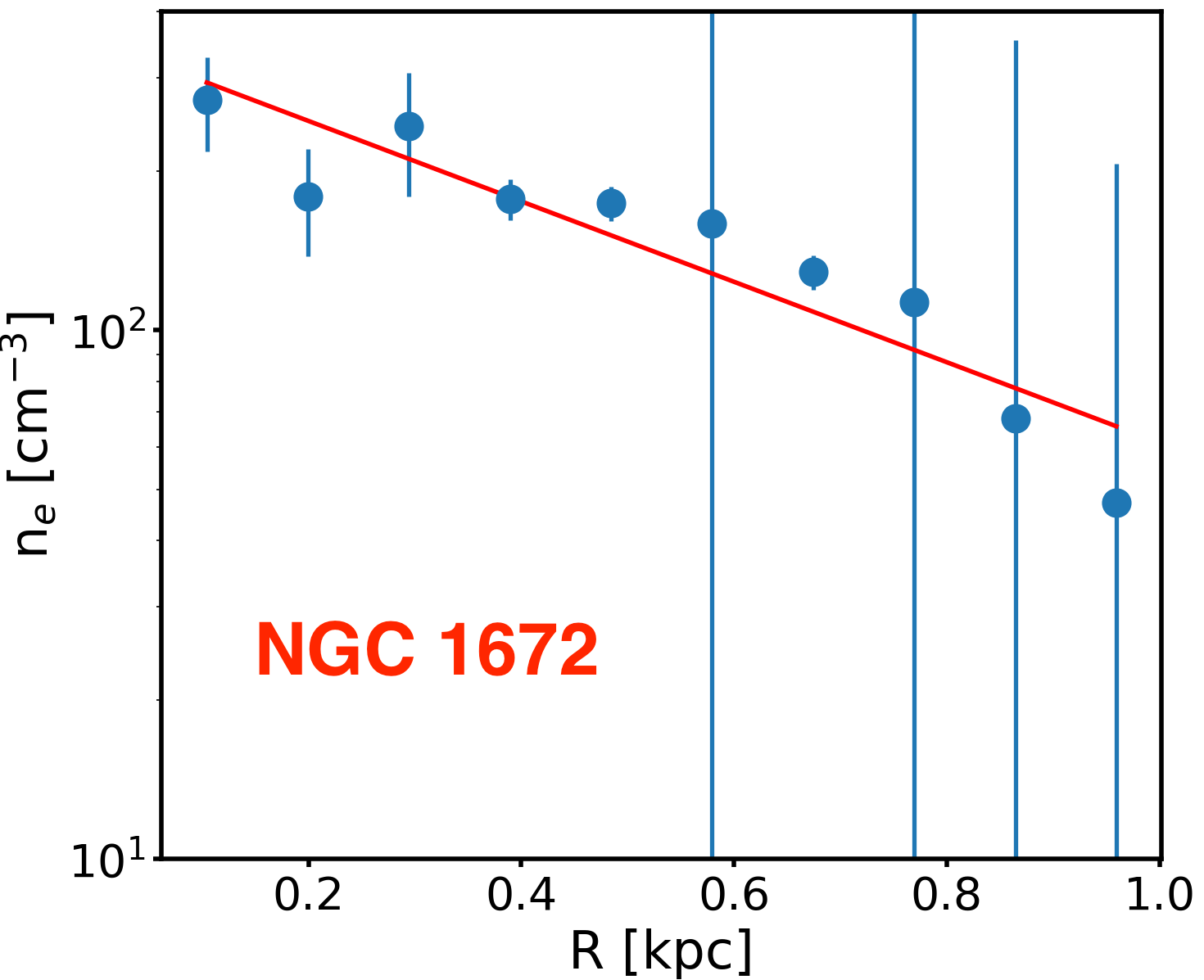}}
\subfloat{\includegraphics[width=4.5cm, height=4cm]{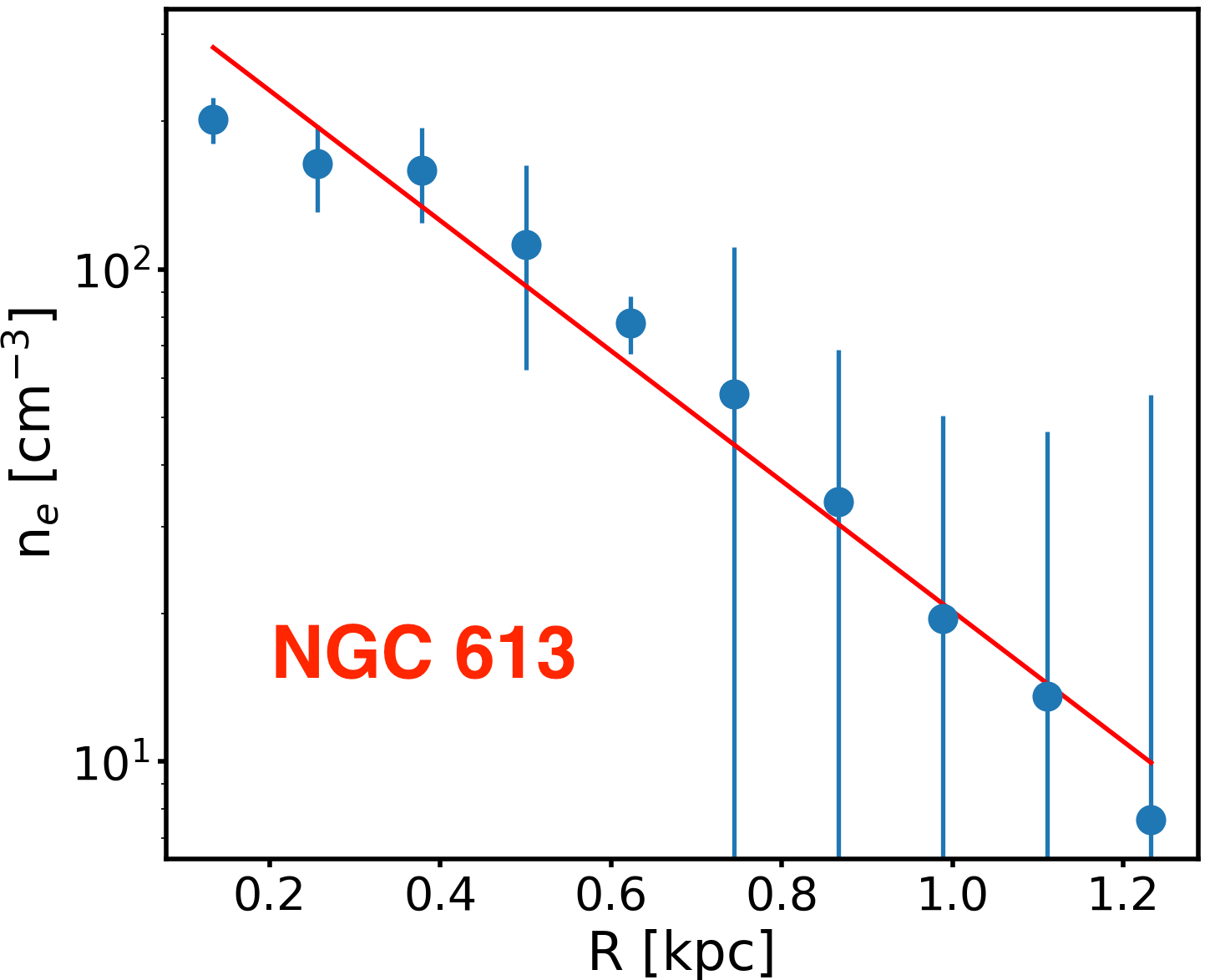}}
\subfloat{\includegraphics[width=4.5cm, height=4cm]{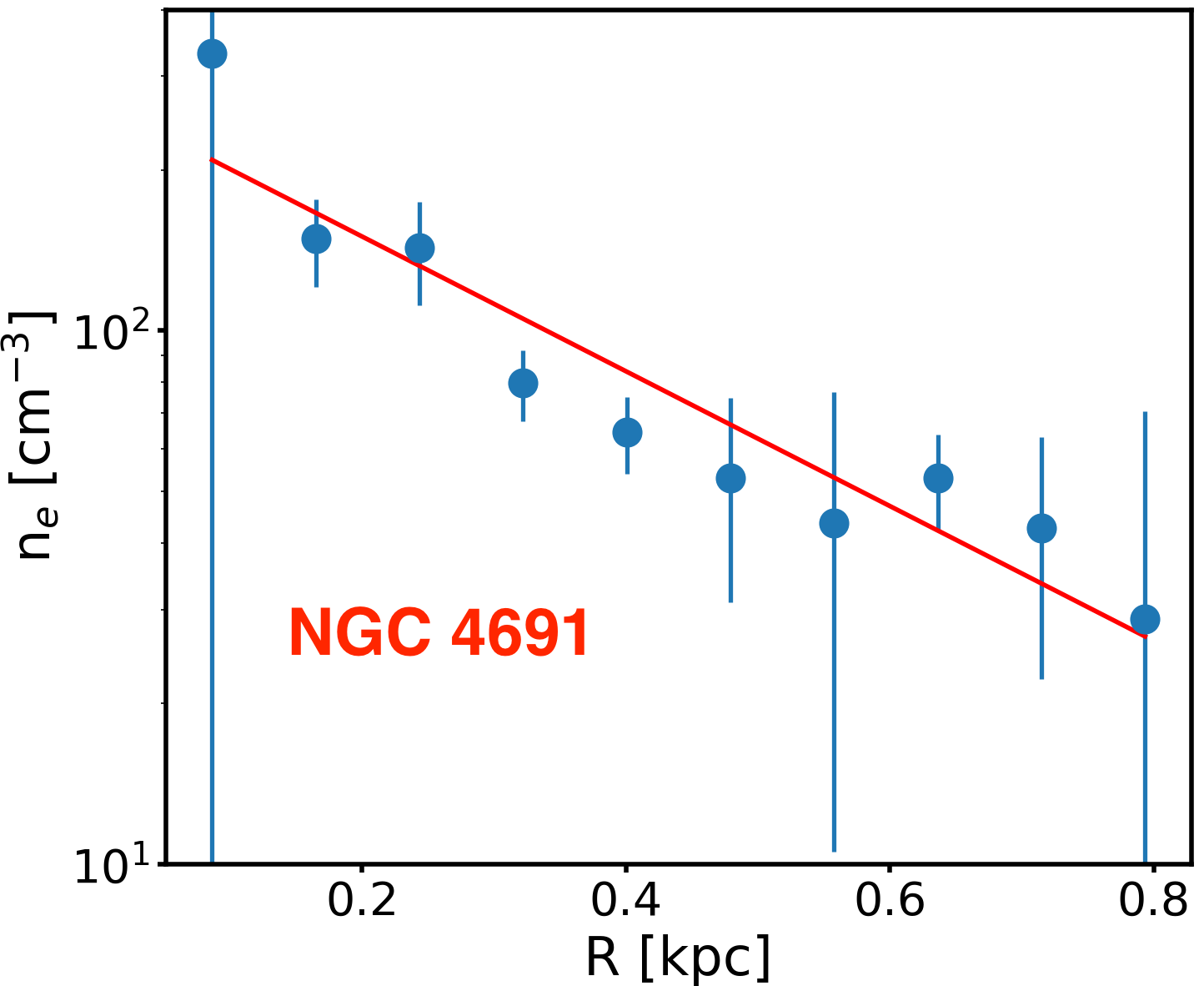}}\\
\subfloat{\includegraphics[width=4.5cm, height=4cm]{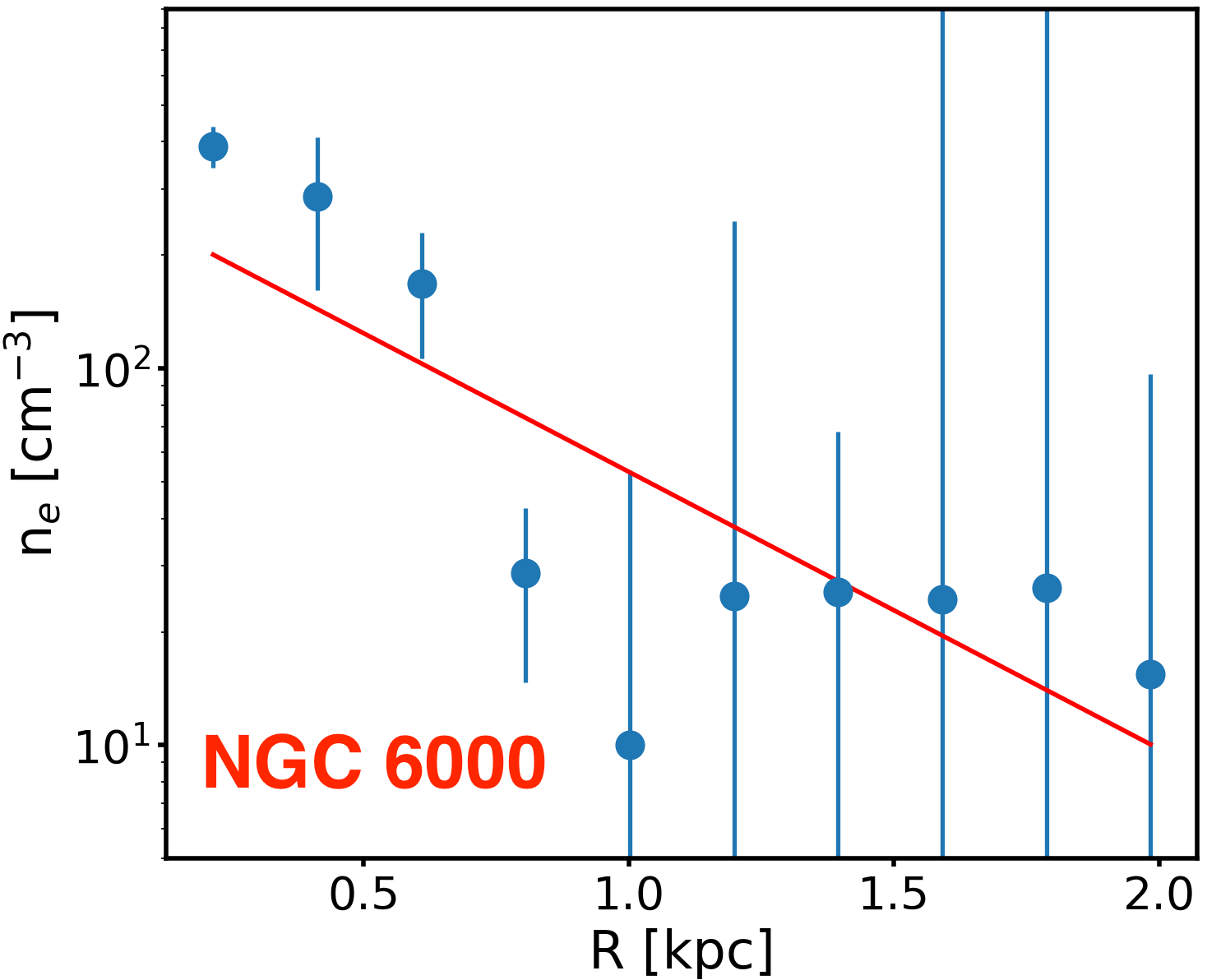}}
\subfloat{\includegraphics[width=4.5cm, height=4cm]{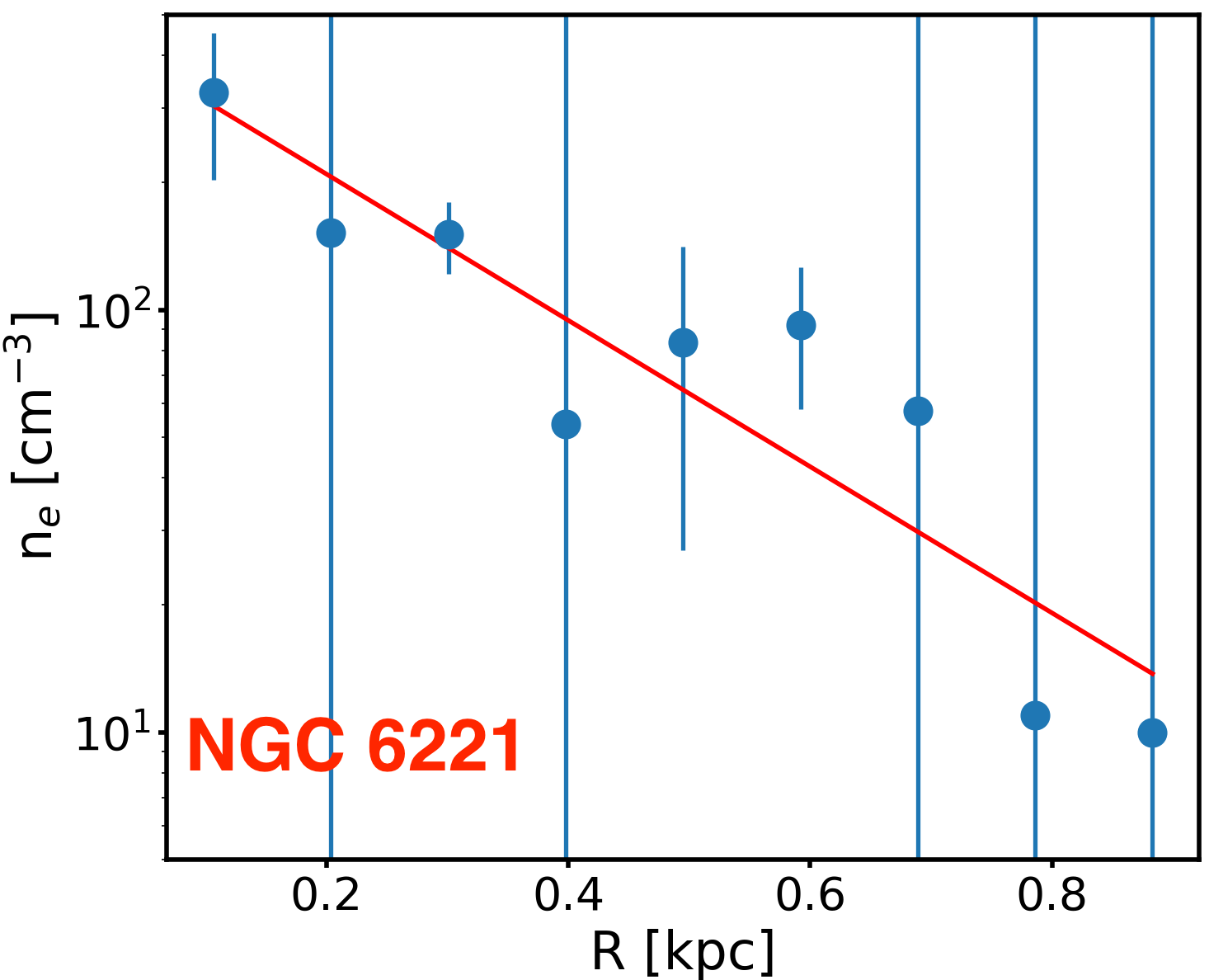}}
\subfloat{\includegraphics[width=4.5cm, height=4cm]{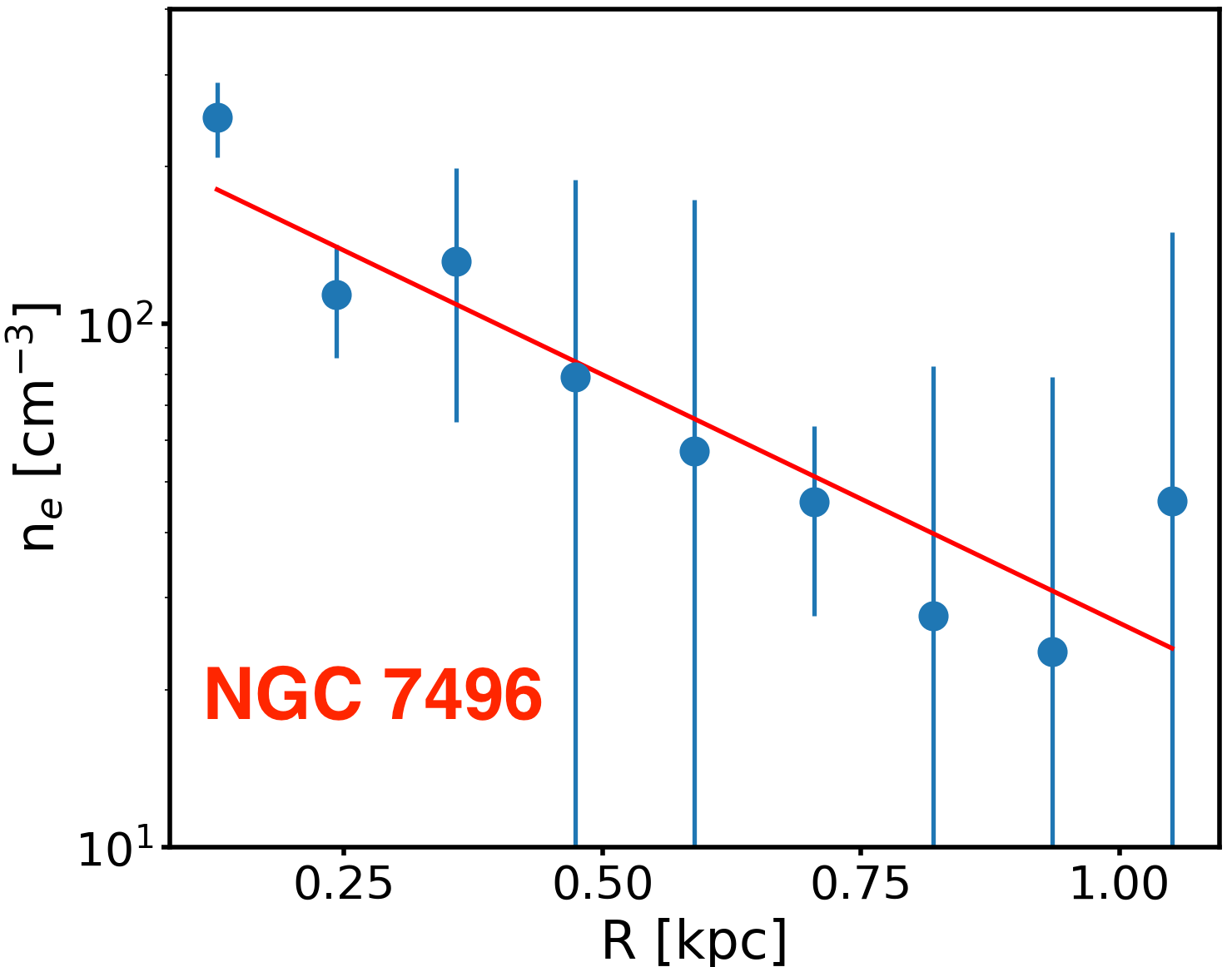}}\\
\subfloat{\includegraphics[width=4.5cm, height=4cm]{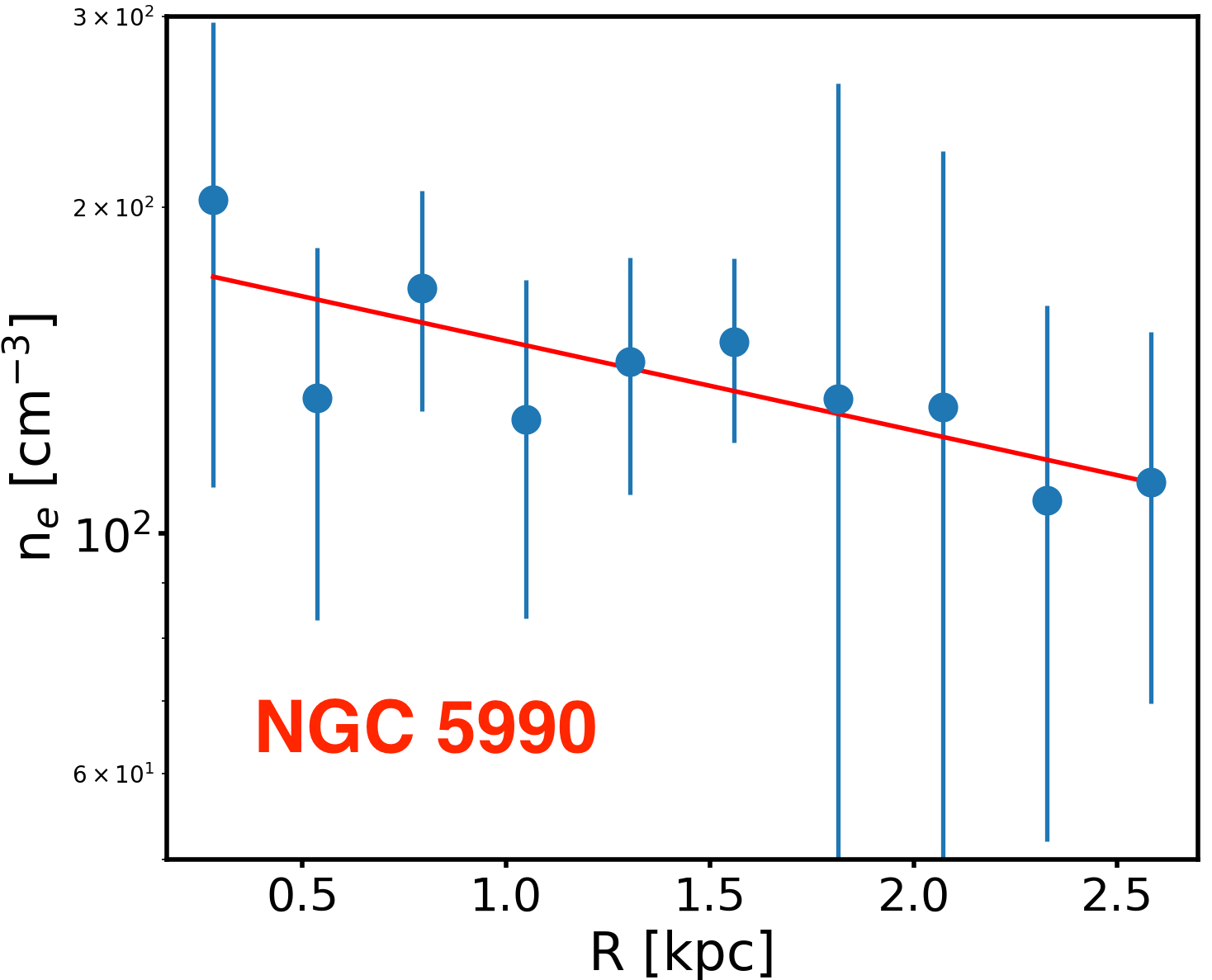}}
\subfloat{\includegraphics[width=4.5cm, height=4cm]{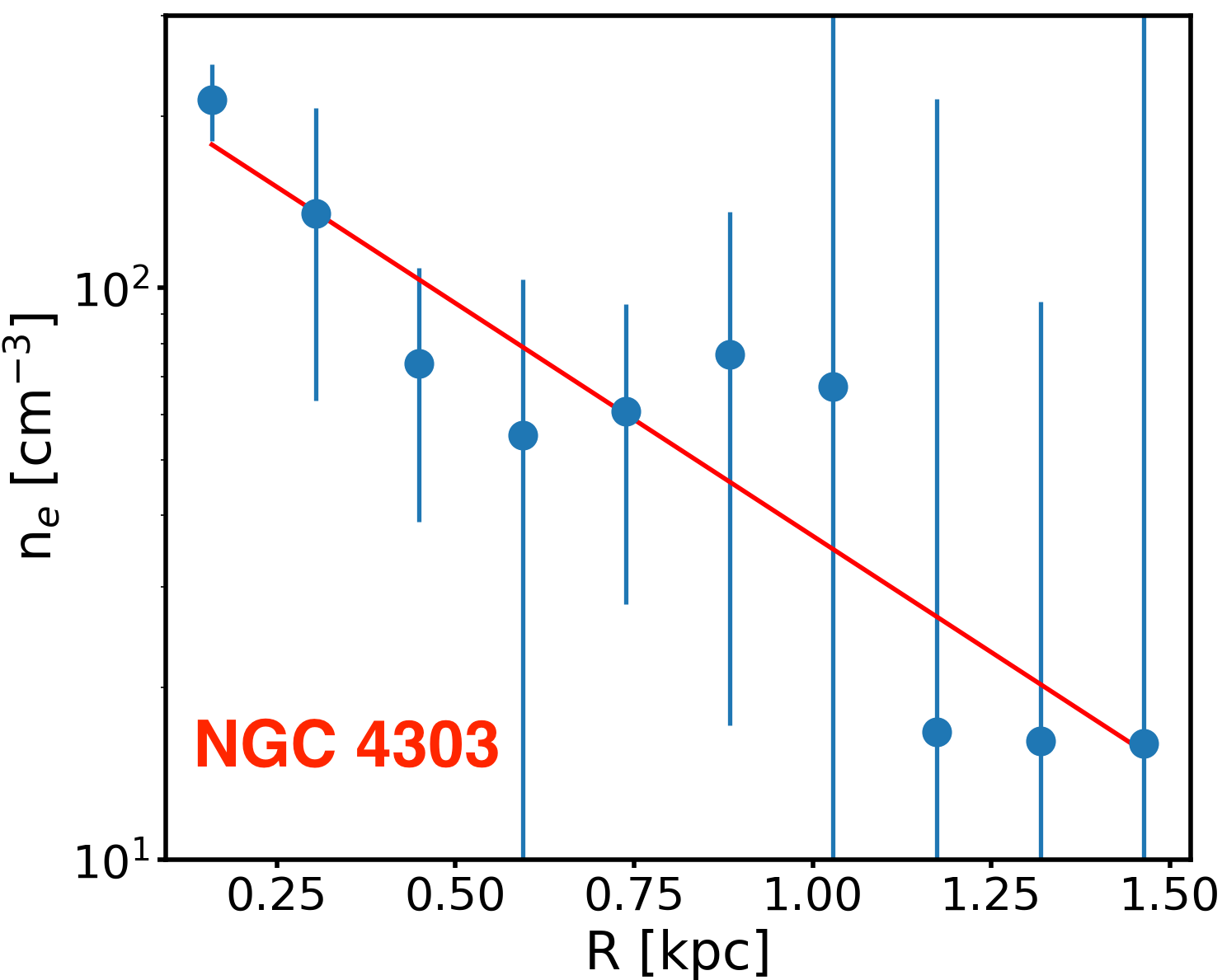}}
\subfloat{\includegraphics[width=4.5cm, height=4cm]{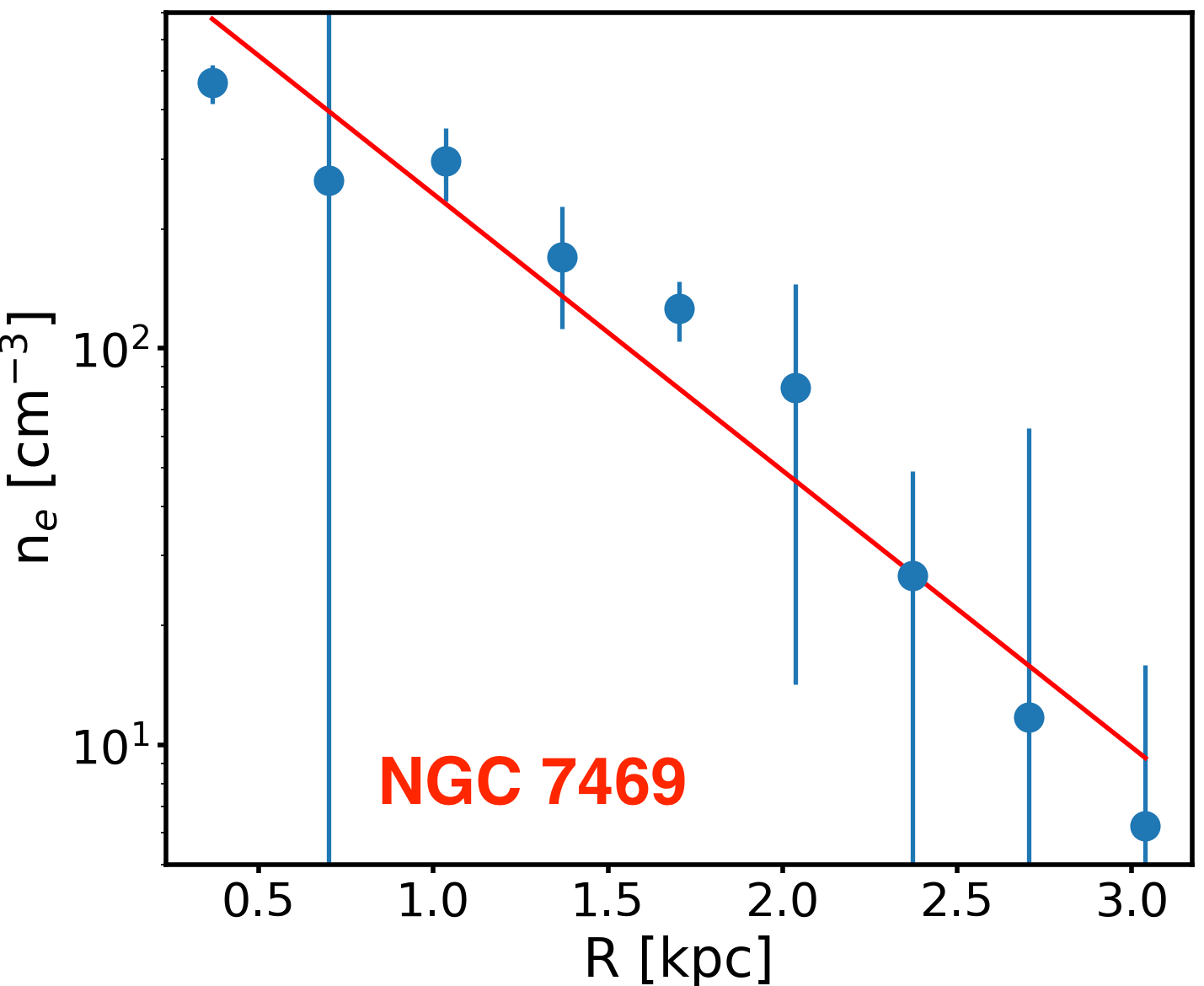}}\\
\subfloat{\includegraphics[width=4.5cm, height=4cm]{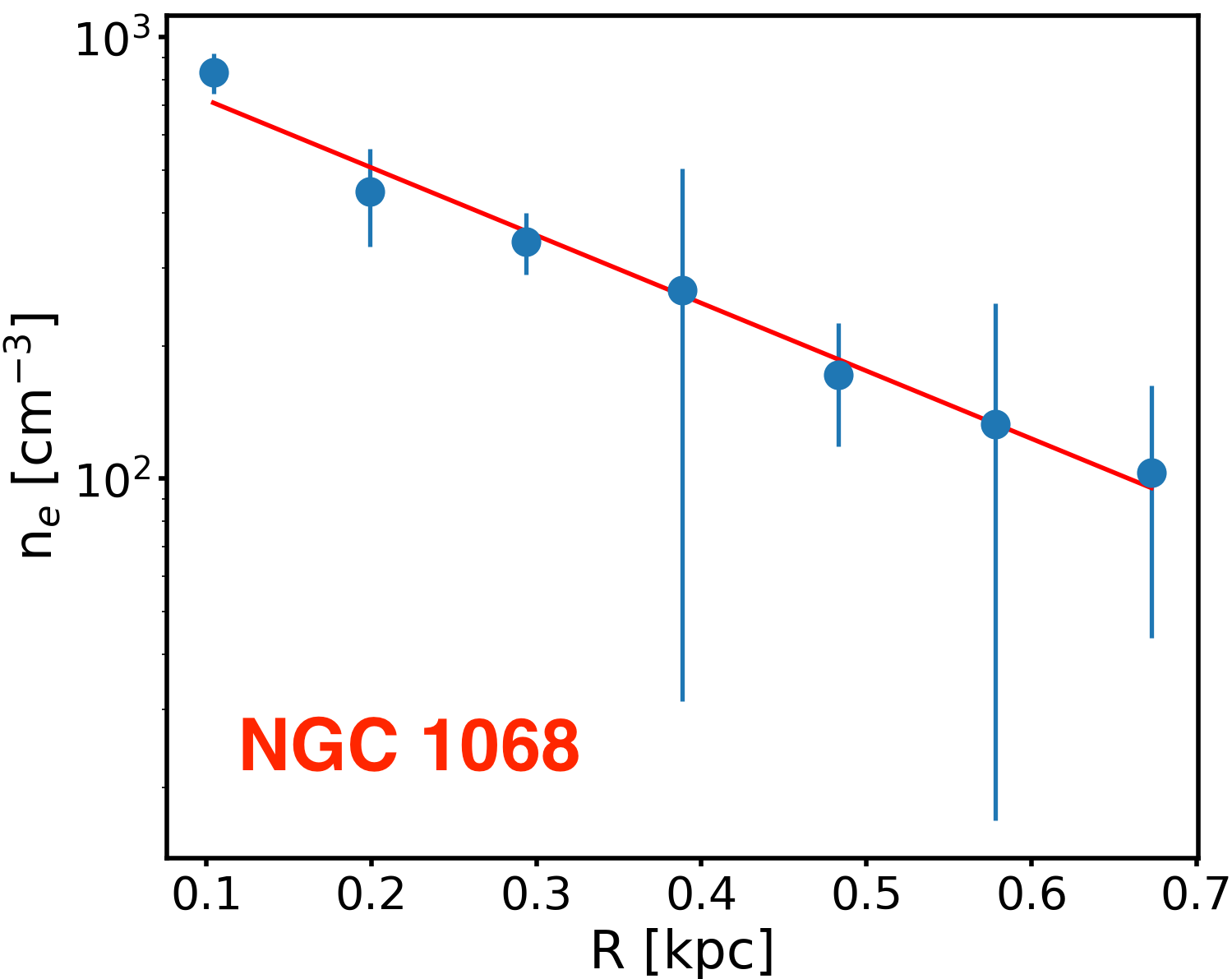}}
\subfloat{\includegraphics[width=4.5cm, height=4cm]{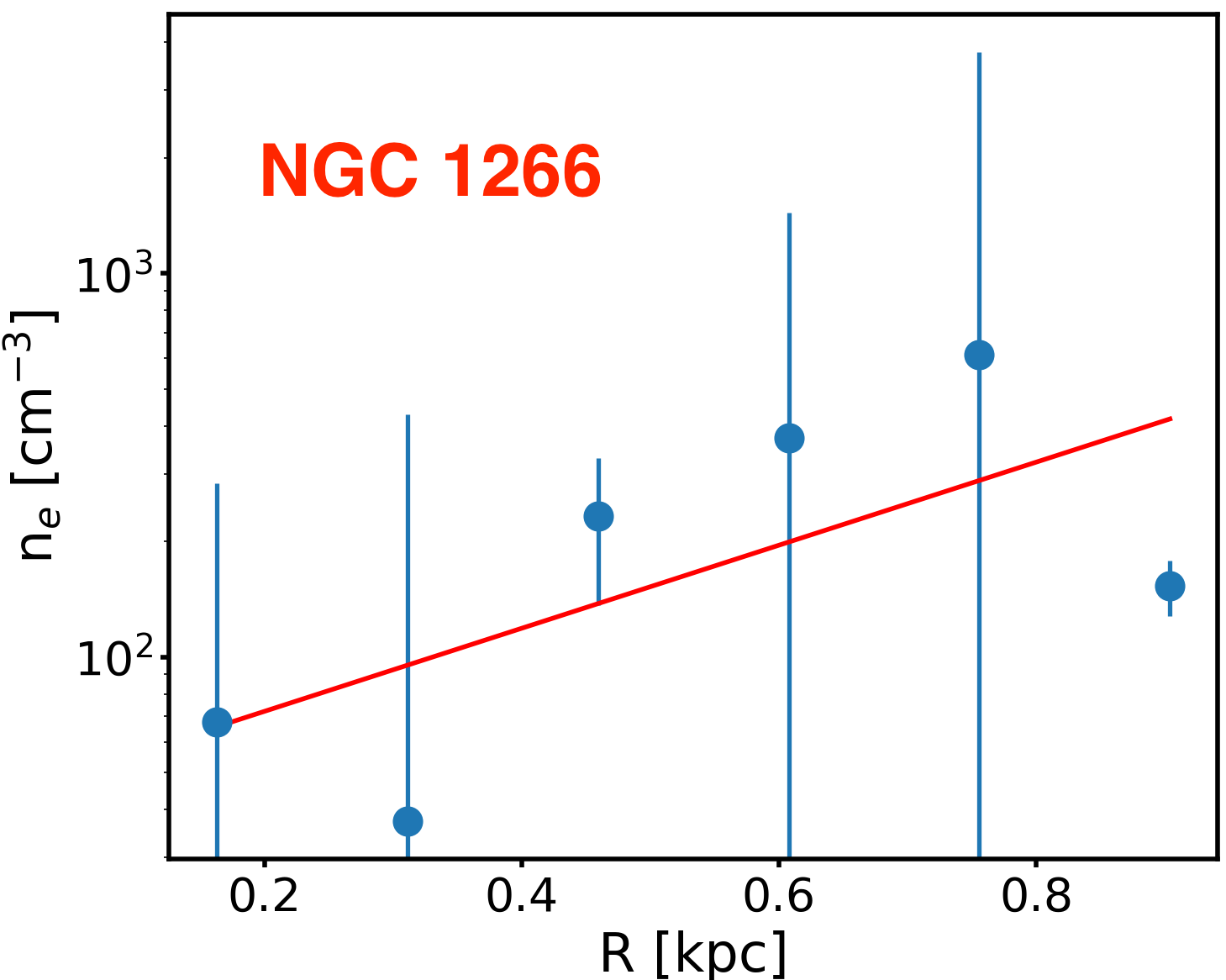}}
\subfloat{\includegraphics[width=4.5cm, height=4cm]{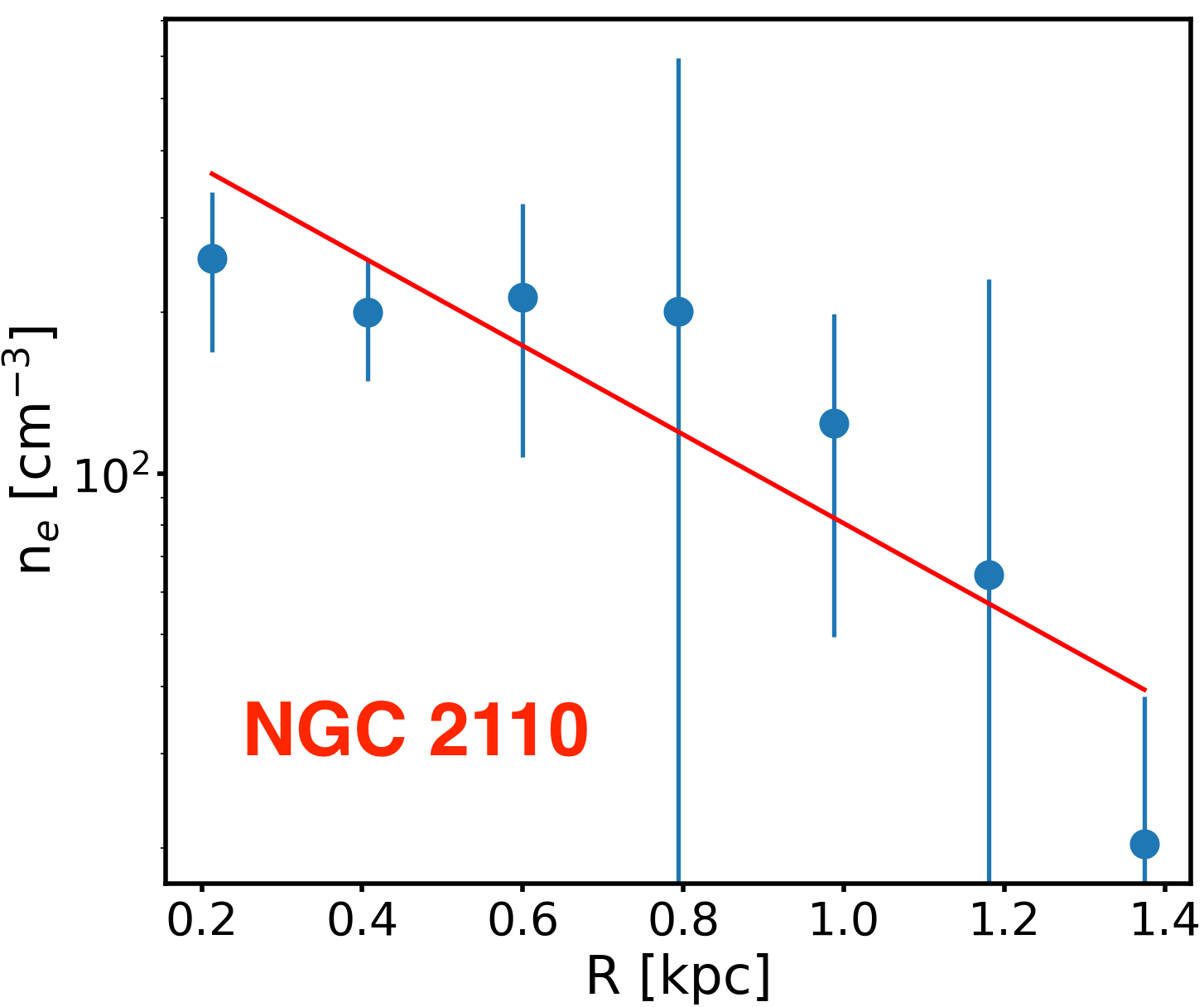}}\\
\caption{Electron density profiles of the targets presented in this paper. Blue data points show the electron density values obtained from the maps in Figs. \ref{fig:density_maps_correlation} and \ref{fig:density_maps_2} while the solid red line shows the exponential fit to the profile (See text for details). The exponential index are reported in Table. \ref{table:results}. For the targets with ionized outflows i.e. NGC 1068, NGC 1266 and NGC 2110 shown in the bottom panels, the density profiles have been obtained for the maps presented in Fig. \ref{fig:density_ngc1068}.
\label{fig:profiles_all}}
\end{figure*}

With the coverage of H$\beta$ and H$\alpha$ emission lines with WiFeS, we were also able to construct reddening maps (or dust extinction maps) using the Balmer Decrement ($\mathrm{H_{\alpha}/H_{\beta}}$) as shown in Fig. \ref{fig:maps_NGC1365}(e). The extinction map is useful to convert the narrow H$\alpha$ map to star formation, correcting for any attenuated \hii ~region. We do not expect any significant changes to electron density maps due to this correction as the \sii ~doublet lines are closely spaced leading to similar correction factor in both the fluxes and therefore not a significant change in the flux ratio, R in Eq. \ref{eq:density_sii}. The extinction maps were obtained assuming a \citet{calzetti00} dust attenuation law with $\mathrm{R_V}$ = 4.05 and a fixed temperature of 10,000 K which is the same as that derived for the electron density maps. Note that the extinction maps have been obtained using only the narrow component of $\mathrm{H_{\alpha}}$ and $\mathrm{H_{\beta}}$ which avoids flux contributions from BLR. For NGC 1365, the extinction map shows high obscuration in the north with respect to the central AGN where there is also a bright \hii ~clump suggesting that the star formation in this clump might be obscured. In fact, the presence of dust in NGC 1365 has also been confirmed by 24 $\mu$m Herschel and SPIRE (Spectral and Photometric Imaging Receiver) observations in \citet{alonso-herrero12} where a comparison between the SFR derived from the 24 $\mu$m continuum and the narrow H$\alpha$ line shows that nearly 85\% of the ongoing star formation in the central regions of NGC 1365 is taking place in the dust obscured regions. The extinction maps for the rest of the targets are shown in Figs. \ref{fig:density_maps_correlation}, and \ref{fig:density_maps_2}. Most extinction maps show high extinction at the edges of the \hii ~clumps. 

Fig. \ref{fig:maps_NGC1365}(f) shows the density map for NGC 1365 obtained by the pixel-by-pixel analysis of the \sii ~doublet with narrow H$\alpha$ and \oiii ~contours from the respective maps in panels (a) and (b) overlaid in magenta and orange colours respectively. From the electron density map of NGC 1365, it is clear that the electron density is high in sites of star forming regions and is spatially uncorrelated with the \oiii ~emission. This correlation between the star formation and high density is also seen in the maps of the majority of the targets shown in the right panels of Figs. \ref{fig:density_maps_correlation} and \ref{fig:density_maps_2} except NGC 5990 and NGC 4303. In the case of NGC 7469, shown in the second row from top in Fig. \ref{fig:density_maps_2}, in addition to the spatial correlation with the \hii ~regions, the electron density in this target also shows an extension towards the South-East with respect to the central AGN. This site also seems to be a region of high extinction, indicating the possibility of obscured star formation in this region. This suggests that, in cases like NGC 7469, density maps could also possibly hint towards sites of obscured star formation as well in some objects. As mentioned earlier, exceptions to this spatial correlation does exist in targets like NGC 4303 and NGC 5990 as shown in the bottom two panels of Fig. \ref{fig:density_maps_2} which will be discussed in Sect. \ref{sect5}.

We can infer the pressure at the center of galaxies using:

\begin{equation}
p/k ~[K cm^{-3}] \sim 2.4 ~n_e ~T
\label{eq:pressure}
\end{equation}

\noindent
where 2.4 is a correction factor for the total number of ions and the temperature T is assumed to be 10,000 K (same as the assumed temperature for the derivation of electron density for consistency) and n$_e$ is the electron density at the center. The pressures for the sample in this paper are reported in Table \ref{table:results}. 

Fig. \ref{fig:density_broad_maps} show the electron density in the outflowing medium, derived from the flux ratio of the broad components of the \sii ~doublet. The red contours in Fig. \ref{fig:density_broad_maps} represent the broad \oiii ~emission tracing the outflowing gas in the NLR. For completeness, we also show the density maps derived from the flux ratio of the total flux of the individual lines of the \sii ~doublet in Fig. \ref{fig:density_ngc1068} i.e. summing over both narrow and broad Gaussian components, which is often used in literature. A first look at both set of maps suggests that the clear spatial correlation between the density and star forming sites (as in case of non-outflowing targets) ceases to exist for these outflowing targets, partly due to high AGN contamination to the narrow H$\alpha$ regions inferred from the BPT diagram. Although star formation may also increase the density as the narrow H$\alpha$ emission is co-spatial with the AGN ionization region, the range of density is significantly higher than the range observed in the star forming regions in the non-outflowing targets. The density maps from the outflowing component in Fig. \ref{fig:density_broad_maps} suggests that there is no uniform density within the outflowing medium which is often assumed in the mass outflow rate calculations. The density in fact can take a wide range of values ranging from, for example in NGC 1068, < 50 cm$^{-3}$ to > 2000 cm$^{-3}$. These results from spatially resolved maps are in contrast with the high electron density ($\sim$1000 cm$^{-3}$) reported in outflowing medium from integrated spectrum analysis \citep[e.g. NGC 1068 and NGC 1266 in this paper and the sample in][]{perna17}. In fact for targets like NGC 1266, the density at the outflow location traced by \oiii ~is less than 50 cm$^{-3}$ throughout which further reinforces the need for cautiousness while interpreting results obtained from integrated spectrum analysis.

The implications of these observations are further discussed in Sect. \ref{sect5}.

\subsection{Electron density profiles} \label{sect4.3}

The electron density maps presented above are also useful to quantify how the density drops as a function of the distance from the central AGN. To do this, we constructed electron density profiles by extracting integrated density values from circular annuli centered on the AGN and with a width of one pixel, which roughly matches the resolution of the observations. We did not do this exercise for the density measured in the outflowing medium (Fig. \ref{fig:density_broad_maps}) as there is no obvious pattern across the three targets. Therefore, we only report the density profile results in the cases where the density maps were obtained from the ratio of the total flux of the individual emission lines of the  \sii ~doublet. The data points obtained from the electron density map of NGC 1365 are shown as blue circles in panel (g) of Fig. \ref{fig:maps_NGC1365} as an example, while for the rest of the targets the profiles are shown collectively in Fig. \ref{fig:profiles_all}. The red line in all the plots show the model exponential profile as explained below. 

From the plots in Fig. \ref{fig:profiles_all}, it is clear that the electron density decreases as a function of the inclination corrected distance from the center of the galaxy (except for NGC 1266, see Sect. \ref{sect5}). The drop is density is not necessarily uniform due to the clumps like in case of NGC 1365 (Fig. \ref{fig:maps_NGC1365}(g)). Nevertheless, to a first order, these profiles can generally be fit using an exponential function of the distance from the center of the galaxy which can be mathematically expressed as:

\begin{equation}
n_e = \alpha\cdot e^{-\beta\cdot R}
\label{eq:power_law}
\end{equation}

where $\alpha$ is a normalization constant, $\beta$ is the exponential index and $R$ is the distance (in kpc) from the center, corrected for inclination. The exponential fit for all the targets is shown as a solid red line in Fig. \ref{fig:profiles_all}. The exponential index for most of these targets lie between 0.1-0.2 as reported in table \ref{table:results}. Such an exponential function fit gives the least $\chi^{2}$ estimate compared to other models such as an inverse-square drop with distance. 

It is apparent from the profiles that in most of the cases,the electron density inferred from the ratio of the total flux of the individual emission lines of the \sii ~doublet drops to $<$50 cm$^{-3}$ for distances larger than 1 kpc from the center. \oiii ~emitting regions in targets such as NGC 1365 (Fig. \ref{fig:maps_NGC1365}) and NGC 613 (Fig. \ref{fig:density_maps_correlation}) also show low densities of $\leq$50 cm$^{-3}$. This is valid even in cases of outflowing targets NGC 1068 and NGC 2110 (Fig. \ref{fig:density_ngc1068}) where the density drops to the lower limits for distances larger than 1 kpc from the center. NGC 1266 is an exception as the presence of outflow seems to increase the density in one direction and we are limited by the S/N of the \sii ~doublet for distances larger than 1 kpc.  

The above results of the density profile are not necessarily true for density within the outflowing medium. This is clear from Fig. \ref{fig:density_broad_maps} where the density either increases or decreases from the center which is possible due to a turbulent outflowing medium and the presence of shocks.

\section{Discussion} \label{sect5}

One of the main motivation of this study is to constrain the density values in the NLR of AGNs traced by the \oiii ~emission line and how the presence of an ionized outflow affects the density values and its morphology in an attempt to reduce the systematic uncertainties in the estimation of mass outflow rates. Studies often assume a uniform density in the outflowing medium which leads to systematic errors spanning multiple orders of magnitude \citep[e.g.][]{veilleux05, cano-diaz12, genzel14, mcelroy15, husemann16, kakkad16, leung17}. Although a few studies have calculated electron densities based on the integrated \sii ~doublet ratio \citep[e.g.][]{brusa15a, perna15a}, the analysis in this paper highlights the complexity as well as the differences in the electron density values when comparing information from the integrated spectrum and a spatially resolved data set. 

For the AGNs presented this paper, we find a mean electron density of $\sim$160 cm$^{-3}$ calculated from the ratio of the total flux of each component of the \sii ~doublet in the spectrum binned over the entire S7 FOV. These binned values are consistent with a recent study by \citet{rupke17} who also find the spatially averaged value of electron density to lie between 50-410 cm$^{-3}$ with a median value of 150 cm$^{-3}$, in a sample of Type 1 radio quiet quasars at z<0.3. Compared to the integrated electron density values in the local star forming population derived from SDSS, DR7 sample in \citet{sanders16} and \citet{kaasinen17}, the electron density in the local AGN host galaxies are a factor of $\sim$5 higher which might suggest  that AGNs are increasing the density or that the high density gas is associated with triggering the AGNs in the first place. However, note that the above comparison between the density in local star forming population and the AGN host galaxies is only suggestive, as the physical scales sampled by S7 and SDSS spectra are different, and the S7 spectra might be biased to higher density present in the nuclear regions.

The mean electron density in the outflowing medium calculated from the flux ratio of the broad components of the \sii ~doublet in the binned spectrum for the targets showing ionized outflows is $\sim$700 cm$^{-3}$. The analysis on the integrated spectrum is therefore indicative of the electron density in the outflowing medium to be higher by a factor of $\sim$5 compared to the electron density in the non-outflowing medium. The presence of high electron density in the outflowing medium in AGN host galaxies has also been inferred in \citet{perna17} who reported enhanced densities of $\sim$1200 cm$^{-3}$ in the outflowing gas for a z$<$0.8 sample of AGN host galaxies selected from the SDSS survey using stacked spectrum. High density outflows >1000 cm$^{-3}$ have also been predicted indirectly in \citet{chamberlain15} for a Broad Absorption Line (BAL) Quasar using column density ratio of S{\sc iv} and S{\sc iv}*. However, as reported in Sect. \ref{sect4} spatially resolved maps of electron density in the outflowing medium shows that this scenario of dense ionized outflow is not always true.

Spatially resolved electron density maps for the targets not having an active ionized outflow suggests that the high-density sites for most of the targets correlate with the location of the \hii ~regions \citep[or star forming regions, e.g.][]{westmoquette11, westmoquette13, mcleod15}. This observation from the local star forming clumps reconciles well with recent results of \citet{shimakawa15} and \citet{kaasinen17} who report an evolution of the overall electron density with star formation rate for star forming galaxies, providing a link between local cloud-scale and global-scale properties within a galaxy. Also, the electron density in the NLR ($\lesssim$ 100 cm$^{-3}$) is much less than the density in the star forming regions ($\sim$200-600 cm$^{-3}$). This is very clear in case of NGC 1365 in Fig. \ref{fig:maps_NGC1365}. This might be a consequence of the gas density and SFR correlation given by Kennicutt-Schmidt relation \citep{kennicutt98} while the clumps in the extended-NLR traced by the \oiii ~emission is ionized by the AGN in an ionization cone. In targets such as NGC 5990 and NGC 4303, the correlation between gas density and SFR is absent which might be due to obscuration from the dust or the host galaxy itself or if the target went through an outflowing phase from the AGN or starburst previously \citep[e.g.][]{riffel16}.

However, in the presence of an outflow, the correlation between the high density and star forming sites ceases to exist for the targets presented in this paper. As mentioned earlier, most of the previous studies on spatially resolved electron density maps of AGN host galaxies showing an outflow using the \sii ~doublet \citep[e.g.][]{sharp10, westmoquette12, cresci15b, rupke17, freitas18} use the total flux from each component of the \sii ~doublet to calculate the density, as shown for outflowing sample in this paper in Fig. \ref{fig:density_ngc1068}. However, this has the limitation that one includes the contribution from the host galaxy as well as the outflowing medium, making it hard to infer the density within the outflow. Fig. \ref{fig:density_broad_maps} shows the density derived from the flux ratio of only the broad Gaussian components of the \sii ~doublet, which are believed to trace the density of the outflowing medium. In targets with an active outflow inferred from the broad \oiii ~profiles such as NGC 1068 and NGC 2110, the density can reached values > 1000 cm$^{-3}$ in certain locations supporting a dense outflow scenario as predicted from the integrated spectrum analysis described earlier. However, this is not true for the entire spatial extent of the outflowing medium as apparent from the density maps in Fig. \ref{fig:density_broad_maps}. The maps show that the density can be between < 50 cm$^{-3}$ to $\sim$2000 cm$^{-3}$ with high density clumps most likely arising as a result of a turbulent medium which gives rise to shocks and possibly a pressurized medium within the outflow. From the observations presented above, an outflowing medium therefore does not have a uniform density and rather shows a wider range in density compared to star forming regions which is in contrast with the uniform density assumption often used in literature for mass outflow rate estimations. Therefore, our results suggest that the outflow models should include variable electron density  instead of uniform density assumption.

Based on the electron density profiles of all the targets but NGC 1266, we do not expect very high densities for distances larger than 1 kpc from the central AGN. Since the density measurements from the \sii ~doublet ratio has the limitation of measuring densities upwards of 50 cm$^{-3}$, the maps constrain the density values to $\leq$50 cm$^{-3}$ in all the galaxies for r $\geq$ 1 kpc. Note that the S7 field of view only allows us to probe the inner few kpc of the host galaxies. If this trend continues, one might expect much lower densities in the outskirts of the galaxy. Assuming that a trend in the drop in electron density observed within the S7 FOV continues across the entire scale of the galaxy, we would expect the electron density to remain in the lower limit of < 50cm$^{-3}$ in the outskirts of the galaxy, which can be probed by instruments with wider FOV such as MUSE. This, however, might not be true in case of high redshift galaxies as discussed later. The drop is density profile is absent in one outflowing target NGC 1266 which could be attributed to a combination of high nuclear obscuration and the presence of shocks from outflows in multiple phases \citep[e.g.][]{alatalo15, glenn15}. Also, we are limited by the S/N ratio of the \sii ~to probe the density in the outer regions in NGC 1266. As mentioned earlier in Sect. \ref{sect4.3}, we did not construct such profiles for the outflowing medium as there is no trend as apparent from Fig. \ref{fig:density_broad_maps}.

For galaxies at high redshift, the results obtained in this paper might not hold true due to the changing ISM conditions. Observations and theoretical models over the past decade have shed light into the history of star formation and ISM gas and dust content as a function of cosmic time \citep[e.g.][]{silk12, madau14, popping17}.  It is now accepted that the volume averaged star formation rate was at its maximum in the universe at 1$<z<$3 and that the gas content is also higher by an order of magnitude at this epoch \citep[e.g.][]{scoville14}. Taking into account the evolution of star formation rate along with our observation that the higher density sites spatially correlate with the \hii ~regions, we would expect the average electron density to also increase as a function of redshift. As mentioned earlier, this has been true in case of star forming galaxies in recent papers by \citet{shimakawa15}, \citet{sanders16}, \citet{kaasinen17} who show an increase in density by a factor of $\sim$5-10 when compared with the average electron density of the local samples. 

Observations have shown that high redshift galaxies have the potential to host more powerful and extended AGN-driven outflows compared to the local counterparts due to higher mass accretion rates. Therefore, the conditions of the outflowing medium might be significantly different from the galaxies presented in this paper. Current observational capabilities does not allow us to have enough spatial resolution to do a similar study at redshift higher than $\sim$1. With upcoming facilities like James Webb Space Telescope (JWST) and the Narrow Field Mode (NFM) on board MUSE/VLT, it would be possible to probe the ISM conditions within AGN host galaxies to much wider redshift space.

\section{Summary and Conclusions}      \label{sect6}

In this paper, we presented spatially resolved electron density maps of a sample of low redshift AGN host galaxies derived from the S7 survey using the optical \sii$\lambda$6716, 6731 diagnostic lines which trace the gas in the warm ionized phase. Being a very crucial parameter in the estimation of physical quantities such as the mass outflow rates and energy associated with outflows, we have attempted to understand how the values of electron density and it's morphology is affected by the presence of AGN ionization, star formation or outflows. Following points summarize the main results of this paper.

\begin{itemize}
\item[1.] We find a mean electron density of $\sim$160 cm$^{-3}$ calculated using the ratio of the total flux of each component of the \sii ~doublet for the low redshift AGN sample presented in this paper. This value is consistent with recent literature results on electron density calculated from integrated spectrum for low redshift AGNs.

\item[2.] For AGN host galaxies which do not show any presence of an ionized outflow, the electron density maps spatially correlate with narrow H$\alpha$ maps tracing star formation. For these non-outflowing targets, we find that the electron density in the NLR ($\lesssim$ 100 cm$^{-3}$) is less compared to the density in the star forming regions ($\sim$200-600 cm$^{-3}$). Such observations on small-scale reconciled well with recent results suggesting the evolution of electron density with star formation from integrated spectrum at high redshift.

\item[3.] We find a non-uniform distribution of electron density in the outflowing medium calculated from the flux ratio of the broad component of the \sii ~doublet with values ranging from $\lesssim$50 cm$^{-3}$ to $\gtrsim$2000 cm$^{-3}$. Our results therefore suggest the need to include variable electron density within the outflowing medium instead of the uniform density assumption. Unlike the non-outflowing targets, the high density does not show a correlation with the star forming sites for the targets presented in this paper.

\item[4.] Radial density profiles of most of the AGN host galaxies suggest a drop in density to <50 cm$^{-3}$ for distances larger than 1 kpc from the central AGN. The radial density profiles are best fit with an exponential function with a mean exponential index of $\sim$0.15. On the other hand, the density within the outflowing medium do not show any common pattern, most likely due to the presence of shocks arising out of a turbulent medium.
\end{itemize}

Due to different ISM conditions at high redshift i.e. AGNs being more luminous and powerful and extended outflows are more common, the results of the current work may not hold true at the high redshift. Therefore, it is essential to have a spatially resolved electron density study for galaxies at high redshift as well, where the effect of radiation pressure on the surrounding ISM seems dominant due to the peak activity of accretion rate of black holes. With upcoming instruments such as the Narrow Field Mode on MUSE and Near-Infrared Spectrograph (NIRSpec) and High Angular Resolution Monolithic Optical and Near-infrared Integral field spectrograph (HARMONI) on board James Webb Space Telescope (JWST) and European Extremely Large Telescope (E-ELT) respectively which will provide an unprecedented angular resolution, such a study can be extended to wider redshift space. While WiFeS is ideal to probe the central few-kpc of the host galaxies in nearby AGNs, future studies with MUSE with it's wide FOV will be able to probe the outskirts of these low redshift galaxies as well. 

\noindent
{\it Acknowledgements:} The authors thank the anonymous referee for the very constructive comments to improve the paper. B.G. acknowledges the support of the Australian Research Council as the recipient of a Future Fellowship (FT140101202). MD acknowledges the support of the Australian Research Council (ARC) through Discovery project DP16010363. Parts of this research were conducted by the Australian Research Council Centre of Excellence for All Sky Astrophysics in 3 Dimensions (ASTRO 3D), through project number CE170100013.

\bibliographystyle{aa}
\bibliography{reference.bib}

\begin{appendix}

\section{S7 spectra}           \label{appendix1}
\begin{figure*}
\centering
\subfloat{\includegraphics[width=4.5cm, height=4cm]{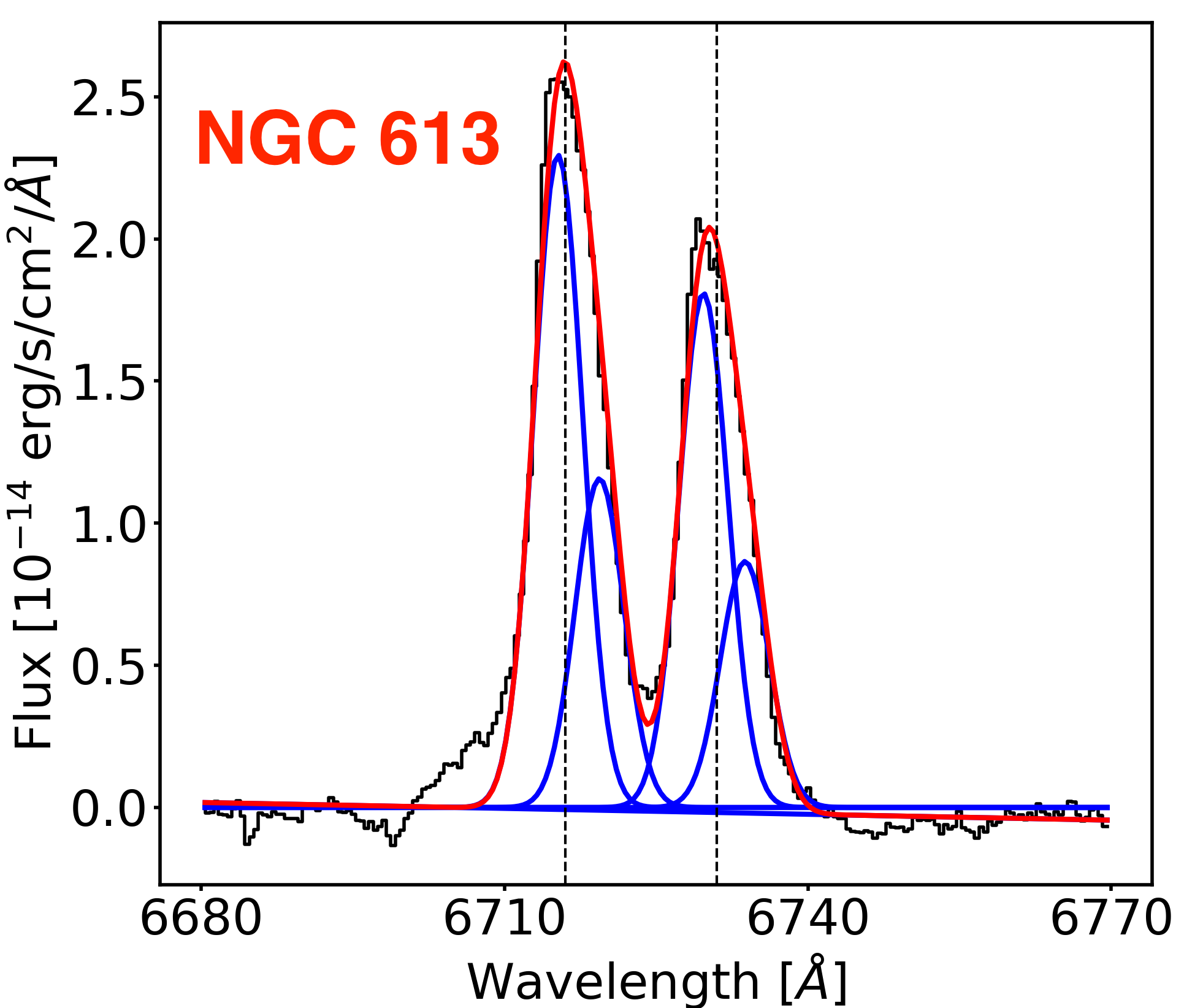}}
\subfloat{\includegraphics[width=4.5cm, height=4cm]{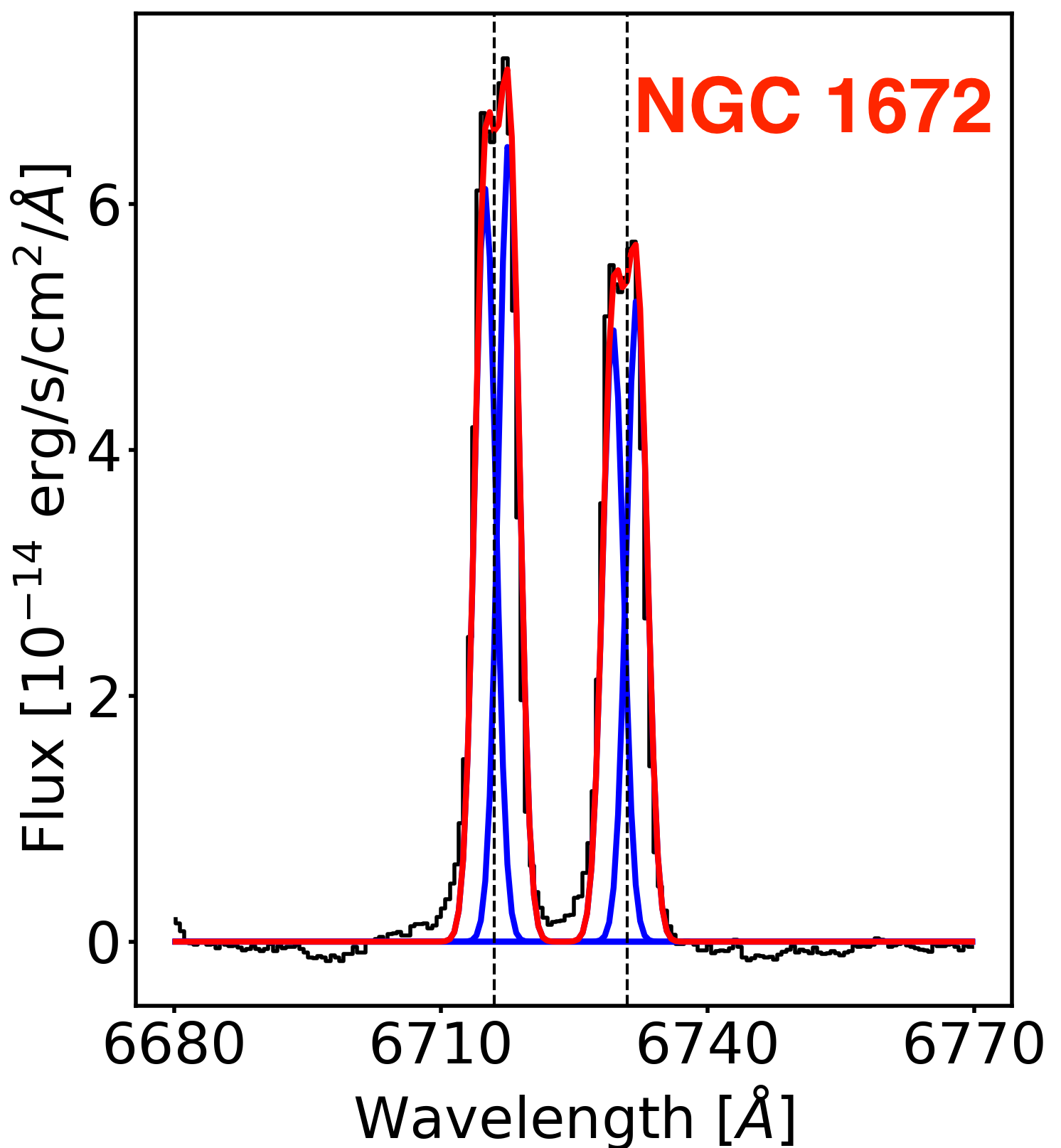}}
\subfloat{\includegraphics[width=4.5cm, height=4cm]{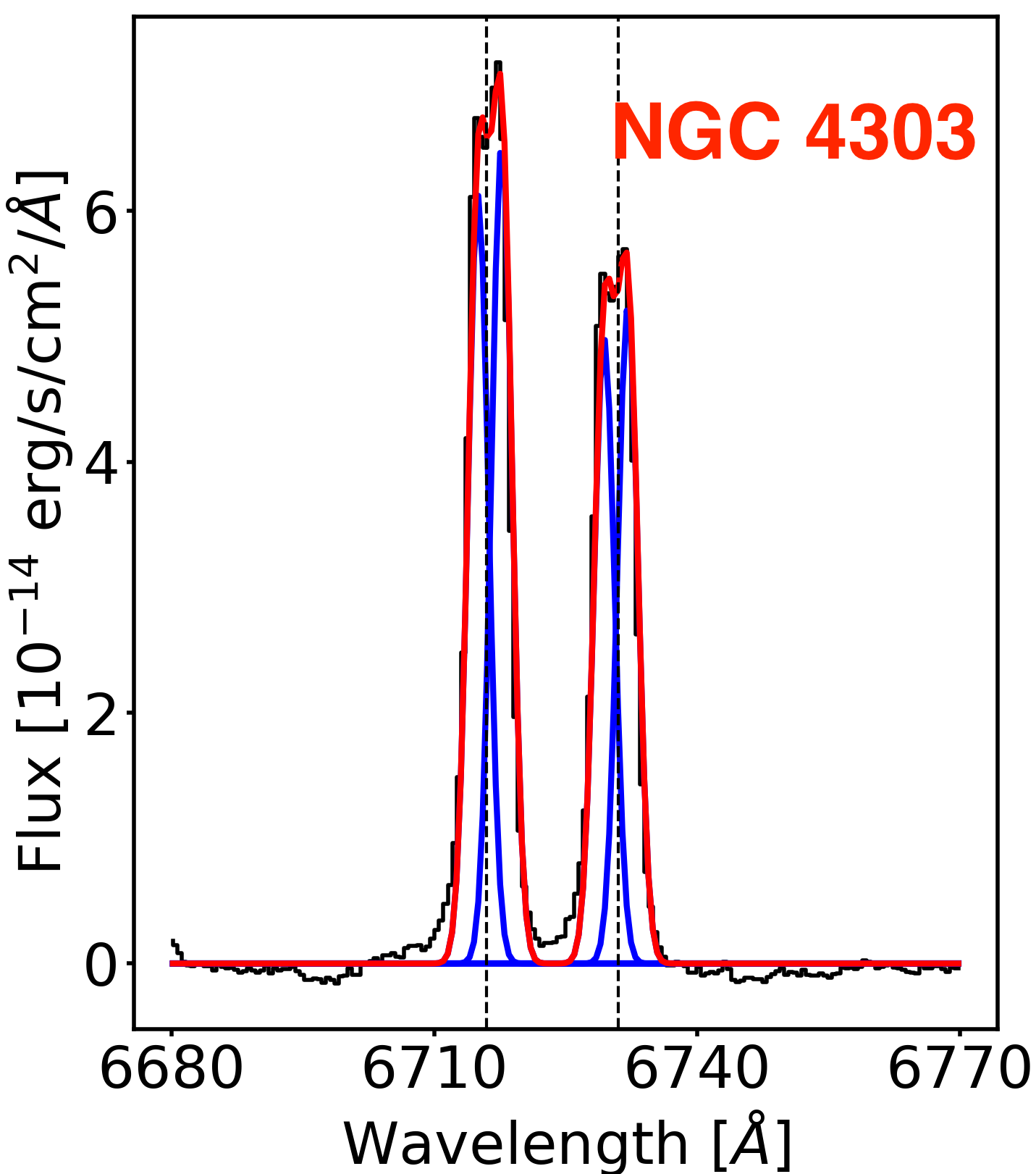}}\\
\subfloat{\includegraphics[width=4.5cm, height=4cm]{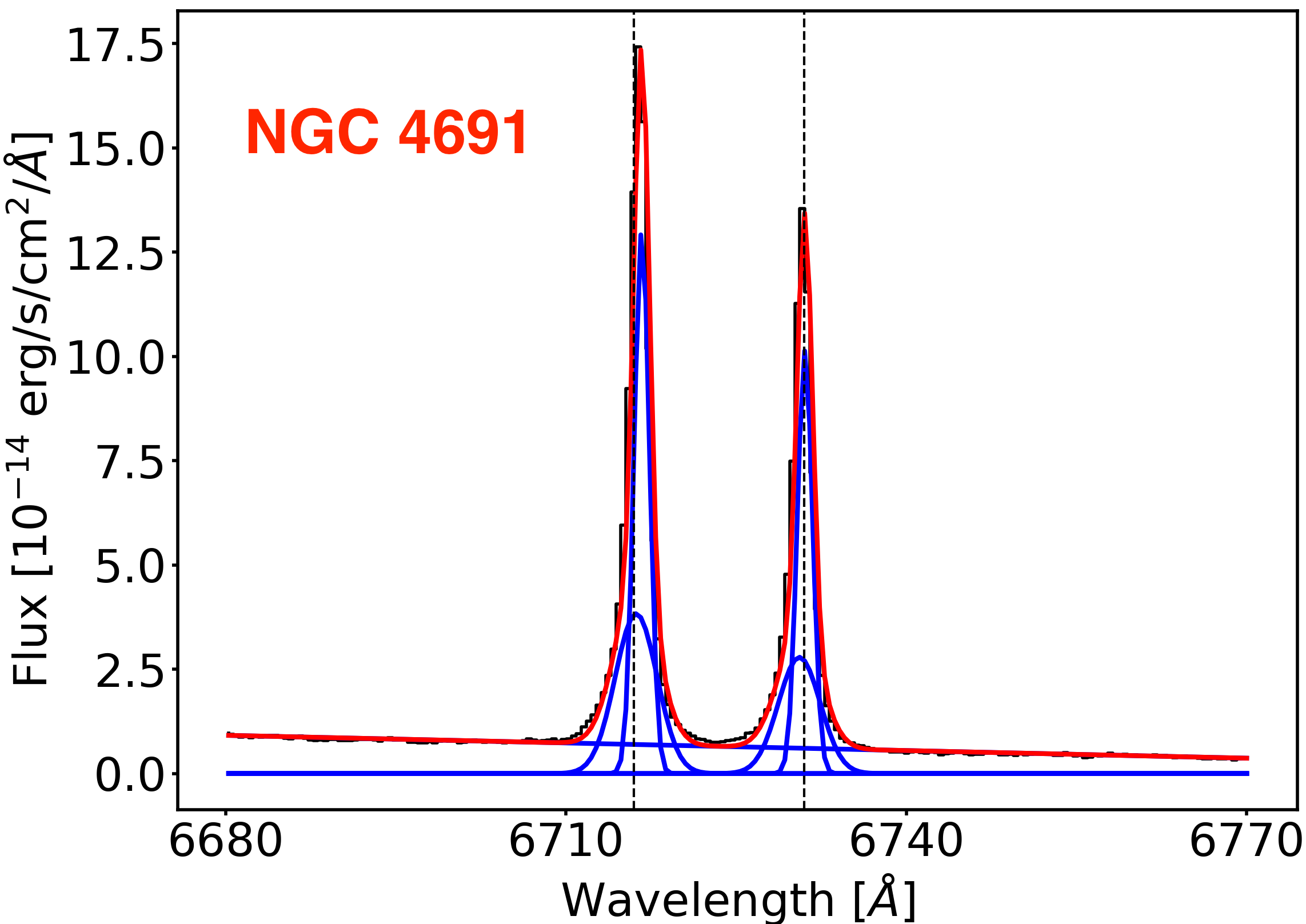}}
\subfloat{\includegraphics[width=4.5cm, height=4cm]{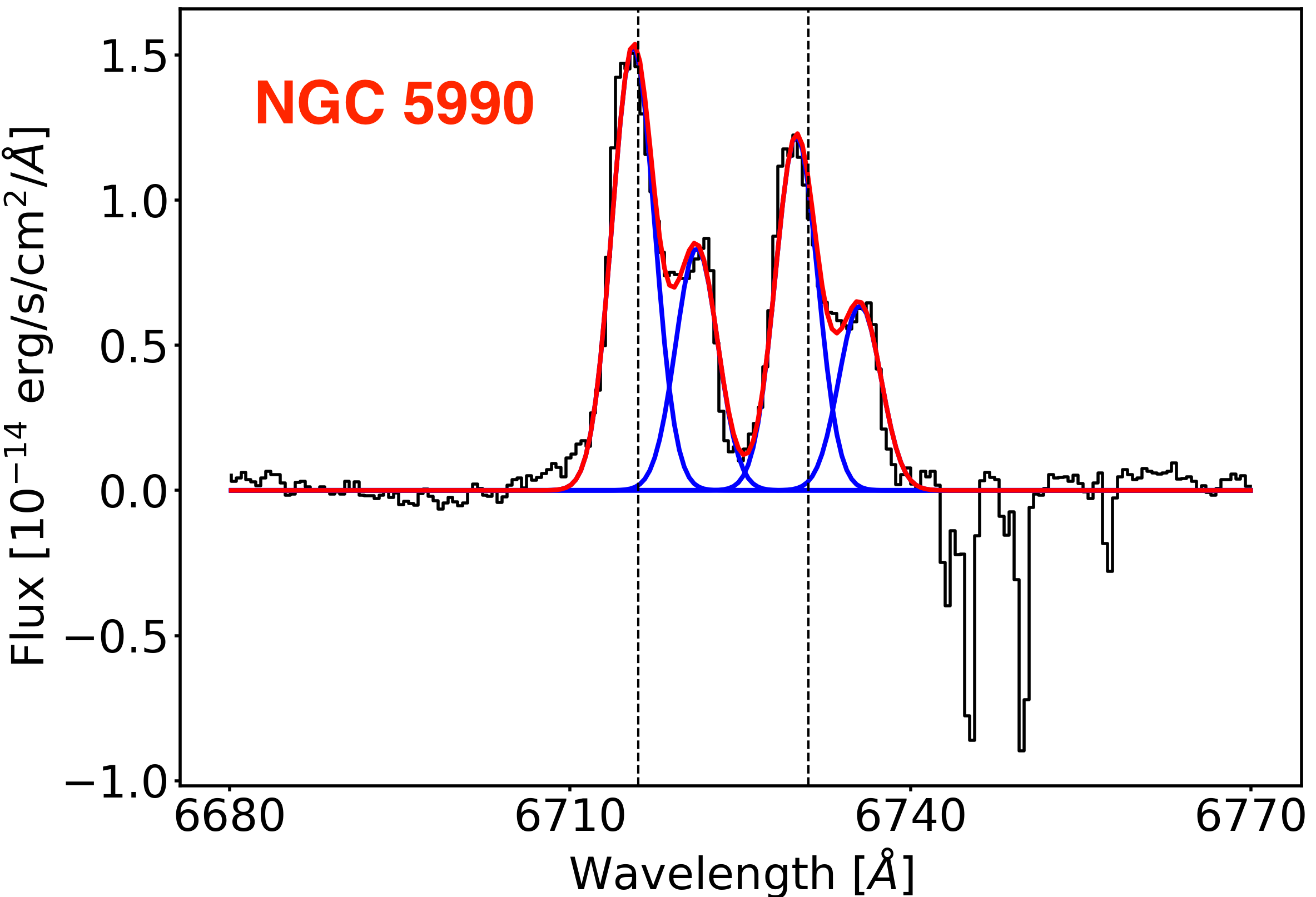}}
\subfloat{\includegraphics[width=4.5cm, height=4cm]{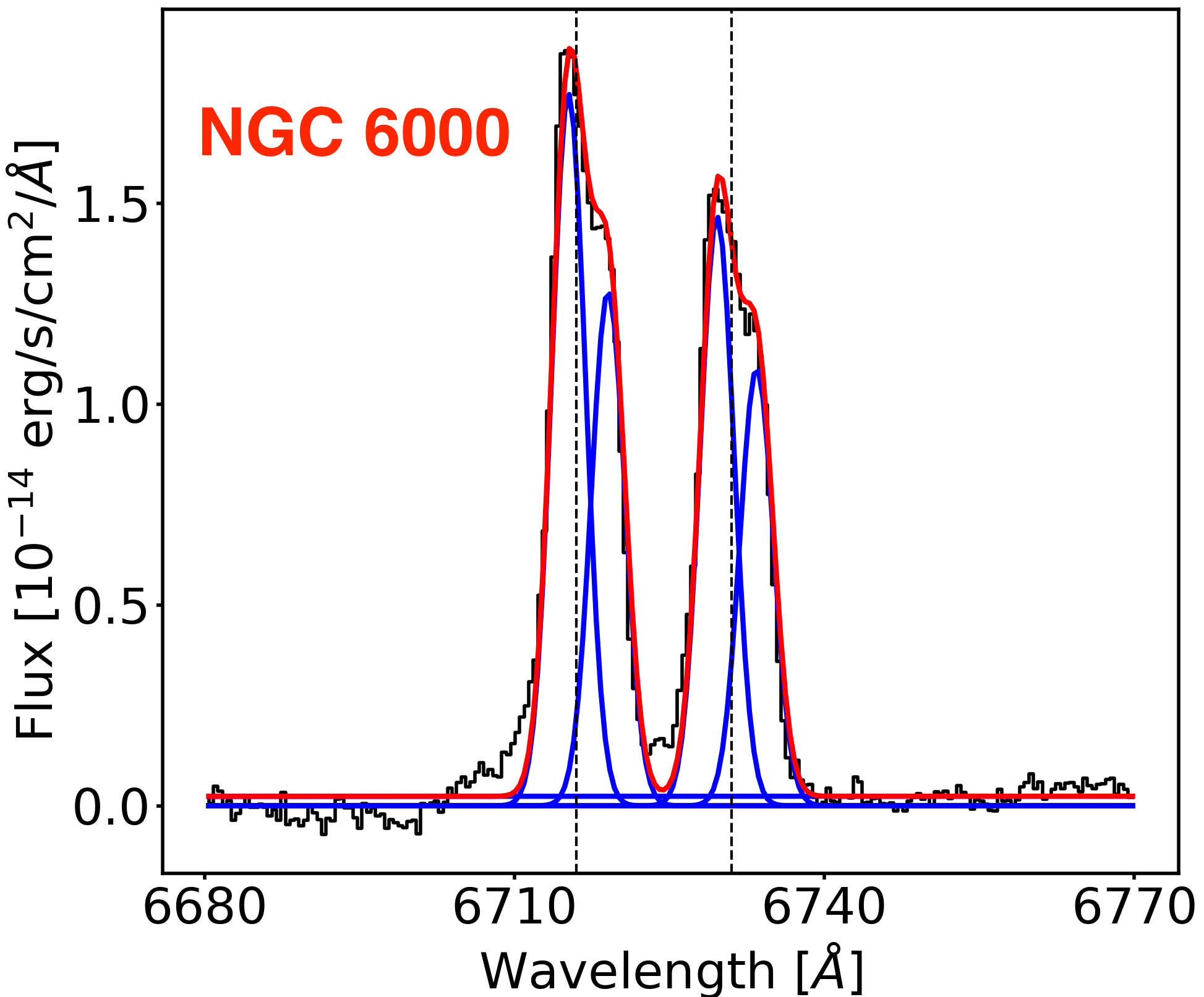}}\\
\subfloat{\includegraphics[width=4.5cm, height=4cm]{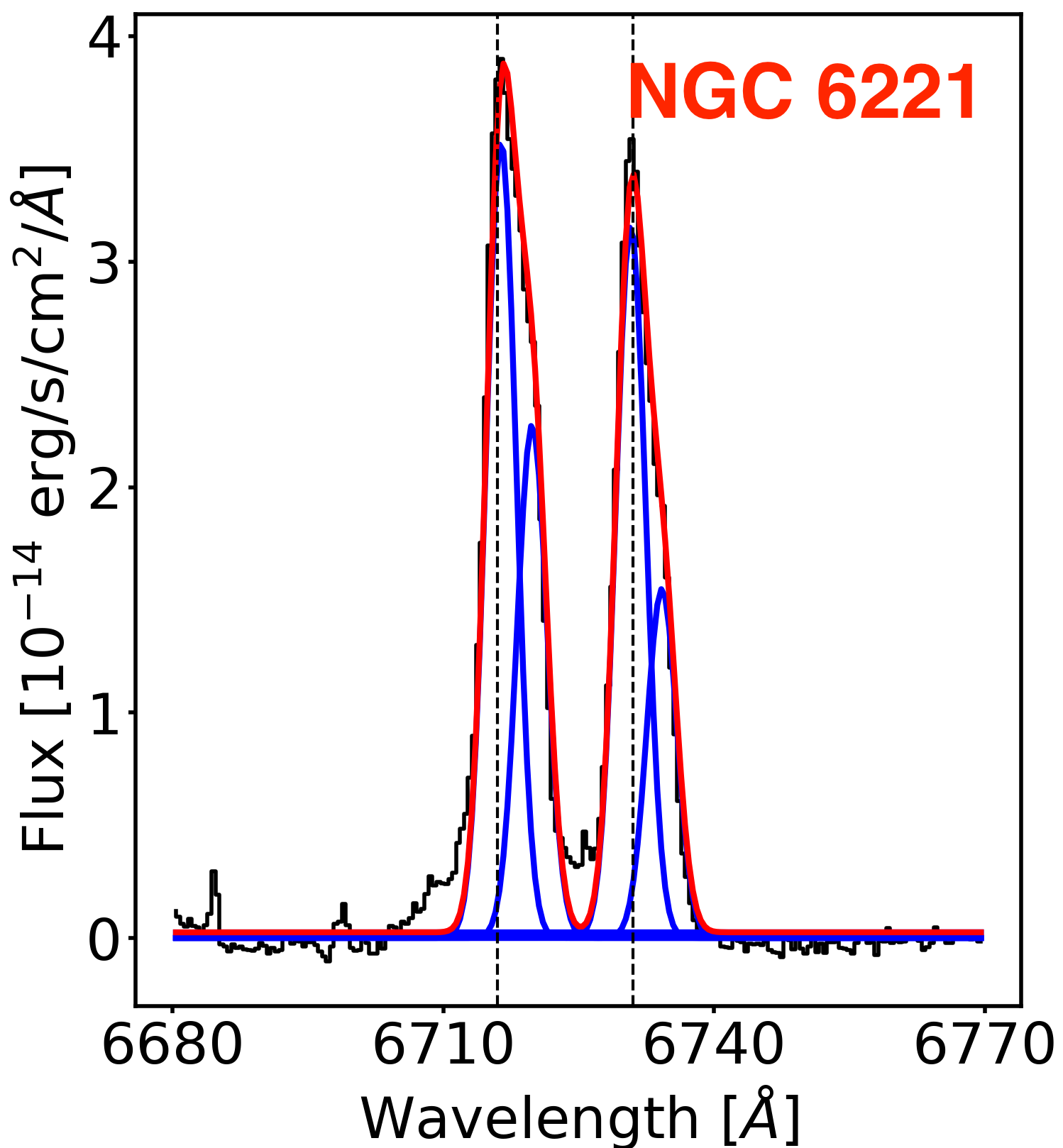}}
\subfloat{\includegraphics[width=4.5cm, height=4cm]{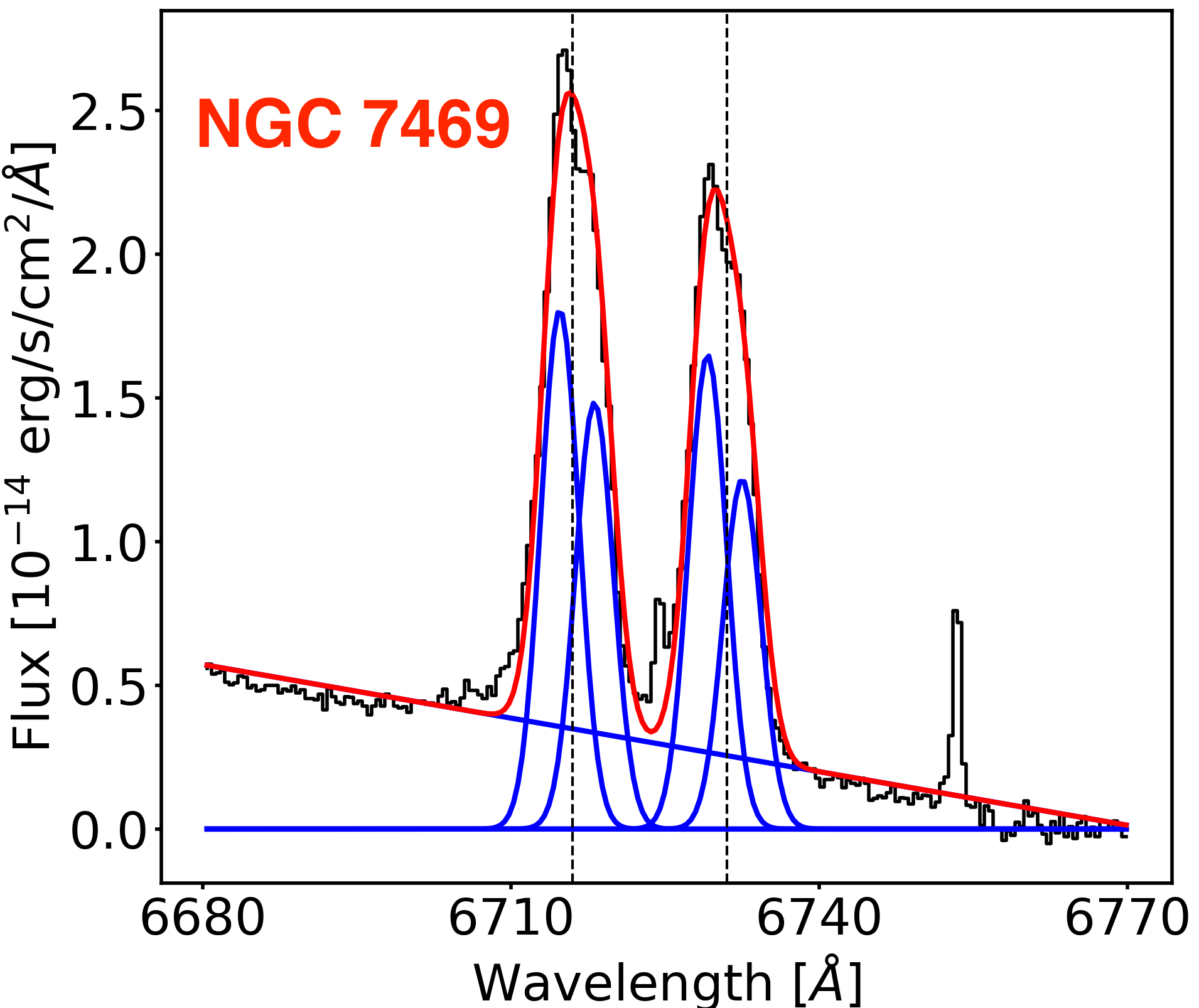}}
\subfloat{\includegraphics[width=4.5cm, height=4cm]{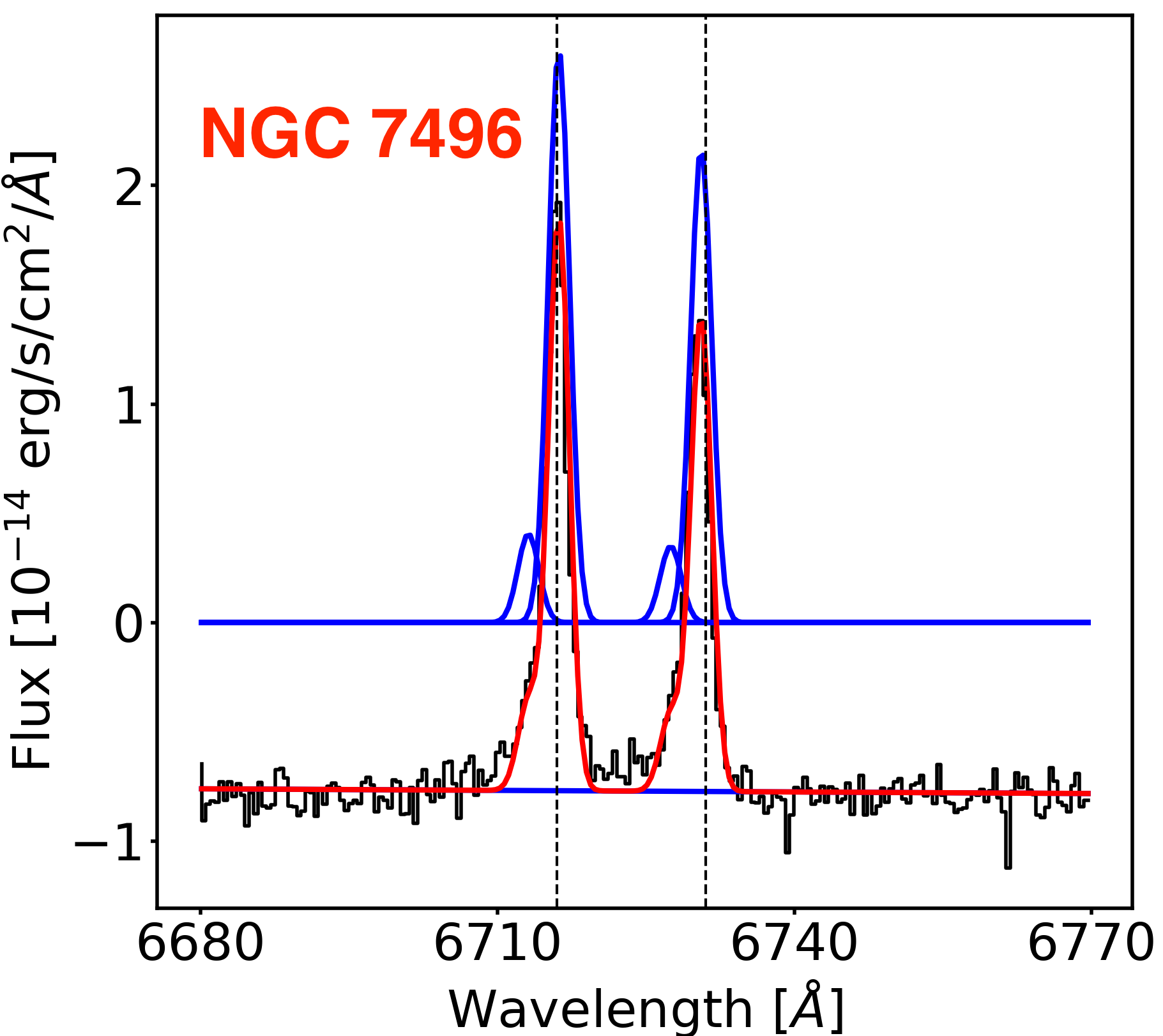}}\\
\caption{Spectra showing \sii ~doublet and the line fitting for the targets presented in this paper.}
\end{figure*}
\end{appendix}

\end{document}